\documentclass[crop]{CSL}%%%%where CSL is the template name

\jname{\textbf{Authors' accepted manuscript.}\\
Accepted for publication in the
International Journal of Astrobiology
on 2nd November 2026.}%

\usepackage{amsmath} %for dealing with mathematics,
\usepackage{amsthm} %for dealing with theorem environments,
\def\Gyr{\mathrm{Gyr}}
\def\Myr{\mathrm{Myr}}
\def\yr{\mathrm{yr}}
\def\mg{\mathrm{mg}}
\def\gram{\mathrm{g}}
\def\kg{\mathrm{kg}}
\def\Msun{M_{\odot}}
\def\um{\mu\mathrm{m}}
\def\cm{\mathrm{cm}}
\def\meter{\mathrm{m}}
\def\km{\mathrm{km}}
\def\pc{\mathrm{pc}}
\def\kpc{\mathrm{kpc}}
\def\sec{\mathrm{s}}
\def\Kelv{\mathrm{K}}
\def\kms{\km\ \sec^{-1}}

\DeclareMathOperator{\sgn}{sgn}

\def\bOpt{b_{\mathrm{opt}}}
\def\betaOpt{\beta_{\mathrm{opt}}}
\def\etaBeta{\eta_{\beta}}
\def\etaIncl{\eta_{\ell}}
\def\etaTheta{\eta_{\theta}}
\def\FluxEnc{F_{\mathrm{enc}}}
\def\Fluxinf{F_{\infty}}
\def\FluxRec{F_{\mathrm{rec}}}
\def\iinf{\ell_{\infty}}
\def\iinfmax{\ell_{\infty,\mathrm{max}}}
\def\mRef{\bar{m}_G}

\def\nEncSlow{n_{e,\mathrm{slow}}}
\def\ninf{n_{\infty}}
\def\nOpt{n_{\mathrm{opt}}}
\def\Qrad{Q_{\mathrm{rad}}}           % This is the former \eta_opt, where "Q" is in line with Draine (2011), chapter 22. Additionally, I use "rad" instead of "opt" because "opt" might be confused with optimum.
\def\rEarth{r_{\oplus}}               % Position of Earth, not its radius
\def\rEarthVec{\mathbf{r}_{\oplus}}
\def\rhoinf{\rho_{\infty}}
\def\rhoOpt{\rho_{\mathrm{opt}}}
\def\tint{t_{\mathrm{int}}}
\def\thetaEarth{\theta_{\oplus}}
\def\thetaOpt{\theta_{\mathrm{opt}}}
\def\vEarth{v_{\oplus}}
\def\vEarthVec{\mathbf{v}_{\oplus}}
\def\vEnc{v_{\mathrm{enc}}}
\def\dvEncMax{\Delta v_{\mathrm{enc,max}}}
\def\vEncTan{v_{\mathrm{enc,tan}}}
\def\vEncVec{\mathbf{v}_{\mathrm{enc}}}
\def\vinf{v_{\infty}}
\def\xNorm{\hat{\mathbf{x}}}
\def\yNorm{\hat{\mathbf{y}}}
\def\zNorm{\hat{\mathbf{z}}}

\def\iinfOLD{i_{\infty}}

\newcommand{\Mean}[1]{{\langle #1 \rangle}}

\usepackage{rotating}  % for sidewaystable

\usepackage{graphicx} % should already be loaded by CSL, but harmless

\usepackage{ragged2e} % for \justifying

\usepackage{tikz}

\usepackage{hyperref} % for \nameref

\newcommand{\doi}[1]{%
  \href{https://doi.org/#1}{doi:\,\nolinkurl{#1}}%
}
\hypersetup{breaklinks=true}

\newif\ifdraftnumbers
\draftnumberstrue % switch to \draftnumbersfalse later

\newcommand{\secref}[1]{%
  \ifdraftnumbers
    Section~\ref{#1}%
  \else
    \textit{\nameref{#1}}%
  \fi
}
\newcommand{\secreftwo}[2]{%
  \ifdraftnumbers
    Sections~\ref{#1} and~\ref{#2}%
  \else
    \textit{\nameref{#1}} and \textit{\nameref{#2}}%
  \fi
}

\ifdraftnumbers

\makeatletter
\renewcommand\thesection{\arabic{section}}

\def\@seccntformat#1{\csname the#1\endcsname\quad}

\def\@sect#1#2#3#4#5#6[#7]#8{%
  \ifnum #2>\c@secnumdepth
    \let\@svsec\@empty
  \else
    \refstepcounter{#1}%
    \protected@edef\@svsec{\csname the#1\endcsname\quad}%
  \fi
  \@tempskipa #5\relax
  \ifdim \@tempskipa>\z@
    \begingroup
      #6\relax
      \@hangfrom{\hskip #3\relax\@svsec}{\interlinepenalty \@M #8\par}%
    \endgroup
    \csname #1mark\endcsname{#7}%
    \addcontentsline{toc}{#1}{%
      \ifnum #2>\c@secnumdepth \else \protect\numberline{\csname the#1\endcsname}\fi
      #7}%
  \else
    \def\@svsechd{%
      #6\hskip #3\relax\@svsec #8%
      \csname #1mark\endcsname{#7}%
      \addcontentsline{toc}{#1}{%
        \ifnum #2>\c@secnumdepth \else \protect\numberline{\csname the#1\endcsname}\fi
        #7}}%
  \fi
  \@xsect{#5}}
\makeatother

\fi

\newif\ifreview
\reviewfalse     % clean version

\usepackage{xcolor}
\usepackage[normalem]{ulem}

\ifreview
  \newcommand{\added}[1]{{\color{blue}#1}}
  \newcommand{\deleted}[1]{{\color{red}\ifmmode [\textit{Deleted:}~#1]\, \else \footnotesize \itshape[Deleted: #1] \fi}}
  \newcommand{\changed}[2]{{\color{gray}[{\ifmmode \textit{Changed:} \else \footnotesize \itshape{Changed:} \fi}} {\color{red} #1} {\color{gray} \ifmmode \rightarrow \else $\rightarrow$ \fi} {\color{blue}#2}{\color{gray}]}}
  \newcommand{\shortchanged}[2]{{\color{blue}#2}}
\else
  \newcommand{\added}[1]{#1}
  \newcommand{\deleted}[1]{}
  \newcommand{\changed}[2]{#2}
  \newcommand{\shortchanged}[2]{#2}

\fi

\usepackage[most]{tcolorbox}

\newtcolorbox{briannote}{
  enhanced,
  breakable,
  colback=yellow!20,
  colframe=yellow!60!black,
  coltext=black,
  boxrule=0.7pt,
  arc=1mm,
  left=5pt,
  right=5pt,
  top=4pt,
  bottom=4pt,
  before skip=8pt,
  after skip=8pt,
  fonttitle=\bfseries,
  fontupper=\small,
  title={Query for Brian}
}

\newtcolorbox{resolvedbriannote}{
  enhanced,
  breakable,
  colback=yellow!10,
  colframe=yellow!30!black,
  coltext=yellow!60!black,
  boxrule=0.7pt,
  arc=1mm,
  left=5pt,
  right=5pt,
  top=4pt,
  bottom=4pt,
  before skip=8pt,
  after skip=8pt,
  fonttitle=\bfseries,
  fontupper=\small,
  title={RESOLVED Query for Brian}
}

\newtcolorbox{notefrombrian}{
  enhanced,
  breakable,
  colback=magenta!20,
  colframe=magenta!60!black,
  coltext=black,
  boxrule=0.7pt,
  arc=1mm,
  left=5pt,
  right=5pt,
  top=4pt,
  bottom=4pt,
  before skip=8pt,
  after skip=8pt,
  fonttitle=\bfseries\small,
  fontupper=\small,
  title={Query for Lewis from Brian}
}

\usepackage{enumitem}

\makeatletter
\renewcommand{\address}[1]{%
  \gdef\@address{%
    \leftskip=\parindent
    \parindent=0pt
    \noindent\ignorespaces#1\par
  }%
}
\makeatother

\begin{document}

\supertitle{Research Paper}

\title[]{Micron-Scale Technosignatures: How a Cubic Metre
of Lunar Regolith May Begin to Constrain the Number of Past Technological Civilisations in the Galaxy}

\author[]{Lewis J. Pinault$^{1,2,3}$, Brian C. Lacki$^{4}$, Ian A. Crawford$^{2,3}$, and Andrew P. V. Siemion$^{1,4,5,6,7}$}

\address{
\add{1}{SETI Institute, 339 Bernardo Ave, Mountain View, CA 94043, USA}
\add{2}{School of Natural Sciences, Birkbeck College, London WC1E 7HX, UK}
\add{3}{Centre for Planetary Sciences, UCL/Birkbeck, Gower Street, London WC1E 6BT, UK}
\add{4}{Breakthrough Listen, Department of Physics, Denys Wilkinson Building, Keble Road, Oxford OX1 3RH, UK}
\add{5}{Department of Astronomy, University of California Berkeley, Berkeley, CA 94720, USA}
\add{6}{Department of Physics and Astronomy, University of Manchester, Oxford Road, Manchester M13 9PL, UK}
\add{7}{Institute of Space Sciences and Astronomy, University of Malta, Msida MSD 2080, Malta}
}

\corres{\name{Lewis J. Pinault} \email{lpinault@setiap.org}}

\begin{abstract}
Micron-scale technosignatures provide a largely unexplored route to detecting extraterrestrial technological activity. Building on Arkhipov’s proposal that technogenic artefacts may survive natural interstellar transport and accumulate on \deleted{airless Solar System bodies}\added{Solar System surfaces}, we examine the prospects for identifying engineered particulate material within the lunar regolith. We do not claim that artificial particles of non-human origin exist on the Moon, nor do we report evidence of their existence. But whereas electromagnetic SETI is restricted to contemporaneous transmitters, searches for artificially generated particles integrate over gigayear timescales and are therefore sensitive to technomaterial released by civilisations long extinct. The Moon’s minimal geochemical alteration and $\sim4\,\Gyr$ exposure history make it an exceptionally stable long-duration collector of such material. We analyse the transport of micron and submicron grains through the interstellar medium, including gas drag, sputtering, and ISM phase-dependent survival, and show that refractory particles with characteristic radii of order $0.3\,\um$ may traverse kiloparsec scales over residence times of $0.1$--$1\,\Gyr$. Solar radiation pressure and heliospheric filtering define a dynamically constrained slow-arrival channel in which a small but non-zero fraction of grains reach the Earth-Moon system at relative velocities compatible with partial or intact survival upon impact. Combining these transport properties with regolith-mixing constraints yields quantitative upper limits on the cumulative undirected technomaterial output of large-scale spacefaring civilisations: a null detection in a cubic metre of regolith excludes scenarios in which Solar-type stars typically disperse more than \changed{\ensuremath{\sim0.09\,M_{\oplus}}}{\ensuremath{\sim0.1\,M_{\oplus}}} of long-lived artificial particulate debris over Galactic history. Deliberate targeting of the inner Solar System with artificial particulate matter defines a complementary regime characterised by the visitation frequency and deposited mass of such releases, for which the probabilities of detection may be orders of magnitude higher. We outline a multi-modal detection strategy integrating machine-vision triage with laboratory forensic techniques to identify anomalous grains within a well-characterised natural background. Particulate technosignatures thus establish an experimentally accessible form of exo-archaeology, capable of placing meaningful constraints -- and, in favourable cases, yielding direct material evidence -- on the technological history of the Galaxy.
\end{abstract}

%BCL (2026-03-05): Originally, the comment (%) started immediately
% after "\keywords", preventing compilation. I put it after an empty
% argument {} instead.
%Technosignatures 1, SETA 2, SETI 4, Interstellar Dust Particles 3, Extraterrestrial Materials 4, Extraterrestrial Artefacts 5, Spacecraft Debris 6, Moon 7}

\maketitle
\Fpagebreak

\section{Introduction}
\label{sec:introduction}

The prevalence, or otherwise, of advanced technological entities in the Universe remains one of the major uncertainties in astrobiology (for discussion, see, e.g., \citealt{Webb2015}, \citealt{Cirkovic2018}, and \citealt{CrawfordSchulzeMakuch2024}, and references therein). One way to address, or at least constrain, this uncertainty is to direct the search for extraterrestrial intelligence (SETI) toward the discovery of material artefacts within the Solar System. 
\added{While SETI has traditionally focused on electromagnetic signals from
technological civilisations, the field has broadened into the search for
technosignatures: observable manifestations of technology spanning a wide range of spatial and temporal scales \citep{LingamLoeb2021,WrightSETI2026}.
Physical artefacts provide a complementary observational channel because they may survive long after the originating civilisation has ceased transmitting} \citep{HaqqMisraKopparapu2012,HaqqMisra2022}.

Early discussions of such searches focused primarily on macroscopic objects. \citet{Clarke1951} introduced the idea of a beacon on the Moon awaiting humanity’s first forays into space, while \citet{Bracewell1960} suggested scrutinising the Solar System for a population of radio-transmitting probes hardened against radiation and impacts. \citet{Sagan1963} proposed that forthcoming orbital surveys of the Moon might reveal bases or installations, particularly on its farside, and noted that such structures could be automated.

Following the Apollo missions, \citet{Foster1972} offered a detailed assessment of where intentional installations—such as robotic devices (``automata''), monitoring stations, monuments, or beacons—might plausibly be located in the Solar System, with particular emphasis on the Moon. \citet{FreitasValdes1985} subsequently introduced the term SETA (the Search for Extraterrestrial Artefacts), emphasising that co-orbital objects sharing Earth's orbital period and circumlunar objects of possible extraterrestrial origin might be identified through systematic surveys. With the advent of high-resolution lunar orbital imaging, \citet{DaviesWagner2013} advocated the use of such datasets to search for anomalous surface features potentially indicative of artificial origin. These early SETA efforts collectively framed the search for extraterrestrial artefacts primarily in terms of deliberately emplaced installations within the Solar System.

In the 1990s \citet[and subsequent works]{Arkhipov1993,Arkhipov1994a} contributed a critical insight. While similarly focussed on macroscopic artefacts and deliberate arrivals in the Earth-Moon system, Arkhipov also suggested that micron-scale particles of technological origin could be passively transported through the Galaxy as interstellar dust and accumulate on inert surfaces such as the Moon.

These fragments -- here termed \emph{Arkhipov Particles} (APs) -- could arise
as the unintended by-products of spacefaring activity, in a manner broadly analogous to Earth's orbital debris. In this way, Arkhipov reframed the search for technosignatures in material terms, treating \deleted{airless bodies} \added{Solar System surfaces} as long-duration collectors of interstellar technological artefacts \deleted{\citep{Arkhipov1994b,Arkhipov1996,Arkhipov1998,Crawford2006}}\added{\citep{Arkhipov1994b,Arkhipov1995,Arkhipov1996,Arkhipov1998,Crawford2006,CrawfordPinault2023}}.
\added{We adopt Arkhipov's reframing here, and note that the microscopic scale of possible search targets need not correspond to the original scale of the source technology. Some candidate particles may have been manufactured and transported directly as grains, whereas others may represent the surviving fragments of larger probes, thin films, sails or other engineered structures subsequently disrupted by impact and prolonged regolith processing. Our objective is therefore not to infer a particular original spacecraft architecture, but to identify durable microscopic materials that may persist as the archaeological residue of technological activity after extended physical and geological evolution.} 

While \citet{Foster1972} had acknowledged that tools, materials, or waste products might accompany extraterrestrial exploration, he argued that advanced spacefaring civilisations would tend to minimise the uncontrolled abandonment of large quantities of propellant, structural material, or hardware; the emphasis remained firmly on deliberate, macroscopic installations as the primary SETA targets.

Arkhipov effectively reversed this logic. Directly contrasting Foster’s analysis, he argued that residual material—rather than intentional construction—might provide the most durable and widely distributed extraterrestrial artefacts, precisely because such by-products are generated inevitably and redistributed by natural dynamical processes beyond the control of their makers.
Subsequent progress in exoplanet discovery \citep[e.g.][]{WinnFabrycky2015,ZhuDong2021}, astrophysical modelling of Galactic dust dynamics \citep{Sterken2019}, and direct measurements of interstellar dust entering the Solar System \citep{Sterken2023,Totani2023} strengthen the plausibility of Arkhipov’s hypothesis. It is now well established that micron-scale grains can survive transport through the interstellar medium and reach the inner Solar System \citep[e.g.][]{Taylor1996,Landgraf2000,Sterken2019,Sterken2023}.

In this paper, we examine
the potential sources of APs and the mechanisms by which such particles may reach and be preserved on the lunar surface. We also consider a more speculative extension, clearly distinguished from the non-intentional case: micron-scale artefacts with functional intent, including passive logging, environmental sensing, or constrained self-replication, deliberately seeded across interstellar distances. Such objects may be viewed as microscopic analogues of Bracewell probes and are referred to here as \emph{Bracewell Particles} (BPs) \changed{\citep{Pinault2014APEX,PinaultCrawford2015ELS,Pinault2024a,Pinault2025a}}
    {\citep{Pinault2014APEX,PinaultCrawford2015ELS,Pinault2024a}}. \added{Where such particles incorporate organisms, biological precursors or encoded biological information, they would also overlap with the longstanding concept of directed panspermia \citep{GinsburgLingam2021}, although neither biological construction nor the propagation of life is required by our definition of a BP.} This distinction allows us to treat non-intentional technogenic debris and deliberately engineered particulate probes within a common physical framework, while maintaining a clear separation of assumptions.

The discovery of micron-scale technosignatures on the Moon -- particularly BPs or their diagnostic residues -- would have implications for the longstanding Hart-Tipler conjecture \citep[e.g.][]{Hart1975,Tipler1980}, which argues that if self-replicating probes are technologically inevitable, widespread Galactic dispersal would be expected. A range of alternative perspectives and critiques have been proposed (e.g. \deleted{\citealt{PhysicsToday1982,Burchell2006,CrawfordSchulzeMakuch2024}}\added{\citealt{Ball1973,PhysicsToday1982,Burchell2006,Osmanov2020,CrawfordSchulzeMakuch2024}}), and analyses by \citet{Tough1998} and \citet{Ellery2022,Ellery2025} emphasise that small, durable, and industrially tractable systems may represent the most realistic engineering pathway for sustained Galactic dispersal.

From this perspective, particulate technosignatures provide a complementary window onto technological activity integrated over gigayear timescales. Several Solar System bodies could, in principle, act as collectors of such material, including the Earth, the Moon, asteroids, and spacecraft-based free-flight capture systems. Each offers a distinct combination of surface stability, environmental filtering, and accessibility. As summarised in Table~\ref{tab:collectors}, the Moon provides a particularly favourable balance of long integration time, minimal geochemical alteration, and forthcoming sample-return opportunities, making it a compelling site for the potential detection of accumulated APs and BPs.
The following sections examine these ideas in turn, beginning with the possible origin and classification of Arkhipov Particles.

\begin{table*}
\small
\centering
\renewcommand{\arraystretch}{1.45}
\setlength{\tabcolsep}{6pt}

\caption{Comparison of Solar System environments as potential collectors of interstellar or technogenic particulate material, including both
non-intentional Arkhipov Particles and deliberately dispersed Bracewell Particles.}
\label{tab:collectors}

\begin{minipage}{0.98\textwidth}
\centering

\begin{tabular}{p{0.15\textwidth}p{0.385\textwidth}p{0.385\textwidth}}
\hline
Environment & Advantages & Limitations \\
\hline
Earth &
Extremely easy access; very large collecting area; atmosphere can decelerate
fast grains, increasing the survivability of undirected grains and providing useful braking in directed-delivery scenarios; an inhabited world is a possible target of interest for BPs. &
Atmospheric chemistry, weathering, biology, and surface recycling erase
signatures; tectonics restrict integration times; high risk of anthropogenic
contamination; minimum $11~\km\,\sec^{-1}$ entry speed into the atmosphere for undirected grains. \\

Moon &
Airless, stable surface with $\sim$4~Gyr exposure; minimal geochemical
alteration; regolith preserves microcraters, impact melts, and residues;
numerous forthcoming sample-return missions; possible target of interest for BPs because of proximity to Earth. &
Minimum impact speed $2\added{.4}~\km\,\sec^{-1}$ due to lack of atmosphere to decelerate grains; regolith gardening
mixes material; \textit{in situ} access and examination remain logistically challenging. \\

Asteroids\footnotemark[1] &
Pristine, uncontaminated surfaces; minimal geological alteration; slow regolith
evolution may preserve microscopic features; weak gravity allows engineered
low-velocity emplacement and enhanced survival of deliberately delivered
particles\footnotemark[2]. &
Minimal gravitational focussing; no atmosphere to decelerate fast-moving grains;
surface areas are small; access remains difficult; fewer near-term
sample-return missions than for the Moon. \\

Free-flight capture &
Controlled collection geometry; negligible surface alteration; no weathering;
precise calibration of impact velocities and particle trajectories; especially
well suited to detection of recently delivered or deliberately injected
particulate systems. &
Very short integration times; small collecting area; requires dedicated
spacecraft hardware for dust interception; requires the Solar System to be presently within a technograin-laden region for undirected APs and BPs; no possibility of directly emplaced BPs. \\

\hline
\end{tabular}

\footnotetext[1]{Co-orbital asteroids, objects sharing the Earth's orbital period and which approach Earth very closely annually at distances much shorter than anything except the Moon, have been proposed as advantageous sites for technosignature searches owing to their stable vantage points for observing the Earth \deleted{\citep{Benford2019}}\added{\citep{Benford2019,Benford2021}}. While primarily relevant for macroscopic probes, such bodies may in future provide additional and relatively accessible particulate archives as sample-return capabilities expand.}

\footnotetext[2]{Directed or engineered particulate delivery, in which grains are injected into low-velocity orbits around or directly onto small bodies, is discussed in \secref{sec:deliberatetargetingoftheinnersolarsystem}.}

\end{minipage}

\end{table*}
\section{Origin and classification of Arkhipov Particles}
\label{sec:origin}

In this section we outline a classification framework for particulate technogenic material potentially present in the Solar System. The aim is not to catalogue all conceivable extraterrestrial artefacts, but to distinguish broad classes of micron-scale material according to their \emph{origin}, \emph{degree of intentionality}, and \emph{material complexity}. These distinctions are essential for interpreting both detections and null results, and for guiding the transport, survival, and constraint analyses developed in later sections. For clarity, the particulate technogenic material considered in this work is organised into a small number of distinct categories, as summarised in Table~\ref{tab:APtaxonomy}.

Throughout this work we focus on solid particles in the micron-scale range. This regime is physically distinct from macroscopic artefacts: particles of this size can be redistributed by non-gravitational forces such as radiation pressure, plasma interactions, and gas drag, while remaining large enough for long-term survival as solid matter in the interstellar medium. Larger artefacts are not excluded in principle, but they are not treated in the present analysis.

This size range is particularly relevant to Arkhipov’s original insight that technological activity may generate a persistent \emph{technogenic component} of the meteoroid and dust environment, arising inevitably from spacefaring and industrial processes rather than from deliberate signalling or emplacement.

The physical transport of micron-scale particles through interplanetary and interstellar space, and their interaction with heliospheric and planetary environments, is treated quantitatively in \secref{sec:transportprocessesthroughtheinterstellarmedium} and \hyperref[sec:appendix:derivationsoffluxoflowspeedgrains]{Appendix~B}.\deleted{In this section we adopt these transport mechanisms only as a working premise for the purpose of defining a material taxonomy.} The following subsections distinguish three broad classes of particulate technosignatures: non-intentional technogenic debris; weakly intentional particulate systems not designed for interstellar travel; and deliberately dispersed micron-scale probes.

\subsection{Non-intentional technogenic debris}\label{subsec:nonintentional}

The most conservative class of particulate technosignature consists of
non-intentional technogenic debris: solid material generated as an incidental by-product of spacefaring or industrial activity, without any requirement for deliberate dispersal, signalling intent, or functional design. This category corresponds most closely to the non-intentional debris scenario articulated by Arkhipov in the early 1990s.

Arkhipov emphasised that routine spacecraft operations, collisions, explosions, and long-term orbital evolution would generate debris across a wide range of size scales \citep{Arkhipov1993,Arkhipov1994a,Arkhipov1997}.
A fraction of this material could escape its parent planetary system through gravitational interactions, radiation-pressure effects, or dynamical perturbations. Once in interstellar space, sufficiently durable particles could persist for gigayear timescales and eventually intersect other planetary systems, including the Earth-Moon system.

In this framework, micron-scale fragments are not assumed to be probes, messages, or functional devices. Rather, they represent the
technological analogue of natural meteoroids within the solid-matter environment: passively transported material whose existence follows from industrial and exploratory activity alone. Arkhipov explicitly framed such material as unintentional and emphasised its potential accumulation on geologically stable surfaces such as the Moon, where it might remain detectable long after the originating civilisation had ceased to exist.

Subsequent developments in astrophysics suggest that Arkhipov’s mechanism may operate at substantially larger mass scales than originally envisaged. Modern treatments of large orbital infrastructures -- such as distributed energy-harvesting systems commonly referred to as Dyson swarms rather than rigid spheres or shells \citep{Dyson1960,Wright2014} -- imply sustained processing, redistribution, and long-term degradation of planetary masses of material within circumstellar space. Even in the absence of deliberate signalling, the collisional evolution, maintenance, and eventual disassembly of such systems would be expected to produce extensive debris populations spanning macroscopic fragments down to dust-sized particles \citep{Lacki2025}. We adopt the non-intentional Arkhipov Particle population as our baseline case (Table~\ref{tab:APtaxonomy}),
underpinning the transport, survival, and null-detection constraints described in the sections that follow.

\subsection{Weakly intentional particulate systems and smart dust}
\label{subsec:weaklyintentionalparticulatesystemsandsmartdust}

Between purely non-intentional technogenic debris and deliberately engineered interstellar probes lies an intermediate class of particulate artefacts: small, engineered systems that are intentionally manufactured and deployed, but not necessarily designed for long-range interstellar travel. These systems may be
released for local or circumstellar purposes, yet still possess physical
properties that allow them to escape their parent system and enter \deleted{broader
interplanetary or }interstellar circulation. Such systems occupy a regime between non-intentional debris and deliberately dispersed probes, and are treated here as a distinct category of distributed smart materials (Table~\ref{tab:APtaxonomy}).

Terrestrial research on \emph{smart dust} provides a useful analogue
for this category. Originally conceived as swarms of millimetre-scale computing
nodes \citep{Warneke2001}, smart-dust research has expanded to encompass a wide
range of micro- and nanoscale devices integrating sensing, communication,
actuation, and limited information processing within compact physical
substrates \citep{Meiser2022,Mondal2025,Wang2025}. Contemporary designs include
micron-scale structures with embedded photonic, electronic, or chemical
functionality, often optimised for low mass, low power consumption, and passive
operation.

In an extraterrestrial context, similar particulate systems could plausibly be
manufactured for applications such as circumstellar monitoring, debris
management, radiation sensing, materials processing, or distributed
environmental diagnostics. While such systems need not be designed with
interstellar dissemination in mind, their small mass-to-area ratios render them
susceptible to radiation pressure, stellar winds, and gravitational scattering.
As a result, a fraction of these particles could escape their host system and
become incorporated into the ambient interstellar dust population.

Related ideas have been explored in a SETI context by \citet{Lacki2016}, who
considered the deliberate deployment of large populations of small artificial
particles—at millimetre scales—for sensing or control purposes within the
interstellar medium. In this work we categorise engineered particulates as a technological class distinct from
both macroscopic artefacts and accidental debris.

From a materials perspective, weakly intentional particulate systems may also
exhibit designed responses to environmental stress without requiring active
control. Experimental work has demonstrated, for example, that ultrathin
semiconductor and photonic structures can recover from radiation damage through
thermally or optically driven annealing processes
\citep[e.g.][]{Herasimenka2023}. Such properties suggest that durability and
functional persistence at micron scales can be achieved through material choice
and structure rather than continuous energy input.

Particles in this category are not assumed to carry explicit messages,
perform autonomous replication, or target specific planetary systems. Yet they
are also not purely accidental by-products. Instead, they represent engineered
materials whose original purpose may have been local or circumstellar, but
whose subsequent dispersal and long-term survival are governed by the same
natural transport processes that affect non-intentional technogenic dust.

\subsection{Deliberately dispersed Bracewell Particles}
\label{subsec:deliberatelydispersedbracewellparticles}

Bracewell Particles differ fundamentally from weakly intentional systems in that their dispersal beyond the parent planetary system is not an
incidental outcome, but an explicit design objective. Their small size confers
several advantages for such a role: low mass-to-area ratios facilitate passive
acceleration by radiation pressure or stellar winds; compact geometries reduce
energetic requirements for launch; and passive architectures allow persistence
over long interstellar timescales without continuous power input.

\added{BPs also offer comparative advantages as information carriers. \citet{RoseWright2004} showed that, where rapid exchange is not required, physically transported ``inscribed matter'' can be more energy-efficient than electromagnetic communication over interstellar distances. \citet{Hippke2018} subsequently compared such probes with photon communication in terms of energy per bit, information density, bandwidth, durability and latency, finding that the advantages of inscribed matter can become pronounced over kiloparsec distances and at relatively low transit velocities, although with correspondingly long communication delays \citep[Chapter~9]{LingamLoeb2021}. If a technological civilisation wished to transmit or archive knowledge, establish a durable record of its existence, or distribute information across many planetary systems, as microscopic artefacts BPs could represent one limiting implementation of this broader inscribed-matter paradigm.}

The possible internal architectures of BPs span a wide design space\added{, including inert inscribed matter, biological or bio-hybrid information carriers, smart materials, nanoscale electronic systems and more complex autonomous devices. The present framework intentionally remains agnostic regarding the particular technological implementation and instead concentrates on the common observational consequence that durable engineered materials may survive within planetary archives.}
\deleted{At the simplest level, they may consist of inert but information-bearing substrates, such as inscribed matter in the sense described by \citet{RoseWright2004}, in which structured physical patterns encode data without requiring active computation.}
\added{At the simplest level, they may consist of inert information-bearing substrates in which structured physical patterns encode data without requiring active computation \citep{RoseWright2004}.}

\deleted{Biological or bio-hybrid implementations also merit consideration.} \added{One particularly interesting class of information-bearing substrate comprises biological macromolecules such as DNA. Quantitative analyses have shown that DNA possesses exceptionally high mass-specific information-storage density, making it one of the most compact known physical media for long-term information storage \citep{Hippke2018}. More generally, advances in synthetic biology and molecular information storage have demonstrated that biological polymers can encode large quantities of digital information with extraordinary physical efficiency \citep{Church2012}.} \added{Thus} nucleic acids and related biomolecular systems are exceptionally compact and robust information carriers, and laboratory studies demonstrate that digital information can persist through mutation, replication, and partial degradation. While no assumption is made here that extraterrestrial civilisations employ biological substrates, such systems illustrate that self-assembly, redundancy, and error tolerance can be achieved at microscopic scales\deleted{using naturally available chemistry.} \citep{Yachie2007,Deamer2011,Davies2012,WalkerDavies2013,Deamer2024}.

At greater levels of complexity, BPs could incorporate limited sensing, data logging, or environmental response, implemented through solid-state, photonic, or chemical architectures analogous to advanced smart materials. More speculative designs might include probes capable of limited local
self-assembly or resource utilisation upon encountering suitable environments.
Such behaviour need not entail unrestricted self-replication of the
\deleted{von~Neumann type;}\added{von~Neumann type \citep{VonNeumann1966};} instead, it may involve modest amplification, repair, or
re-release of particulate material into the surrounding medium. Theoretical
analyses of minimal-mass or highly distributed probes suggest that such systems
could operate effectively under severe mass and energy constraints
\citep[e.g.][]{Tough1998,Ellery2022,Ellery2025}.

The deliberate dispersal of BPs \added{also need} not imply precise targeting
of specific planetary systems. Statistical seeding of large numbers of
particles, relying on Galactic dynamics to ensure eventual interception, may be
energetically favoured over targeted delivery. Conversely, nothing in the
definition of BPs excludes deliberate targeting of particular systems,
including the Solar System,\deleted{and the Moon itself}should an advanced civilisation choose to do so.
Both targeted and untargeted dispersal scenarios are therefore encompassed
within this category.

\added{The distinction between microscopic and macroscopic architectures may also be evolutionary rather than fundamental. A civilisation could deploy thin films, sails or larger probes that subsequently generate large numbers of microscopic fragments through impact and long-term degradation. The resulting particles would enter the same planetary archive as material originally released at micron scale, even though their original functions and dimensions were very different. Whether such residues are classified as APs or BPs depends on whether their dispersal was incidental or deliberate, rather than on the size of the precursor structure. The framework developed here therefore encompasses both primary particulate technologies and the secondary microscopic residues of larger engineered systems.}

\added{Deliberate particulate dispersal also intersects with the longer history of directed panspermia \citep{CrickOrgel1973}: the intentional transport of organisms, biological precursors or encoded biological information to other planetary environments. Although directed panspermia is commonly associated with later scientific formulations, versions of the concept pre-date Bracewell’s interstellar-probe proposal and have since been developed as a technological counterpart to natural lithopanspermia \citep{GinsburgLingam2021}. Biological or bio-hybrid BPs could therefore be regarded as one possible implementation of directed panspermia, particularly where microscopic carriers are designed to preserve, deliver or release biological information.}

\added{The correspondence is not exact. BPs as defined here may be wholly inorganic and may serve reconnaissance, archival, communicative or environmental functions unrelated to propagating life. The shared principle is instead the deliberate interstellar distribution of compact material carriers, whether through precise targeting or statistical seeding. The quantitative AP constraints developed in this work remain independent of any biological payload or panspermia hypothesis.}

BPs constitute a distinct technosignature
population whose arrival statistics, spatial distribution, and survivability
may differ markedly from those of non-intentional technogenic dust. The conservative baseline flux constraints developed later in this work apply directly
only to the non-intentional Arkhipov Particle population. Deliberately dispersed
BPs represent an additional and potentially more detectable class of
microscopic technosignatures whose presence would carry correspondingly
stronger implications for extraterrestrial technological activity. \deleted{Deliberately dispersed BPs constitute a separate class of particulate technosignature (Table~\ref{tab:APtaxonomy}), whose arrival statistics need not follow the isotropic, time-integrated flux assumptions adopted here for non-intentional APs.}

\begin{table*}
\centering
\small
\renewcommand{\arraystretch}{1.35}
\setlength{\tabcolsep}{6pt}
\caption{Classification of micron-scale technogenic particulate material considered in this work.
The categories are ordered from non-intentional debris to deliberately dispersed interstellar probes.
\deleted{Only the non-intentional Arkhipov Particle (AP) population is used to derive conservative flux and
null-detection constraints in later sections; deliberately dispersed systems represent an extension
beyond the baseline model.}}
\label{tab:APtaxonomy}
\begin{tabular}{p{0.16\textwidth} p{0.23\textwidth} p{0.20\textwidth} p{0.33\textwidth}}
\hline
\textbf{Category} & \textbf{Origin and Intent} & \textbf{Representative Examples} & \textbf{Key Assumptions / Notes} \\
\hline

\textbf{Non-intentional APs}
& Incidental by-products of spacefaring or industrial activity; no signalling or probe intent
& Spacecraft debris; collision fragments; erosion products; engineered alloys
& Baseline case for transport modelling and null constraints. Includes both simple and structurally complex fragments. Passively transported by Galactic dynamics and subject to isotropic flux assumptions. \\
\hline

\textbf{Distributed APs}
& Engineered for local or circumstellar function but not explicitly designed for interstellar dispersal
& Smart dust; dipole sensors; responsive or self-healing materials
& Weakly intentional systems. May escape planetary systems incidentally. Not assumed to encode deliberate interstellar signalling, but may preserve functional or structured microarchitecture. \\
\hline

\textbf{Bracewell Particles}
& Explicitly designed for interstellar dispersal; targeted or statistically seeded
& Inscribed matter; passive probes; bio-hybrid artefacts; self-assembling systems; systems with limited self-replication
& Distinct population not constrained by the isotropic debris-flux model. Arrival statistics are governed by engineered delivery strategies. May imply interstellar travel capability or large-scale astroengineering. \\
\hline
\end{tabular}
\end{table*}

\section{Transport processes through the interstellar medium}
\label{sec:transportprocessesthroughtheinterstellarmedium}

The feasibility of detecting APs or BPs in the lunar regolith depends critically on whether micron-scale grains can traverse interstellar distances and arrive within the inner Solar System at velocities conducive to survival upon impact. The motion of such grains through the ISM is governed by a combination of gas drag, magnetic coupling, radiation pressure, Lorentz forces and large-scale gravitational perturbations, operating within a heterogeneous, turbulent, and evolving Galactic environment (\citealp{FrischRedfieldSlavin2011,Draine2011}; \deleted{Hopkins et al., 2016}\added{\citealp{HopkinsLee2016,LeeHopkinsSquire2017}}; \citealp{Sterken2019,Sterken2023}).

For grains in the submicron to micron size range, interactions with the ambient ISM play a dominant role in shaping their trajectories and velocity distributions. Rather than propagating ballistically with the kinematics of their parent stars, such grains are expected to become dynamically coupled to the surrounding gas and magnetic fields over characteristic length scales and timescales that are short compared with Galactic orbital periods. As a result, their transport through the Galaxy is more naturally described in terms of ISM flow properties and phase structure than stellar velocity dispersions.

In this section, we outline the principal physical processes governing the transport and survival of micron-scale grains in the ISM, examine the velocity regimes relevant for their eventual entry into the heliosphere, and assess the degree to which their spatial distribution may become homogenised or remain locally clustered prior to interception by the Solar System.

Symbols and notation used in the equations below are defined in \hyperref[sec:appendix:notationandsymboldefinitions]{Appendix~C}.

\subsection{Kinematic regimes of micron-scale grains}
\label{subsec:kinematicregimesofmicronscalegrains}

Micron-scale grains interact strongly with their interstellar
environment, coupling to gas, magnetic fields, and radiation
(e.g.\ \citealt{Draine2011}; \deleted{Hopkins et al., 2016}\added{\citealt{HopkinsLee2016,LeeHopkinsSquire2017}}).
Unlike macroscopic meteoroids, which are effectively collisionless, submicron particles experience substantial drag and Lorentz forces, placing them in a dynamical regime governed primarily by the properties of the surrounding interstellar medium (ISM) rather than by the velocity dispersion of their parent stars.

Throughout this work we assume that submicron grains are efficiently coupled to the ambient ISM and therefore approach the heliosphere with velocities characteristic of the local interstellar flow, rather than retaining the kinematics of their source stellar populations. This assumption is well supported for grains in the $\sim$0.1--1\,$\mu$m size range, for which gas drag
and magnetic coupling act on timescales short compared with Galactic orbital periods.

\subsubsection{Velocity distributions}
\label{subsubsec:velocitydistributions}
The kinematics of interstellar APs depend sensitively on grain size and on the ISM phase through which they propagate. Large grains
($r_G \gg 10\,\mu\mathrm{m}$) may retain the velocity dispersion of their parent stellar populations, with characteristic Solar-neighbourhood values $\sigma_v \sim 30$--50~km\,s$^{-1}$ 
\deleted{(Anguiano et al., 2018)} \added{\citep{Anguiano2020}}.

In contrast, submicron grains are expected to be dynamically locked into the bulk flow of the warm and cold ISM through gas drag and magnetic coupling, yielding relative velocities of only a few km\,s$^{-1}$ with respect to the local standard of rest (LSR).

Because the surrounding ISM flows past the heliosphere at
$\approx 17$~km\,s$^{-1}$ \citep{FrischRedfieldSlavin2011},
typical approach speeds of 20--30~km\,s$^{-1}$ relative to the Sun are therefore plausible for AP-sized grains. To bracket uncertainties in the degree of ISM coupling, we adopt two limiting cases for later calculations:
 (i) a `fast' Maxwellian distribution with $\sigma_v = 40$~km\,s$^{-1}$, appropriate if grains retain a significant memory of stellar motions; and
(ii) a `slow' mono-kinetic distribution with
$v_\infty = 20$~km\,s$^{-1}$, representative of grains fully entrained in the local ISM flow.

\subsubsection{Gas drag and coupling scales}
\label{subsubsec:gasdrag}

The transition between ballistic and ISM-coupled behaviour may be characterised by a gas-drag coupling timescale. \changed{For a spherical grain of radius $r_G$ and density $\rho_G$ moving through ISM gas of density $\rho_{\mathrm{ISM}}$ with relative speed $v_{\mathrm{rel}}$, a characteristic gas-grain coupling timescale may be written as}{For a spherical grain of radius $r_G$ and density $\rho_G$, the coupling time is obtained by dividing the grain momentum by the rate at which collisions with the surrounding gas transfer momentum. The grain mass is $(4/3)\pi\rho_G r_G^3$, while its geometric cross-section is $\pi r_G^2$. For a single-species neutral gas, the characteristic gas-grain coupling timescale is}
\begin{multline}
t_{\mathrm{drag}} =
\shortchanged{\frac{\sqrt{8}}{3}}{\frac{4}{3}}
\frac{r_G\rho_G}
{\rho_{\mathrm{ISM}}
\sqrt{v_{\mathrm{rel}}^2+
\added{\frac{128}{9\pi}}c_{s\added{,t}}^2}}
\sim
\shortchanged{11}{15}\,\Myr
\left(\frac{r_G}{1\,\um}\right)
\\
\times
\left(\frac{\rho_G}{3\,\gram\,\cm^{-3}}\right)
\left(\frac{\rho_{\mathrm{ISM}}/m_H}{0.5\,\cm^{-3}}\right)^{-1}
\left(\frac{v_{\mathrm{rel}}}{10\,\kms}\right)^{-1},
\end{multline}
where $c_{s\added{,t}}$ is the local \added{isothermal} sound speed
(e.g.\ \added{\citealp{DraineSalpeter1979}; }\citealp{Draine2011}
\deleted{; Hopkins et al., 2016}
\added{; \citealp{HopkinsLee2016}}).
\added{The coefficient \(4/3\) follows from the ratio of the numerical
coefficients in the spherical grain mass,
\((4\pi/3)\rho_G r_G^3\), and the projected collisional area,
\(\pi r_G^2\). The thermal term is the standard approximation to the
full neutral-drag expression presented by
\citet[their equation~5a]{DraineSalpeter1979} and summarised by
\citet{Draine2011}; the exact collisional-drag expression is more
complicated, but this form provides an accurate interpolation between
thermal and supersonic drift. In the highly supersonic limit, the thermal
term becomes negligible and the expression reduces to the geometric-drag result. The numerical expression on the right evaluates that limit for
representative warm neutral-medium conditions.}
This expression captures the timescale over which grains are dynamically captured by the surrounding gas flow; more detailed formulations introduce velocity-dependent corrections but do not alter the qualitative behaviour for submicron grains.

\changed{The corresponding coupling length,}{Multiplying the drag time by the initial grain-gas drift speed gives the characteristic distance over which the grain loses memory of its initial motion and becomes entrained in the ambient flow:}
$s_{\mathrm{drag}}\sim v_{\mathrm{rel}}t_{\mathrm{drag}}$. \added{This gas-grain coupling distance} is typically of order tens to $\sim100\,\pc$ in the warm neutral and warm ionised media for $r_G\sim0.1$--$1\,\um$ grains. As a result, submicron grains do not respond to small-scale ISM turbulence but remain entrained in the largest-scale flows, effectively sampling the Galactic ISM rather than individual stellar trajectories. This size-dependent coupling plays a central role in shaping the velocity distribution of grains entering the heliosphere, and therefore the fraction of APs capable of surviving lunar impact (\secreftwo{sec:deliveryandsurvivalonthelunarsurface}{subsec:fluxformalismandconstraintmodelling}).

\subsection{ISM phases and grain survival}
\label{subsec:ismphasesandgrainsurvival}

The long-term survival of micron-scale grains in the interstellar medium is governed by the balance between dynamical residence times and destructive processes such as thermal sputtering, non-thermal erosion and grain-grain collisions. Although the ISM is heterogeneous, the broad picture that emerges from modelling and laboratory constraints is that refractory grains of the sizes relevant to APs enjoy survival times of several hundred million years to $\gtrsim1\,\Gyr$, depending on their trajectories through the various ISM phases \deleted{\citep{Draine2011,Hirashita2015,Hirashita2017,Heck2020}}\added{\citep{HirashitaYan2009,Draine2011,Hirashita2015,Heck2020}}. Such lifetimes exceed the $\sim10^8\,\yr$ intervals between the Sun's spiral-arm passages or vertical oscillations through the Galactic disc (\secref{subsec:spatialmixingversuslocalclustering}), and therefore do not preclude Galactic-scale transport.

\changed{For a grain of radius $r_G$ embedded in hot, low-density plasma, the sputtering erosion rate may be expressed approximately as}{When a grain is embedded in hot plasma, energetic ions remove atoms from its exposed surface. At temperatures above $\sim10^6\,\Kelv$, the adopted order-of-magnitude sputtering law reduces the grain radius at the rate}
\begin{equation}
\frac{\mathrm{d}r_G}{\mathrm{d}t}
\simeq
-10^{-2}\,\um\,\Myr^{-1}
\left(\frac{n}{0.01\,\cm^{-3}}\right),
\end{equation}
appropriate for $T\gtrsim10^6\,\Kelv$ gas, where $n$ is the hydrogen number density \citep[see e.g.\ \S 25 of][]{Draine2011}. \added{The negative sign denotes erosion.}
\changed{The corresponding sputtering timescale is}{Dividing the initial radius by the magnitude of this radius-loss rate gives a sputtering timescale that scales linearly with $r_G$ and inversely with the ambient number density:}
\begin{equation}
t_{\mathrm{sput}}
\sim
30\,\Myr
\left(\frac{r_G}{0.3\,\um}\right)
\left(\frac{n}{0.01\,\cm^{-3}}\right)^{-1}.
\end{equation}
Carbonaceous grains typically erode faster, whereas refractory silicates and SiC survive longer. \added{This estimate assumes that the local plasma conditions and sputtering yield remain approximately constant.} The hot ionised medium (HIM) therefore represents the most aggressive sputtering environment. Although the HIM occupies approximately half the volume of the ISM, it contains only a small fraction of its mass. If dust broadly traces the gas mass on large scales, grains are therefore expected to spend only a small fraction of their lifetimes there.

In the warm neutral and warm ionised media, with densities of order $0.1$--$0.5\,\cm^{-3}$ and temperatures near $10^4\,\Kelv$, thermal sputtering is far less efficient. Instead, grain destruction in these phases is thought to be dominated by grain-grain shattering and vaporisation in supernova-driven shocks (e.g.\ \changed{\citealt{DraineSalpeter1979,Jones1996,Slavin2015}}
    {\citealt{DraineSalpeter1979b,Jones1996,Slavin2015}}; \changed{Zhukovska et al. (2016)}{\citealt{Zhukovska2017}}). Characteristic survival times in the ISM nevertheless remain long, $\tau_{\mathrm{ISM}}\sim0.1$--$1\,\Gyr$ for submicron grains (\added{\citealt{HirashitaYan2009}; }\citealt{Hirashita2015}\deleted{; \citealt{Hirashita2017}}). As these warm phases constitute the majority of the ISM by volume, they are expected to be the primary transport channels for micron-scale APs.

Cold molecular clouds provide the gentlest physical environment for grain survival. Thermal sputtering is negligible at $T\lesssim50\,\Kelv$, but strong drag forces can immobilise submicron grains, trapping them within the cloud until dispersal by star-formation activity or turbulent fragmentation. This delayed release adds stochasticity to AP arrival times but does not diminish overall survival probabilities.

Taking these contributions together, the survival of AP-sized grains reflects a weighted integral over residence times in each ISM phase. For refractory materials of radius $r_G\sim0.1$--$1\,\um$, effective lifetimes of several hundred Myr to beyond $1\,\Gyr$ are expected. These values are comparable to, or exceed, the characteristic coupling and drag timescales discussed in \secref{subsec:kinematicregimesofmicronscalegrains}, and allow transport over hundreds of parsecs to kiloparsec scales when grains are advected with large-scale ISM flows. ISM processing therefore constrains the \emph{condition} in which APs arrive at the Solar System, but does not preclude their long-distance survival or accumulation within the lunar regolith over $\sim4\,\Gyr$.

\subsubsection{Larger precursor structures}
\added{The present calculation treats the transported material as discrete grains, but the final micron-scale population preserved in regolith need not have begun in this form. Larger engineered structures, including ultralight sails, thin films, chips or more substantial probes, may undergo interstellar transport before fragmenting during arrival, impact or subsequent surface processing. The particles considered here may therefore comprise both primary grains and secondary fragments derived from larger artefacts.

The survival of ultralight spacecraft during interstellar transit has been studied particularly in connection with relativistic lightsail concepts. \citet{Hoang2017}, for example, quantified damage caused by interstellar gas and dust to spacecraft travelling at velocities of approximately \(0.05c\)--\(0.5c\), including ion penetration, structural damage tracks, melting, cratering and surface erosion. Their results show that survivability is strongly dependent on composition and geometry: graphite performs more favourably than quartz under some forms of bombardment, while dust impacts may remove substantial surface layers unless shielding or a reduced frontal cross-section is employed. \citet{Hippke2018} likewise considered shielding against dust and radiation in evaluating inscribed-matter probes and found the associated mass penalty to be comparatively small for appropriately designed probes travelling below approximately \(0.2c\). These analyses address architectures and velocities different from those assumed for our fiducial particles, but they provide an independent treatment of relevant transport limits and yield broadly comparable maximum distances and timescales.

Accordingly, the survival model developed here should be understood as applying to one limiting transport morphology, while the eventual regolith population may also contain fragments of initially larger technologies whose survivability must be evaluated according to their particular composition, geometry and velocity.}

\subsection{Velocities and maximum transport distances}
\label{subsec:velocitiesandmaximumtransport}

Given the survival times estimated in \secref{subsec:ismphasesandgrainsurvival}, micron-scale grains can, in principle, be transported over large distances through the Galactic disc before significant erosion occurs. Once dynamically coupled to the ambient interstellar medium, such grains are advected with the bulk ISM flow rather than travelling ballistically. A simple upper-bound estimate of the characteristic transport scale is therefore obtained by considering an effective ISM flow speed $v$ acting over a grain survival time $t$.

\deleted{Over typical ISM residence times of $t \sim 0.1$--1\,Gyr, this yields a maximum displacement}\added{If the large-scale flow remains coherent, the maximum advective displacement is the product of the characteristic flow speed and the survival or residence time. Using \(1\,\mathrm{km\,s^{-1}}\!\simeq\!1.02\,\mathrm{pc\,Myr^{-1}}\), a grain carried at \(10\,\mathrm{km\,s^{-1}}\) for \(100\,\mathrm{Myr}\) is displaced by approximately \(1\,\mathrm{kpc}\):}
\begin{equation}
\label{eqn:maxadvectivedisplacement}
s_{\max} \sim v t
       \approx 1~\mathrm{kpc}\,
       \biggl(\frac{v}{10~\mathrm{km\,s^{-1}}}\biggr)
       \biggl(\frac{t}{100~\mathrm{Myr}}\biggr),
\end{equation}
where $s_{\max}$ should be interpreted as an upper limit on the net spatial displacement rather than a \changed{literal path length.}{claim of rectilinear travel. Turbulent reversals and diffusive mixing can reduce the net displacement below $v t$, whereas coherent large-scale advection allows the estimate to be approached.}

In reality, interstellar transport is not strictly laminar. Turbulent motions in the ISM imply that grain trajectories are better described as stochastic or diffusive on small scales, with coherent advection only emerging on scales comparable to the outer scale of turbulence, which is thought to be of order hundreds of parsecs to $\sim$1~kpc. As a result, the net displacement achieved over a given survival time may be smaller than $v t$, depending on the degree of turbulent mixing and the coherence of large-scale flows.

Nevertheless, even under conservative assumptions in which transport proceeds as a random walk rather than rectilinear advection, the displacement scale implied by Equation~\ref{eqn:maxadvectivedisplacement} remains astrophysically significant. A characteristic scale of order $\sim$0.1--1~kpc is comparable to the width of a Galactic spiral arm ($\sim$0.3--0.7~kpc; \changed{\citealt{Vallee2017}}{\citealt{Vallee2014}}) and represents a non-negligible fraction of the radial scale length of the Galactic thin disc (a few kiloparsecs; \citealt{BlandHawthornGerhard2016}). Similar stochastic transport behaviour has been demonstrated in numerical studies of supernova-driven interstellar turbulence, where short-lived radioactive isotopes and passive tracers undergo diffusive–advective mixing over scales of hundreds of parsecs to $\sim$1~kpc on timescales of $\sim10^8$--$10^9$~yr \citep{deAvillezMacLow2002}.

Accordingly, micron-scale APs that remain dynamically coupled to ISM flows with characteristic velocities of $\sim$10--30~km\,s$^{-1}$ over several hundred Myr can plausibly be displaced across substantial portions of the local Galactic disc, even if their actual trajectories are non-ballistic and partially diffusive.

Over the past $\sim$4~Gyr, the Sun has completed $\sim$16--18 revolutions around the Galactic centre in an inertial frame, repeatedly sampling the interstellar medium under a wide range of density and kinematic conditions. However, because the spiral arms rotate with a distinct pattern speed, the Solar System has
crossed the spiral-arm structure only a few times over the same interval, typically $\sim$4--5 passages relative to the spiral pattern \changed{\citep{Vallee2017}}{\citep{Vallee2014}}. Together with vertical oscillations above and below the Galactic midplane and encounters with ISM structures of varying density and ionisation state, this motion implies that the Solar System has sampled a broad range of Galactic environments over gigayear timescales. Because the lunar
regolith integrates material arriving across this entire orbital history, its possible AP inventory represents a time-averaged mixture of grains originating from many distinct Galactic locales. Even if individual release events are spatially or temporally clustered, the combined effects of Galactic rotation, ISM advection, and long integration times act to average their cumulative contribution.

This long-term averaging has two important implications. First, APs of refractory composition and micron-scale size dispersed within the Galactic disc have had ample opportunity to encounter the Solar System at least once during its multi-Gyr orbital history, provided that their release rates are not extraordinarily low. Second, the net AP surface density on the Moon is insensitive to short-term variations in the ISM encountered over $\lesssim10^{8}$\,yr, instead reflecting the aggregate particle flux integrated over the Solar System’s Galactic trajectory. These considerations prompt the
examination of spatial inhomogeneity and localised clustering effects in \secref{subsec:spatialmixingversuslocalclustering}. Source-population constraints associated with the ability of stars to eject grains of the relevant size are treated separately in \secref{subsec:fluxformalismandconstraintmodelling}.

\subsection{Spatial mixing versus local clustering}
\label{subsec:spatialmixingversuslocalclustering}
A central uncertainty in AP transport concerns the extent to which grains become homogeneously mixed throughout the Galactic disc. One limiting case is a closed-box approximation, in which APs survive for sufficient time to trace the underlying stellar-density distribution on large Galactic scales \citep[e.g.][]{Draine2011}. At the opposite extreme, finite grain lifetimes and incomplete mixing may lead to spatially inhomogeneous distributions, or 'bubbles,' with characteristic radii $R_{B}\sim10^{2}\,\pc$, within which technogenic grains remain locally enhanced around their source regions for $10^{7}$--$10^{8}$\,yr before destruction or dispersal. Such behaviour is expected in a supernova-driven ISM, where dust injection and destruction are episodic and mixing proceeds on hundred-parsec scales rather than instantaneously \citep[e.g.][]{Jones1996,Slavin2015,deAvillezMacLow2002}.

The distinction between these regimes is not that of the total mass of particles in the Galaxy, but of sampling geometry. If technogenic grains are rapidly mixed throughout the Galactic disc, then the Solar System samples a background set by the global stellar density and the total technomaterial budget. In contrast, if grains have finite lifetimes and remain spatially localised for a significant fraction of their survival time, then the probability of detection depends on whether the Solar System intersects one or more such regions at all. In this inhomogeneous limit, the relevant quantity is thus not the total volume of the Galactic disc, but the number density, size, and lifetime of discrete grain-rich regions that the Solar System may encounter along its Galactic trajectory. Our bubble formalism provides a conceptually simple way to quantify this interception probability and to identify the threshold below which the Solar System is unlikely ever to sample technogenic material, regardless of the total mass released elsewhere in the Galaxy.

If release events occur frequently enough, individual bubbles overlap and the Galaxy approaches an effectively homogeneous AP technodust field. If releases are rare, bubbles remain isolated, and the Solar System may or may not intersect one during its orbit through the disc. We now formalise this interception argument by estimating the expected number of such grain-rich regions encountered by the Solar System over its Galactic lifetime. To quantify these regimes, suppose that the grains are released over a lifespan short compared to the star, which we model as a burst. The bursts occur independently of each other among a subset of stars with a number density $n_{\star,B}$ at a rate per star of $\Gamma_B$. The grains then diffuse out to form a bubble with a finite lifespan $\tau_B$. Thus, the volumetric rate of AP-bubble generating events is $n_{\star,B} \Gamma_B$. \changed{The mean filling factor of the bubbles (total volume of bubbles per unit volume) is}{At any instant, the number density of surviving grain-rich regions is $n_{\star,B} \Gamma_B \tau_B$, the volumetric event rate multiplied by their mean lifetime $\tau_B$. Multiplying this number density by the mean spherical volume per region, \(\tfrac{4\pi}{3}\langle R_B^3\rangle\), gives the dimensionless mean filling factor:}
\begin{multline}
\Phi_B = \frac{4}{3} \pi n_{\star,B} \Gamma_B \tau_B \Mean{R_B^3} = 0.4 \left(\frac{n_{\star,B}}{10^{-3}\,\pc^{-3}}\right) \\
\times \left(\frac{\Gamma_B}{10^{-3}\,\Gyr^{-1}}\right)\left(\frac{\tau_B}{100\,\Myr}\right) \left(\frac{\Mean{R_B^3}}{(100\,\pc)^3}\right) .
\end{multline}
\added{Thus, \(\Phi_B\!\ll\!1\) corresponds to isolated regions, \(\Phi_B\!\sim\!1\) to substantial overlap, and \(\Phi_B\!\gg\!1\) to an effectively volume-filling particulate background.} The number density of stars is scaled to the populations of stars that are nearly sunlike. We use the spherical volume $(4/3)\pi R_B^3$ to represent the effective region within which the presence or absence of technogenic grains is correlated with a given release event (not the volume of the Galactic disc as a whole). Thus we use this spherical geometry as a bookkeeping device for estimating interception probabilities, not as an assertion that bubbles are isolated physical objects embedded in an otherwise empty disc. Because APs survive for long periods and can travel considerable distances before destruction, the bubble populations can overlap even if only one in a hundred sunlike stars is the site of a release event in its $\sim 10\,\Gyr$ lifespan.

The total number of AP impacts on the Moon is the integrated contribution from all such regions along the Solar System's path. Even if APs are strongly confined within bubbles, the \emph{total} AP mass in the Galaxy remains conserved: clumping affects temporal variability, not the time-integrated fluence. Provided the Solar System has entered a representative sample of bubbles over the integration time $t_{\mathrm{int}} \sim 4\ \Gyr$, the cumulative surface density of APs on the Moon converges to the homogeneous expectation. The only failure mode arises if releases are so rare that the Solar System has \emph{never} encountered a bubble.

The threshold between these two regimes is set by the mean number of bubbles intercepted over the Solar System's lifetime, $\Mean{N_{B,\odot}}\sim1$. At any instant, the number density of extant bubbles is $n_{\star,B}\Gamma_B\tau_B$, and each presents a mean projected cross-section $\pi\Mean{R_B^2}$. We describe the relative speed between the Solar System and the ambient ISM in the Galactic frame by a distribution $f_v(v)$. If $v$ is very small, bubbles may appear and disappear around the Solar System; for typical values of $v$, however, the Solar System crosses a bubble in a time much shorter than its lifetime, so the bubble size may be treated as fixed during the encounter. \changed{When all the bubbles have the same size, the mean number of bubbles encountered by the Solar System during the integration is}{In this fast-crossing regime, the expected number intercepted follows the usual ``number density $\times$ cross-section $\times$ path length'' construction:}
\begin{equation}
\label{eqn:NBubblesEncountered}
\begin{aligned}
\Mean{N_{B,\odot}}
&\approx
\pi n_{\star,B}\Gamma_B\Mean{R_B^2}
\Mean{v}\tau_B t_{\mathrm{int}}
\\[2pt]
&\approx
\changed{2.6}{3}
\left(\frac{n_{\star,B}}{10^{-3}\,\pc^{-3}}\right)
\left(\frac{\Gamma_B}{10^{-5}\,\Gyr^{-1}}\right)
\\
&\quad\times
\left(\frac{\Mean{R_B^2}}{(100\,\pc)^2}\right)
\left(\frac{\Mean{v}}{20\,\kms}\right)
\\
&\quad\times
\left(\frac{\tau_B}{100\,\Myr}\right)
\left(\frac{t_{\mathrm{int}}}{4\,\Gyr}\right).
\end{aligned}
\end{equation}
\added{Appendix~A gives the corresponding exact expression, valid in both the slow- and fast-crossing regimes.} Below a threshold release rate -- \changed{of order one event per few hundred thousand sunlike stars over Galactic history}{of order one event per \(10^4\)--\(10^5\) Sun-like stars over \(10\,\Gyr\)} -- the Solar System becomes unlikely to encounter even a single AP-bearing bubble. In this case, no matter how much material each event releases, the Moon would simply never intercept it, and constraints on AP mass budgets cannot be set. If $\Gamma_B$ is higher, the Moon receives either a constant AP flux (when bubbles overlap) or a sequence of discrete showers (when they do not). A high variance in the AP fluence over the integration time results only in a transitional regime $\Mean{N_{B,\odot}}\sim1\text{--}10$, with most particles arriving in a few short episodes, or if there is additional clustering, for example, induced by the spiral arm structure of the Galaxy.

Note this does not assume any particular number of grains released per event; rather, it concerns only whether the Solar System intersects regions that contain technogenic material at all. The total number of particles deposited on the Moon is treated separately in \secref{subsec:fluxformalismandconstraintmodelling} through the mass-per-event and grain-size distribution, once an interception has occurred.

The structure of the Galactic disc introduces additional modulation of the expected AP flux, primarily affecting its spatial and temporal distribution rather than the long-term integrated fluence. The Solar System crosses
a spiral arm approximately every 200--300 Myr, depending on the uncertain
pattern speed of the spiral density wave
\deleted{\citep{Vallee2017}}\added{\citep{Vallee2014}}. While dust lanes in external galaxies appear strongly
arm-traced in optical and far-infrared images, quantitative studies of dust
extinction and emission in the Milky Way present a more nuanced picture.
Three-dimensional dust maps derived from stellar photometry show only weak
arm signatures locally \citep{Green2019}, whereas recent Gaia-based
extinction reconstructions reveal clearer spiral structure further afield
\citep{Barbillion2025}. Models of the Milky Way disc suggest that the dust-to-gas
ratio does not vary strongly across arm and interarm regions
\citep{DrimmelSpergel2001}, and observations of M31 indicate that dust surface
density may indeed be elevated in spiral arms, but primarily because it
follows the gas distribution rather than tracing a separate dynamical component
\citep{Draine2014}. Because the lunar regolith integrates particulate influx over multiple
spiral-arm passages, vertical oscillations, and ISM environments spanning
$\sim$4~Gyr, these structural variations are largely averaged out and do not
dominate the long-term AP surface density relevant for our modelling.
We note in passing that the lunar geological record itself may preserve evidence of the changing Galactic environment of the Solar System over time, including variations in cosmic-ray flux and episodic nearby supernova activity, providing an independent archive of Galactic-scale processes
\citep{Crawford2021}.

Moreover, vertical gradients in ISM density and dust distribution on scales of tens to hundreds of parsecs can introduce fluctuations in the local dust column by factors of order unity \citep{Guo2025}, comparable to the expected variation from spiral-arm passages.

Given that the lunar regolith integrates grain influx over $\sim$4 Gyr, and
thus across many such arm and midplane passages, these effects average out to
first order. As a result, spiral-arm structure and vertical oscillations modulate the AP
flux but do not dominate it, and are largely subdominant to transport-phase
destruction and heliospheric filtering.

These considerations apply to undirected, passively drifting APs; directed or
actively targeted particulate probes (BPs) need not follow the spatial
statistics of the background ISM.

In addition to the large-scale mixing and interception effects described above,
the local delivery of interstellar grains to the inner Solar System may be
temporarily modulated by episodic astrophysical events without altering their
global abundance. Supernova-driven disturbances can restructure the heliosphere
on timescales of $10^{5}$--$10^{6}$~yr, reducing its effectiveness as a filter
and permitting enhanced penetration of submicron interstellar dust into the
Earth-Moon system \citep{JenniskensPinault2025}. Shock fronts propagating
through the ambient ISM may also sweep up and compress dust, producing transient
enhancements in the local particulate flux. Terrestrial detections of
$^{60}\mathrm{Fe}$ at 2--2.5~Ma and 9~Ma indicate that such events have occurred
in the recent geological past and imply that corresponding stratigraphic
horizons should be preserved in the lunar regolith
\deleted{\citep{Crawford2016,Crawford2021}}\added{\citep{Crawford2017,Crawford2021}}.

Passages of the Solar System through dense interstellar clouds provide a
complementary mechanism for episodic exposure. For cloud densities
$n \sim 10^{2}$--$10^{3}\,\mathrm{cm^{-3}}$, heliospheric models predict
compression to radii of order 1~AU or smaller, leaving the Earth-Moon system
directly exposed to interstellar gas and dust for the duration of the encounter
\citep{Mueller2006,FrischRedfieldSlavin2011}. Such episodes may enhance near-surface
deposition and influence the stratigraphic expression of incoming material, but
they do not change the long-term, time-integrated fluence of technogenic grains
accumulated by the Moon. Instead, they primarily affect the temporal structure
and preservation context of AP deposition. These episodic modulations therefore
provide additional context for the mechanisms discussed below in
\secref{subsec:velocityfilteringandslowarrivaldynamics} and do not modify the statistical constraints on cumulative
technomaterial production developed in \secref{subsec:fluxformalismandconstraintmodelling}.

\subsection{Velocity filtering and slow-arrival dynamics}
\label{subsec:velocityfilteringandslowarrivaldynamics}

Only a small fraction of undirected interstellar grains are expected to arrive at the lunar surface with relative velocities low enough to survive impact in recognisable form. The Moon orbits the Sun at $v_\oplus \simeq 30$~km\,s$^{-1}$, while a grain falling freely from rest at infinity under Solar gravity alone would reach $\simeq 42$~km\,s$^{-1}$ at 1~AU. Hypervelocity experiments show that micron-scale grains are largely melted or vaporised at encounter speeds $\gtrsim 5$~km\,s$^{-1}$, with partial preservation possible only below this threshold \citep{Burchell1999,Burchell2008}.\footnote{Bacteria of similar scale belonging to the genus \textit{Rhodococcus} have also been tested for their survivability in hypervelocity impacts at $5.1 \pm 0.1$~km\,s$^{-1}$, with indications that up to this threshold they can still survive and subsequently grow \citep{Burchell2001,Burchell2004}, suggesting that BPs integrating or exploiting biological systems may be subject to similar impact survival constraints.} Any mechanism that allows the heliocentric velocity of an incoming grain to closely match $v_\oplus$ therefore plays a critical role in determining detectability.

Micron-scale grains experience substantial drag within the ISM and are therefore typically slowed to velocities characteristic of the ambient gas rather than the velocity dispersion of their parent stars. As discussed in \secref{subsec:kinematicregimesofmicronscalegrains}, this implies that the approach-speed distribution at heliospheric entry is governed primarily by the relative Sun-ISM motion, yielding characteristic encounter speeds of $\sim$20--30~km\,s$^{-1}$ for grains that have equilibrated with the local flow.

\changed{Solar radiation pressure modifies the effective gravitational potential experienced by small grains through the dimensionless parameter}{Radiation pressure and Solar gravity both scale approximately as \(r^{-2}\), so their ratio is approximately independent of heliocentric distance for a grain with fixed optical properties:}
\begin{equation}
\beta = \frac{F_{\mathrm{rad}}}{F_{\mathrm{grav}}},
\end{equation}
which depends on grain size, density, and optical properties. \added{The grain consequently moves in an effective central potential proportional to \(1\!-\!\beta\): gravity is reduced for \(0\!<\!\beta\!<\!1\), cancelled at \(\beta\!=\!1\), and reversed into a net outward force for \(\beta\!>\!1\).} For submicron- and micron-scale grains, $\beta$ can approach or exceed unity \citep{Sterken2012}. For grains with $\beta \ge 0.5$, the combination of radiative deceleration and a favourable encounter geometry permits a subset of particles to enter the inner Solar System with heliocentric velocities at 1~AU that are arbitrarily close to the Earth's orbital speed, despite much larger speeds at infinity.

The resulting \emph{slow-arrival} channel occupies a narrow but
non-zero region of the relevant parameter phase space and provides a physical mechanism by which grains may impact the lunar surface with velocities below the survivability threshold. This slow-arrival subset underlies the survivability fraction adopted in  \secref{subsec:fluxformalismandconstraintmodelling}.\footnote{The competing roles of radiation pressure and gravity acting on dust grains have been recognised since the early development of Solar System dynamics; \changed{an explicit treatment of this balance appears, for example, in \citet{Feynman1964}, and in a closely related radiation-pressure exercise in \citet{FeynmanExercises1964}, where the \(\beta\) parameter emerges naturally as a control variable for grain trajectories.}{Exercise~34-9 in \citet[p.~34-2]{FeynmanExercises1964} for example asks the reader to show that the ratio of Solar gravity to radiation pressure on a spherical interplanetary dust grain is independent of heliocentric distance, to determine the grain radius at which the two forces balance, and to consider whether its effective optical cross-section can exceed \(\pi R^2\). Although the exercise does not introduce this notation, it presents the same force balance represented here by
\(\beta=F_{\mathrm{rad}}/F_{\mathrm{grav}}\).}}

Quantitatively, the slow-arrival channel corresponds to a restricted region in the space of incoming grain parameters, characterised by the speed at infinity $v_\infty$, impact parameter $b$, inclination $i_\infty$, and radiation-pressure response $\beta$. The admissible region is an approximately ellipsoidal subset of the full $(v_\infty,b,i_\infty)$ parameter space near an ``optimal trajectory'' where the approach geometries yield near-tangential velocities at 1~AU.

The probability of a slow encounter scales approximately as
$(\shortchanged{\Delta v_{\max}}{\dvEncMax}/v_\oplus)^4$, where $\shortchanged{\Delta v_{\max}}{\dvEncMax}\approx5\,\kms$ is the maximum survivable encounter speed and $v_\oplus$ is the Earth's orbital velocity. \changed{This steep scaling reflects the fact that three independent constraints must be satisfied simultaneously—radiation-pressure response (mass-to-area ratio), impact parameter, and inclination—while the fourth power arises from the conversion of a restricted phase-space density into a flux, since slow-moving grains are sampled over a longer encounter time.}{The steep scaling arises because the phase-space volume admitting slow trajectories grows cubically with the maximum allowed speed. In our Solar-modulation model, the three tolerances concern $\beta$, the Earth's location in its orbit, and the relative inclination of the Earth's orbit, as derived in Appendix~B. A fourth power enters when this restricted phase-space density is converted into a flux, which introduces one additional factor of relative speed.}
\changed{Because the allowed region in this multi-parameter space is better approximated as an ellipsoid than as a rectangular prism, we introduce a geometric correction factor $\xi \simeq 0.5$, corresponding to the ratio of the volume of an inscribed ellipsoid to that of a bounding prism with the same characteristic diameter.}{A trajectory displaced strongly from the optimum in one coordinate has less remaining tolerance in the others. We therefore approximate the admissible neighbourhood as an ellipsoid rather than the rectangular prism obtained by multiplying the three full tolerance widths. The correction factor $\xi$ is taken to be of order $0.5$; its precise value is not consequential given the accuracy of the model.}

\deleted{Additional modulation occurs once grains enter the Earth-Moon system. For a grain approaching the Moon with speed \(v\), lunar gravitational focussing enhances the collision cross-section by the factor \(1+v_{\mathrm{esc}}^2/v^2\). Averaging this enhancement over the slow-arrival speed distribution defines the local modulation factor \(\eta_{\mathrm{lm}}\), where \(v_{\mathrm{esc}}=2.38\,\kms\) is the lunar escape speed.}

\added{Additional modulation occurs once grains enter the Earth-Moon system. For a grain approaching the Moon with speed \(v\), lunar gravitational focussing enhances the collision cross-section by the factor \(1+v_{\mathrm{esc}}^2/v^2\). Appendix~B defines \(\eta_{\mathrm{lm}}=1+2(v_{\mathrm{esc}}/\dvEncMax)^2\) as the focusing enhancement averaged relative to the corresponding no-focusing flux under the adopted slow-arrival speed distribution. The separate mean-speed average contributes a factor \(3/4\), so the combined local correction is \((3/4)\eta_{\mathrm{lm}}\simeq1.1\) for \(\dvEncMax=5\,\kms\), with \(v_{\mathrm{esc}}=2.38\,\kms\).}

\changed{Collecting these effects, the heliospheric modulation factor governing the survivable flux of grains may be written schematically as}{The total modulation is obtained by combining the three phase-space tolerances, the ellipsoidal correction $\xi$, the local lunar factor $\eta_{\mathrm{lm}}$, and the velocity weighting required to convert number density into flux, and then averaging over the incoming speed distribution. We express this relative to the mean isotropic flux through a unit surface in interstellar space:}
\begin{equation}
\added{\Fluxinf = \frac{1}{4}\,\eta_m\,\frac{\rhoinf}{\mRef}\,\Mean{\vinf},}
\end{equation}
\added{where $\rhoinf$ is the mass density of the technograin population in the ISM, $\eta_m$ is a normalisation factor, $\mRef$ is a representative mass for slow-channel technograins with $\beta\sim0.5$--$1$, and $\Mean{\vinf}$ is their mean speed in interstellar space. The factor $1/4$ is geometric: for an isotropic population, the projected collecting area is one quarter of the total spherical surface area. Defining $\eta_{\mathrm{mod}}\equiv\FluxRec/\Fluxinf$,}
\begin{equation}
\eta_{\mathrm{mod}}
=
\shortchanged{\frac{4}{\pi}}{\frac{9}{\pi}}\,
\xi\,\eta_{\mathrm{lm}}\,
\frac{\dvEncMax^4}{\Mean{\vinf}\,v_\oplus^3}
\int_0^\infty
\frac{v_\oplus/v}{1+(v/v_\oplus)^2}\,
f_{\vinf}(v)\,\mathrm{d}v,
\end{equation}
where $f_{\vinf}(v)$ is the distribution of incoming grain velocities at infinity. \added{The expression, including its prefactors, is derived in Appendix~B.}

For an optimistic case of the `slow' distribution, in which the Sun moves at a fixed speed of $\sim20\,\kms$ relative to the ISM, this yields
\[
\eta_{\mathrm{mod}}
\simeq
\shortchanged{7.7\times10^{-4}}{0.003}
\left(\frac{\shortchanged{\Delta v_{\max}}{\dvEncMax}}{5\,\kms}\right)^4.
\]
For a pessimistic case in which the `fast' distribution applies, we obtain
\[
\eta_{\mathrm{mod}}
\simeq
\shortchanged{1.1}{2}\times10^{-4}
\left(\frac{\shortchanged{\Delta v_{\max}}{\dvEncMax}}{5\,\kms}\right)^4.
\]

The effective inflow scale entering the lunar flux is therefore
$\eta_{\mathrm{mod}}\Mean{\vinf}\simeq\shortchanged{0.005\text{--}0.015}{0.01\text{--}0.05}\,\kms$, for which we adopt $0.01\,\kms$ as a \changed{representative}{conservative} value in the calculations made in \secref{subsec:fluxformalismandconstraintmodelling}.

\added{The velocity threshold $\dvEncMax$ is a kinematic parameter in the present calculation rather than a universal, composition-independent boundary for intact survival. Detailed material-dependent impact outcomes, including the behaviour of carbon-fibre composites and graphene, are considered in \secref{subsec:hypervelocityimpactandpartialsurvivability}; here $\dvEncMax$ serves only to define the phase-space fraction admitted by the slow-arrival channel.}

\deleted{These effects imply that the population of grains capable of surviving impact constitutes a small but physically well-defined subset of the total interstellar particulate flux. The slow-arrival channel therefore provides a natural pathway for the delivery of recognisable APs or BPs to the lunar surface and forms a central component of the model developed in \secref{subsec:fluxformalismandconstraintmodelling}.}

\added{These effects imply that a small but physically well-defined subset of the interstellar particulate population can reach the Moon at impact speeds more favourable to preservation than those characteristic of unfiltered interstellar arrival. The slow-arrival channel does not guarantee intact survival; it defines a kinematic regime in which recognisable preservation becomes more plausible. Material-specific impact outcomes are treated in \secref{subsec:hypervelocityimpactandpartialsurvivability}, while their aggregate effect on the constraints enters through the completeness factor in \secref{subsec:fluxformalismandconstraintmodelling}.}

\section{Delivery and survival on the lunar surface}
\label{sec:deliveryandsurvivalonthelunarsurface}

As summarised in Table~\ref{tab:collectors}, the Moon presents one of the most favourable environments in the Solar System for the long-term preservation of exogenous particulate material. Its lack of atmosphere and geological quiescence, together with an exposure history exceeding $4\,\mathrm{Gyr}$, make it an efficient collector of interplanetary and interstellar dust.
The Moon’s negligible global magnetic field increases exposure of surface materials to solar-wind particles and galactic cosmic rays, which can degrade exposed grains over time. However, the absence of an atmosphere, hydrosphere, and large-scale tectonic recycling processes limits physical and chemical alteration and reprocessing, so that once technogenic residues are buried beneath even modest regolith depths, their long-term preservation is favoured.
The physical processes governing dust accumulation, regolith formation, and impact modification on the Moon are well described in the standard lunar reference work of \citet[see especially Chapters~7 and~9]{Heiken1991}.

\subsection{Hypervelocity impact and partial survivability}
\label{subsec:hypervelocityimpactandpartialsurvivability}

\deleted{Given the arrival-speed distribution described in \secref{subsec:velocityfilteringandslowarrivaldynamics}, surviving APs are expected to span a continuum of impact outcomes rather than intact preservation alone. Impacts may generate diagnostic microcraters, melt residues, or spalled fragments whose morphology and composition deviate from those of natural micrometeoroids, even when the primary particle itself is destroyed.}

\added{Given the arrival-speed distribution described in \secref{subsec:velocityfilteringandslowarrivaldynamics}, preservation should be understood as a continuum rather than as a binary distinction between intact survival and complete destruction. Possible outcomes include an intact or deformed particle, larger fragments, dispersed fibres or lamellae, implanted material, melt or vapour-condensate residues, spalled fragments and diagnostic microcraters. Even where the primary particle is destroyed, combinations of crater morphology, residue chemistry, isotopic composition and surviving microstructure may preserve evidence of an anomalous material.}

\added{The position along this continuum is material-specific. Hypervelocity experiments on mineral and silicate projectiles demonstrate that some projectile material may survive as fragments or implanted residues even when the bulk impactor is disrupted \citep{Burchell1999,Burchell2008}. Advanced composites and two-dimensional materials may instead fail through delamination, fibre fracture, perforation and crystallographically directed cracking, as discussed in greater detail below. Exceptional strength under one loading geometry does not by itself imply intact survival as a freely incident lunar projectile.}

\deleted{APs engineered from refractory or non-volatile materials may exhibit enhanced survivability relative to silicate-dominated interstellar grains, increasing the likelihood that recognisable residues or inclusions persist within the regolith. Studies of natural impact products on the Moon \citep[see Chapters~5 and~7 of][]{Heiken1991} provide a baseline against which such deviations may be assessed. In the context we consider here, indicators of artificial origin would not be generic impact features, but anomalies such as microcrater morphologies characteristic of high-velocity impacts by engineered materials, non-chondritic alloy compositions, phase assemblages inconsistent with lunar or meteoritic materials, isotopic ratios outside natural ranges, or internal microstructures indicative of fabrication rather than geological processing.}

\added{Refractory or non-volatile technomaterials may preserve diagnostic residues or inclusions more readily than volatile or weakly bound materials, although the survival probability remains dependent on composition, architecture and impact geometry. Studies of natural impact products on the Moon \citep[see Chapters~5 and~7 of][]{Heiken1991} provide a baseline against which such deviations may be assessed. In the present context, indicators of artificial origin would not be generic impact features, but anomalies such as microcrater morphologies characteristic of engineered impactors, non-chondritic alloy compositions, phase assemblages inconsistent with lunar or meteoritic materials, isotopic ratios outside natural ranges, or internal microstructures indicative of fabrication rather than geological processing.}

Lower effective impact velocities are possible when grains approach from the trailing side of the Moon’s orbit, while oblique impacts reduce the normal component of the encounter velocity and distribute shock energy over a larger area, lowering peak pressures and increasing the likelihood of partial survival or diagnostic residue formation \citep[e.g.][]{Burchell1999,PierazzoMelosh2000}.
\changed{Although intact survival declines sharply above $\sim5\,\mathrm{km\,s^{-1}}$, diagnostic residues may nonetheless persist. The resulting record is therefore not binary but continuous: intact grains, fused fragments, melt residues, and impact-generated features may all contribute to the detectable technosignature population.}{For conventional micron-scale impactors, intact-survival fractions generally decline sharply above $\sim5\,\mathrm{km\,s^{-1}}$, but diagnostic fragments and residues may persist at higher speeds. The resulting archive is therefore continuous rather than binary: intact grains, deformed particles, fragments, implanted material, melt or vapour-condensate residues and impact-generated features may all contribute to the detectable technosignature population.}

\added{Carbon-fibre-reinforced polymer (CFRP) structures provide a useful analogue for hierarchical engineered matter. Hypervelocity testing of CFRP spacecraft panels at velocities of several kilometres per second has shown that the visible entry perforation may be accompanied by a substantially larger subsurface and rear-face damage zone \citep{Taylor1997,Wicklein2008}. Principal failure modes include fibre breakage, matrix cracking and decomposition, ply separation, delamination and the ejection of heterogeneous fragments. These experiments demonstrate that recognisable carbon fibres or composite fragments may persist, but also that the original laminate architecture is unlikely to remain intact under direct hypervelocity impact. For particulate technosignature searches, the relevant target may therefore be an assemblage of fibres, graphitic fragments, polymer-derived carbon, inclusions and anomalous interfacial structures rather than a complete composite particle.}

\added{Graphene requires separate treatment from bulk graphite or carbon-fibre composites. Supersonic microprojectile experiments on multilayer graphene indicate very high energy absorption per unit mass, enabled by strong in-plane bonding and rapid radial propagation of stress away from the impact point \citep{Lee2014}. Atomistic simulations extending into the hypervelocity regime nevertheless predict perforation, bond breaking and preferential cracking along crystallographic directions \citep{Xia2016,Bizao2018}. These properties suggest that graphene-based artefacts could generate distinctive folded, torn, multilayer or edge-structured fragments, but current studies do not establish the fraction of an unsupported graphene projectile that would survive impact into lunar regolith. Nor can macroscale graphene performance be assumed for defective, oxidised, irradiated or multilayer extraterrestrial material after prolonged interstellar exposure.}

\added{Arrival at the Moon may also transform the scale and morphology of the technomaterial. A thin film, sail or larger probe that survives interstellar transport need not remain macroscopically recognisable after impact. Depending on its velocity, orientation and material properties, it may be perforated, melted, vaporised in part, or fragmented into a distribution of smaller residues. Subsequent impact gardening would disperse and further comminute these remnants, progressively incorporating them into the surrounding regolith. The search proposed here may therefore detect not only deliberately manufactured micron-scale particles, but also diagnostic fragments of originally larger engineered structures.}

\added{This broader interpretation reinforces the exo-archaeological character of the search. As with terrestrial archaeological materials, the recovered object need not retain the form or function of the original artefact. Its significance may instead reside in anomalous composition, microstructure, isotopic pattern, fabrication signature or combinations of properties unlikely to arise through known natural processes.}

\added{For the present purpose, the experimentally relevant quantity is therefore not solely the intact surviving mass fraction. It is the fraction of incident technomaterial that remains recognisable through at least one diagnostic property after impact and regolith processing. This may exceed the intact-particle fraction where fragments retain engineered composition, fibre architecture, layering, dopants, inclusions or fabrication-related microstructure. Conversely, it may be smaller where heating, amorphisation or mixing with target material erases the relevant signature. We denote this broader probability of recognisable survival by \(\eta_{\mathrm{surv}}\), which forms one component of the overall detection completeness.}

\added{It is useful to separate the impact-survival and analytical-recognition stages conceptually. We may write}
\begin{equation}
\label{eqn:completenessdecomposition}
\added{\eta_{\mathrm{comp}}
=
\eta_{\mathrm{surv}}\,
\eta_{\mathrm{rec}}\,
\eta_{\mathrm{det}}.}
\end{equation}
\added{Here \(\eta_{\mathrm{surv}}\) is the fraction of delivered particles that retains at least one diagnostic signature after impact and subsequent processing, \(\eta_{\mathrm{rec}}\) is the probability that this signature remains distinguishable from natural and anthropogenic backgrounds, and \(\eta_{\mathrm{det}}\) is the probability that the adopted analytical workflow identifies it. These factors need not be independently measurable at present, and the quantitative calculations retain their product as the aggregate completeness factor \(\eta_{\mathrm{comp}}\). The decomposition nevertheless makes clear where material-specific hypervelocity experiments enter the interpretation of a detection or null result.}

\subsection{Burial, mixing, and post-depositional processing}
\label{subsec:burialmixingandpostdepositionalprocessing}

Once incorporated into the lunar regolith, surviving particles would be redistributed by micrometeorite bombardment and impact-driven regolith gardening. The depth to which material is mixed increases with time but does not do so linearly; instead, the characteristic mixing depth grows approximately as a power law in age, reflecting the cumulative effects of impacts of varying size \citep[e.g.][]{Horz1997,Speyerer2016,Costello2021}. Empirical and modelling studies indicate that material deposited at the surface is progressively mixed to depths of order decimetres on $10^{8}$--$10^{9}$~yr timescales, and to metre scales over several gigayears. Shallower layers may thus be expected to preferentially sample recent influx while deeper layers preserve progressively older contributions, as documented in lunar sample studies and stratigraphic analyses
\citep{Joy2012}.

Burial provides increasing protection against surface-driven degradation processes. Solar-wind sputtering affects only the outermost microns of exposed
grain surfaces and is rapidly suppressed once particles are buried even at millimetre depths \citep{Pieters2000,Grun2011}. Cosmic-ray irradiation penetrates more deeply and can induce \emph{amorphisation}—the disruption of crystalline
lattice order—and can modify isotopic ratios and generate point defects \citep{Burgess2018}. Thermal cycling driven by extreme diurnal temperature
variations imposes mechanical stresses that may fracture brittle grains or weaken composite materials, while secondary micrometeorite impacts can further fragment particles or redistribute them within the regolith \citep{Horz1997}.

Despite these ongoing processes, diagnostic technosignature features \changed{are expected to persist}{may persist}.
Surface erosion does not necessarily eliminate internal structures, isotopic anomalies, or compositional signatures.
Additionally, while fragmentation does not increase the total amount of surviving material, it may redistribute residues over areas larger than the original impact footprint through ejecta dispersal, increasing the chance that a later, spatially independent sample intersects a diagnostic fragment rather than the original impact site.

Even heavily processed grains may retain isotopic or elemental anomalies that are unusual in a lunar context (see, e.g., studies of isotopic signatures in presolar and meteoritic grains by \citealt{Zinner2014}), providing a potential diagnostic even in the absence of intact structures.

\subsubsection{A quantitative model for regolith mixing}
Regolith gardening arises because impact events excavate material from transient craters and redistribute it both vertically and laterally. Larger impacts overturn material to greater depths but occur less frequently, so that the cumulative effect of impacts over time can be characterised statistically.
Thus, it is common to define a mixing depth $\Lambda(t)$ for material of age $t$, the mean depth to which all impact events in the intervening time are expected to overturn the regolith. \changed{Typically, the mixing depth is expressed as a power law:}{Integrating over the crater-size and impact-frequency distributions, the characteristic mixing depth is found to increase with elapsed time $t$ as a sublinear power law:}
\begin{equation}
\Lambda(t) = \Lambda_0 (t/t_0)^q .
\end{equation}
\changed{where $\Lambda_0$ is the mixing depth at a reference time $t_0$, and $0 < q < 1$.}{Here \(\Lambda_0\) is the depth reached at the reference age \(t_0\), while $0 < q < 1$ encodes the sublinear growth of the mixed layer as progressively larger impacts become rarer. The relation is a statistical model of cumulative gardening, not the trajectory of any individual grain.} \changed{This approximation breaks down at ages above around 1--4 Gyr, when the cratering was more intense and regolith gardening faster}{This approximation becomes less reliable at the oldest lunar ages, when impact rates were higher and regolith gardening proceeded more rapidly} \citep{Costello2018,Costello2021}.
In our model, APs arrive at the Moon's surface with a received flux $F_r$ and are distributed randomly through the mixing depth. \changed{A survival fraction $S(t)$ of grains with age $t$ survive to the present. Let $f_z(z | t)$ be the probability density of the depth for material of age $t$. The number density of grains at depth $z$ that have accumulated over a time $t_{\mathrm{int}}$ is then}{Of the grains deposited a time $t$ ago, the fraction recognisably surviving is $S(t)$, and $f_z(z | t)$ is the probability density for the depth $z$ of these grains. Integrating these contributions over all deposition ages gives the present number density}
\begin{equation}
n_{\rm reg}(z) = \int_0^{t_{\mathrm{int}}} F_r S(t) f_z(z|t)\,\mathrm{d}t .
\end{equation}
\added{The dimensions are grains per unit volume: $F_r\,\mathrm{d}t$ supplies grains per unit area, while $f_z$ has dimensions of inverse length.}
\changed{The number of APs per unit area down to a sampling depth $z_s$ is}{Integrating the depth-dependent number density from the surface to the sampling depth $z_s$ gives the number of retained grains per unit sampled area:}
\begin{equation}
\Sigma_{\rm reg}(z_s) = \int_0^{z_s} \int_0^{t_{\mathrm{int}}} F_r S(t) f_z(z|t)\,\mathrm{d}t\,\mathrm{d}z.
\end{equation}
\changed{We can then define an effective sampling time}{Dividing this sampled column density by the constant incident flux defines an effective sampling time,} $t_s = \Sigma_{\rm reg}(z_s)/F_r$. \changed{When we sample down to $z_s$, the number of artificial grains we catch is the same as if we had sampled the entire column of regolith for artificial grains of age $t_s$ or less; it is a convenient way of relating the number of APs discovered to the received flux.}{A sampled column extending to $z_s$ therefore contains the same expected number of grains as an unmixed surface exposed to the same constant flux for a duration $t_s$. This converts the depth-dependent archive into the effective exposure time used in the flux constraints below.} \changed{In our model, we assume the technograin flux is constant. Additionally, we take the survival function $S(t)$ to be $1$ for $t \le t_{\mathrm{int}}$ and $0$ for $t > t_{\mathrm{int}}$. Impact cratering and other processes could destroy APs or render them unrecognizable, which then shortens the integration time.}{For the baseline mixing calculation, the technograin flux is assumed constant and $S(t)=1$ throughout $0\le t\le t_{\mathrm{int}}$. A time-dependent post-depositional survival law could be introduced through $S(t)$; in the quantitative constraints below, losses of recognisability are retained in the aggregate completeness factor $\eta_{\mathrm{comp}}$, avoiding double counting.}
\changed{The simplest depth distribution is uniform down to the mixing depth:}{We adopt the simplest depth distribution, in which material of age $t$ is uniformly distributed between the surface and its characteristic mixing depth \(\Lambda(t)\), and is absent below that depth:}
\begin{equation}
f_z(z|t) = \begin{cases}
	   1/\Lambda(t) = 1/\Lambda_0 \cdot (t/t_0)^{-q} & \text{if}~z \le \Lambda(t)\\
           0                                        & \text{if}~z > \Lambda(t)
           \end{cases} .
\end{equation}
\added{The distribution integrates to unity over depth for every age $t$. It is intentionally idealised: real gardening profiles may be surface-weighted or possess tails extending below the nominal mixing depth.}

\changed{The particles with ages less than $t_{\mathrm{int}}$ have a number density of}{At a fixed depth \(z\), only particles old enough for their mixing depth to have reached \(z\) contribute. Integrating from that minimum age to $t_{\mathrm{int}}$ gives}

\begin{equation}
    \label{eqn:number_density_regolith}
    n_{\rm reg}(z;t_{\mathrm{int}})
    =
    \frac{F_r t_{\mathrm{int}}}{1-q}
    \left[
    \frac{1}{\Lambda(t_{\mathrm{int}})}
    -
    \frac{1}{z}
    \left(
    \frac{z}{\Lambda(t_{\mathrm{int}})}
    \right)^{1/q}
    \right],
    \end{equation}

\changed{when $z \le \Lambda(t)$ and $0$ at greater depths.}{when \(z \le \Lambda(t_{\mathrm{int}})\), and \(n_{\rm reg}=0\) at greater depths.}
\added{In this model, young particles have been mixed only into the uppermost layers, whereas older particles are distributed over a greater range of depths.}

\changed{A further integration and normalization gives us}{Integrating this density from the surface to \(z_s\) and dividing by \(F_r\) yields the effective sampling time}
\begin{equation}
    \label{eqn:effective_sampling_time}
    t_s
    =
    \begin{cases}
    \displaystyle
    \frac{t_{\mathrm{int}}}{1-q}
    \left[
    \frac{z_s}{\Lambda(t_{\mathrm{int}})}
    -
    q
    \left(
    \frac{z_s}{\Lambda(t_{\mathrm{int}})}
    \right)^{1/q}
    \right],
    &
    z_s<\Lambda(t_{\mathrm{int}}),
    \\[6pt]
    t_{\mathrm{int}},
    &
    z_s\geq\Lambda(t_{\mathrm{int}}).
    \end{cases}
    \end{equation}

\changed{When $z_s \ll \Lambda(t_{\mathrm{int}})$, we only skim the surface of the Moon and most of the APs remain buried deeper, with $t_s \approx t_{\mathrm{int}} z_s/[(1 - q) \Lambda(t_{\mathrm{int}})]$.}{When $z_s \ll \Lambda(t_{\mathrm{int}})$, the sample probes only the uppermost part of the mixed layer and most APs remain buried more deeply, giving $t_s \approx t_{\mathrm{int}} z_s/[(1-q)\Lambda(t_{\mathrm{int}})]$.} \changed{On the other hand, if $z_s \ge \Lambda(t_{\mathrm{int}})$, the sampled column exhausts the available grains, giving us $t_s = t_{\mathrm{int}}$.}{Once $z_s \ge \Lambda(t_{\mathrm{int}})$, the sampled column includes the complete surviving population and $t_s=t_{\mathrm{int}}$.}

The convergence of $t_s$ to $t_{\mathrm{int}}$ depends on whether APs are more or less concentrated in the uppermost layers than assumed by the uniform model. \changed{With a distribution with a tail extending to greater depth than $\Lambda(t)$ like an exponential distribution, $t_s$ is somewhat smaller. Contrarily, the grains may be concentrated towards the top of the mixing depth.}{A profile with a tail extending below $\Lambda(t)$, such as an exponential distribution, yields a somewhat smaller $t_s$; conversely, a surface-weighted profile yields a larger value.} In the \citet{Costello2018} model, the number of overturns experienced at a given depth controls the degree of mixing, with full mixing requiring hundreds or even thousands of overturns. \citet{Costello2021} distinguish between an approximately homogeneous reworking zone and a deeper degradation zone.

We adopt the mixing-depth parameterisation from \citet[equation~15]{Costello2020}, based on the \citet{Speyerer2016} crater function: $\Lambda(t)=3.45\times10^{-5}\,\meter\,(t/\yr)^{0.47}$. With a nominal $t_{\mathrm{int}}=4\,\Gyr$, the mixing length of the oldest material is \changed{$\Lambda(4\,\Gyr)=1.12\,\meter$}{$\Lambda(4\,\Gyr)\simeq1.1\,\meter$}. A one-metre column is therefore well matched to the regolith depths within which most technograins are expected to reside. \changed{The sampling time is found to be \(3.95\,\Gyr\) in the uniform distribution model, and \(2.95\,\Gyr\) for an exponential density distribution.}{The effective sampling time in the upper metre is approximately $4.0\,\Gyr$ under the uniform model. For the remainder of the paper, we retain $t_s=3.5\,\Gyr$ as a conservative intermediate fiducial value that allows for more deeply tailed mixing profiles.}

\subsection{Signature preservation and diagnostic features}
\label{subsec:signaturepreservationanddiagnosticfeatures}

The lunar regolith preserves numerous mineralogical and microstructural indicators of past events over gigayear timescales. APs or BPs may leave several classes of preserved technosignatures.
Microcraters produced by natural impactors exhibit a range of well-characterised geometries and melt textures \citep{Horz1997,Burchell1999,Jaeger2021}. Artificial grains would be expected to produce analogous impact features, but potentially with morphologies or residue compositions inconsistent with known meteoritic populations.

Compositional anomalies, such as engineered alloys, advanced ceramics, or non-solar isotopic patterns, may stand out against the backdrop of lunar and meteoritic material.
\deleted{Internal structures, including layered fabrication, patterned voids, or circuitry-like geometries, may be observable using scanning electron microscopy, FIB tomography, or nano-CT \citep{Keller2011}. Encoded architectures—including crystalline doping schemes, molecular encodings, or quantum-dot arrays—may retain detectable order even after partial melting or shock processing.}

\added{Internal structures, including layered fabrication, patterned voids, circuitry-like geometries or deliberately engineered defect centres \citep{Manson2006,McLellan2016}, may be observable using scanning electron microscopy, FIB tomography, or nano-CT \citep{Keller2011}. Encoded architectures—including crystalline doping schemes, molecular encodings or quantum-dot arrays \citep{ToughLemarchand2004,Yin2017,Li2025MEMS}—may retain detectable order even after partial melting or shock processing.}

\changed{Lunar agglutinates have been shown to survive long-term repeated impact cycles \citep{Heiken1991,Yano1994,Kerschmann1995,Yano1997,Joy2012}, and may potentially preserve inclusions of refractory technomaterial entrapped within their glass matrices.}{Lunar agglutinates record repeated impact processing over long timescales \citep{Heiken1991,Yano1994,Kerschmann1995,Yano1997,Joy2012} and may preserve inclusions of refractory technomaterial entrapped within their glass matrices.}

Collectively, these considerations show that the detection space for micron-scale technosignatures extends well beyond intact grains, encompassing a range of durable residues and structural signatures preserved within the lunar regolith. Melt residues, microcraters, spalled fragments, and agglutinate inclusions all represent viable and potentially long-lived indicators of artificial material in the lunar environment.

The impact, burial, and degradation processes described above define the physical basis for the effective sampling times and survivability assumptions used in \secref{subsec:fluxformalismandconstraintmodelling}.

\section{Detection strategies}
\label{sec:detectionstrategies}

Any search for micron-scale technosignatures on the Moon necessarily presupposes access to lunar material, either through the return of samples to Earth or via \textit{in situ} analytical capabilities deployed on the surface. A growing number of governmental and commercial lunar missions now make such access increasingly realistic, including recent and forthcoming sample-return efforts, robotic landers, and sustained surface operations. Recent lunar sample-return missions and planned surface activities— including China’s Chang’e programme, NASA’s Artemis initiative, and commercial lunar payload services (e.g. CLPS)—demonstrate that both returned regolith and \textit{in situ} microscopic analysis are now realistic components of near-term lunar exploration \deleted{\citep{Li2022Change5,NASArtemisScience2022,Colaprete2022CLPS}} \added{\citep{NASAArtemisIII2020,Li2022Change5,NASEMCLPS2021}}.
\subsection{\texorpdfstring{\added{Detection, identification, and evidential standards}}{Detection, identification, and evidential standards}}
\label{subsec:detectionidentificationandevidentialstandards}

\added{Any future detection claim would depend not simply on recovering anomalous particles, but on correctly distinguishing technological materials from natural and anthropogenic alternatives. This distinction is fundamentally an evidential rather than purely analytical problem. Accordingly, we distinguish three successive questions \citep{SheikhAxes2020,LingamFrameworks2023,WrightSETI2026}:}

\begin{enumerate}
\renewcommand{\labelenumi}{(\roman{enumi})}
\item \added{Was an anomalous particle detected?}
\item \added{Can its properties be explained by known natural or human processes?}
\item \added{If not, does the remaining evidence collectively support a technological interpretation?}
\end{enumerate}

\added{The proposed search should therefore be viewed as an exercise in comparative hypothesis testing rather than the identification of any single ``smoking gun.''}

\added{\subsubsection{False positives.} False positives comprise particles that appear anomalous but ultimately prove to have non-technological origins. Several classes deserve explicit consideration.}

\added{The first consists of terrestrial contamination introduced during collection, curation, transport or laboratory analysis. Such contamination is already a central concern within planetary sample science and is addressed through clean-room protocols, procedural blanks, witness plates, contamination-knowledge archives and independent laboratory replication \citep{DworkinContamination2018,McCubbinCuration2019}. Large-scale construction, resource-processing and \textit{in situ} inspection operations may not satisfy sample-return clean-room standards; in those settings, contamination control will depend more heavily on local background libraries, process blanks, materials inventories and repeated measurements than on excluding terrestrial material at source.}

\added{The second comprises anthropogenic materials already present on the Moon. These include spacecraft structural materials, rocket-exhaust residues, metallic fragments, polymers, paints, lubricants, insulation materials, composite fibres and impact-generated debris associated with both historic and future lunar exploration \deleted{\citep{DaviesWagner2013,Li2022Change5,NASArtemisScience2022,Colaprete2022CLPS}}
\added{\citep{DaviesWagner2013,NASAArtemisIII2020,Li2022Change5,NASEMCLPS2021}}. Increasing lunar activity makes comprehensive documentation of these materials progressively more important.}

\added{A third category consists of unusual but ultimately natural materials, including meteoritic alloys, refractory condensates, presolar grains, impact glasses, unusual isotopic assemblages and rare mineral phases \citep{Heiken1991,Joy2012,Zinner2014}. Although individually uncommon, such materials demonstrate that apparent novelty alone is insufficient evidence for technology.}

\added{Accordingly, no single material property should be regarded as diagnostic. Identification instead depends upon combinations of mutually reinforcing characteristics that collectively become increasingly improbable under natural or anthropogenic explanations \citep{SheikhAxes2020,LingamFrameworks2023}.}

\added{\subsubsection{{False negatives.}} False negatives are equally important but have received comparatively little attention within technosignature studies. An originally technological material may lose recognisable characteristics through hypervelocity impact, shock metamorphism, melting, vapour condensation, oxidation, radiation damage, micrometeoroid gardening or prolonged geochemical alteration.}

\added{Advanced materials may also become difficult to recognise if only fragments survive. Carbon-fibre composites may become isolated fibres or graphitic fragments; layered materials may delaminate; ceramics may fracture into apparently ordinary mineral grains; and metallic glasses may partially crystallise. Some engineered materials may ultimately become indistinguishable from naturally occurring phases using conventional analytical techniques \citep{Taylor1997,Wicklein2008,Lee2014,Xia2016,Bizao2018}.}

\added{These processes imply that null detections constrain recognisable technomaterial rather than all technomaterial. The distinction reinforces the importance of experimentally determining survival and recognition efficiencies for candidate material classes.}

\added{\subsubsection{{A hierarchical evidential framework.}} Rather than relying upon any single diagnostic criterion, we propose that candidate technoparticles should be evaluated using multiple independent lines of evidence. Examples include:}

\begin{itemize}[
  leftmargin=2.2em,
  labelsep=0.6em,
  itemsep=0.1em,
  parsep=0pt,
  topsep=0.4em
]

\item \added{elemental composition inconsistent with known geological processes;}

\item \added{anomalous isotopic ratios;}

\item \added{engineered crystallographic or microstructural organisation;}

\item \added{layered, periodic or hierarchical architectures inconsistent with natural growth;}

\item \added{manufacturing residues, dopants or trace-element patterns;}

\item \added{association with diagnostic microcrater morphologies;}

\item \added{spatial clustering inconsistent with expected meteoritic delivery; and}

\item \added{reproducibility across independent samples and laboratories.}

\end{itemize}

\added{Individually, each property may admit natural explanations. Collectively, however, the probability of all occurring simultaneously through unrelated natural processes rapidly becomes much smaller. This cumulative evidential approach mirrors established practice in archaeology, forensic science and planetary sample return, where interpretations are based on converging independent observations rather than any single measurement, and it is consistent with confidence frameworks developed for life and technosignature assessment \citep{GreenCoLD2021,LingamFrameworks2023,McCubbinCuration2019}.}

The detection of micron-scale technosignatures in lunar material will require analytical methods capable of identifying both intact surviving particles and the diverse classes of residues produced by hypervelocity impacts. Unlike previous studies that have applied orbital imagery to the identification of macroscopic surface anomalies or anthropogenic artefacts
\citep[e.g.][]{DaviesWagner2013}, the search for Arkhipov Particles and Bracewell Particles must operate predominantly at the microscale. %Those earlier investigations were not SETA studies, but they established important methodological foundations for recognising artificial features against a complex lunar geological background.

Contemporary laboratory instrumentation now permits systematic analysis of microcraters, melt textures, unusual inclusions, and internal grain architectures using scanning electron microscopy (SEM), energy-dispersive X-ray spectroscopy (EDS), nanoscale X-ray tomography, and focused-ion-beam (FIB) sectioning \citep{Horz1997,Wentworth1999,Keller2011,Jaeger2021,Matsumoto2024,Gu2025}. These tools are capable of resolving structural or compositional anomalies that may indicate technogenic origin even when an incoming particle has been partially or largely destroyed.

\deleted{Machine-learning methods further extend the detection space by enabling high-throughput screening of large grain populations and regolith textures. In this context, supervised learning approaches may be used to identify predefined classes of features based on labelled training data, whereas unsupervised methods search for statistical anomalies without requiring prior specification of artificial signatures. Object-detection pipelines such as YOLO-ET and YOLO-ETA are trained models that learn morphological and textural priors from large image sets, allowing them to discriminate between natural and anomalous features in both laboratory and orbital datasets. YOLO-ET, developed expressly for microparticle analyses in extraterrestrial and analogue samples \citep{Pinault2024a,Pinault2024b}, facilitates rapid triage of grains, microcraters, and agglutinates whose morphologies deviate from expected lunar or meteoritic material.}

\added{The scale of future lunar-regolith surveys makes automated analysis essential, but no single AI architecture is likely to perform every stage of the analytical workflow. Instead, future pipelines will probably combine multiple complementary model classes \citep{LeCun2015,Bommasani2021}.}

\added{Object-detection systems, including architectures such as YOLO, are well suited to rapidly locating candidate particles within large optical and electron-microscope datasets \citep{Redmon2016,Jiang2022}. Foundation vision models can learn general morphological representations from very large image collections without requiring exhaustive manual labelling, improving robustness across previously unseen particle classes \citep{Bommasani2021,Radford2021}.}

\added{Beyond image analysis alone, vision-language models (VLMs) provide a more general framework in which microscopy images, spectroscopic measurements, crystallographic information, elemental abundances, isotopic ratios and associated scientific metadata can be interpreted jointly. Such models can reason across heterogeneous observations, generate interpretable hypotheses, and provide transparent explanations of why particular particles merit further investigation \citep{Radford2021,Bommasani2021}.}

Although this work is centred on micron-scale artefacts, macro-scale detection from lunar orbit remains relevant as a contextual and prioritisation tool. Geological or regolith-level heterogeneities—such as variations in impact history, agglutinate abundance, maturity, or preservation state—may, in some scenarios, give rise to surface-scale anomalies detectable in orbital data. Orbital anomaly detection may thus complement grain-scale analyses by identifying candidate regions for targeted sampling rather than constituting an independent SETA modality. The YOLO-ETA framework \citep{Pinault2026a} provides a practical means of integrating these scales, allowing orbital reconnaissance and regolith-level analysis to be treated as components of a single, hierarchical detection strategy.

The operational context for YOLO-ET and related computer-vision pipelines spans both returned samples and prospective \textit{in situ} analyses. In the near term, such models can be applied to archival Apollo and Luna soils, to the growing suite of Chang’e returned materials, and to samples anticipated from Artemis and commercial lunar missions. Longer-term prospects include autonomous, on-surface implementations integrated into rover or lander payloads, where microscopic imaging can be performed directly within the regolith without the constraints of sample return. Together, these pathways for acquisition and analysis position high-resolution microscopy—whether Earth-based or \textit{in situ}—as a viable means of surveying large grain populations for microstructural or morphological technosignatures as lunar exploration capabilities continue to expand.

\subsection{\texorpdfstring{\added{AI-assisted scientific reasoning}}{AI-assisted scientific reasoning}}
\label{subsec:aiassistedscientificreasoning}

\added{Future analytical workflows are likely to proceed through several successive stages. Candidate particles would first be detected automatically within large imaging datasets. Their morphology, composition, crystallography and spectroscopic signatures would then be characterised using multimodal AI models capable of integrating heterogeneous measurements. The resulting evidence could be compared with reference libraries of natural minerals, meteoritic particles, anthropogenic materials and experimentally generated technomaterials. Particles exhibiting statistically unusual combinations of properties would subsequently be prioritised for detailed laboratory examination.}

\added{This approach shifts the role of AI from simple image classification towards scientific evidence integration. Rather than asking whether an isolated image resembles a known object, the system evaluates the consistency of multiple independent observations before assigning a probabilistic interpretation.}

\added{Unsupervised methods remain particularly important because genuinely novel technomaterials may have no labelled training examples. Autoencoders, variational autoencoders and related representation-learning techniques can identify particles whose combined properties occupy regions of parameter space not represented by known natural or anthropogenic materials \citep{KingmaWelling2014}. Such anomaly detection is complementary to supervised classification and may be especially valuable during the early stages of Solar System exo-archaeology, when the defining characteristics of candidate technosignatures remain uncertain.}

\added{Generative AI provides a complementary capability. Physics-informed generative models can augment scarce experimental datasets by simulating impact morphologies, degradation pathways and preservation states that are difficult to obtain experimentally. They may also assist in image restoration, super-resolution, uncertainty estimation and the exploration of alternative geological histories. Their principal value lies in expanding the range of plausible hypotheses presented to investigators rather than replacing laboratory measurements \citep{KingmaWelling2014,Bommasani2021}.}

\subsection{Grain-scale detection and the role of YOLO-ET}
\label{subsec:grainscaledetectionandtheroleofyoloet}

The most direct route to identifying APs is through the examination of returned regolith samples or by \textit{in situ} microscopy. For this purpose Pinault \textit{et al.} developed YOLO-ET, a convolutional neural-network model adapted for the classification of grains and micro-craters in optical and scanning-electron microscopic images \citep{Pinault2024a,Pinault2024b}. YOLO–ET is designed to operate as a high-throughput classification system,
identifying grains that exhibit morphological, textural, or geometric anomalies
relative to natural silica-aerogel-captured, lunar, or asteroidal/micrometeorite
particles, as well as their telltale impact signatures. Its training set
comprises analogue datasets sourced from the JAXA Astrobiology Project Tanpopo
and its collection of surface contaminants on silica aerogels positioned on the
Kibo module of the International Space Station, and draws additionally from high velocity impact and microcratering studies \deleted{\citep{Horz1997,Yano1994,Yano1997,Burchell1999,Warneke2001}}\added{\citep{Flynn1994,Yano1994,Horz1997,Yano1997,Burchell1999,Warneke2001,Yamagishi2014,Yamagishi2021}}. The model is expressly designed for detection of technodust impacts, trainable on smart-dust analogues and anthropogenic contaminants, all within a compact YOLO-based architecture optimised for the detection of small objects
\citep{Redmon2016,Jiang2022}.

\begin{figure}[t]
  \centering
  \includegraphics[width=\columnwidth,trim=100 50 120 50,clip]{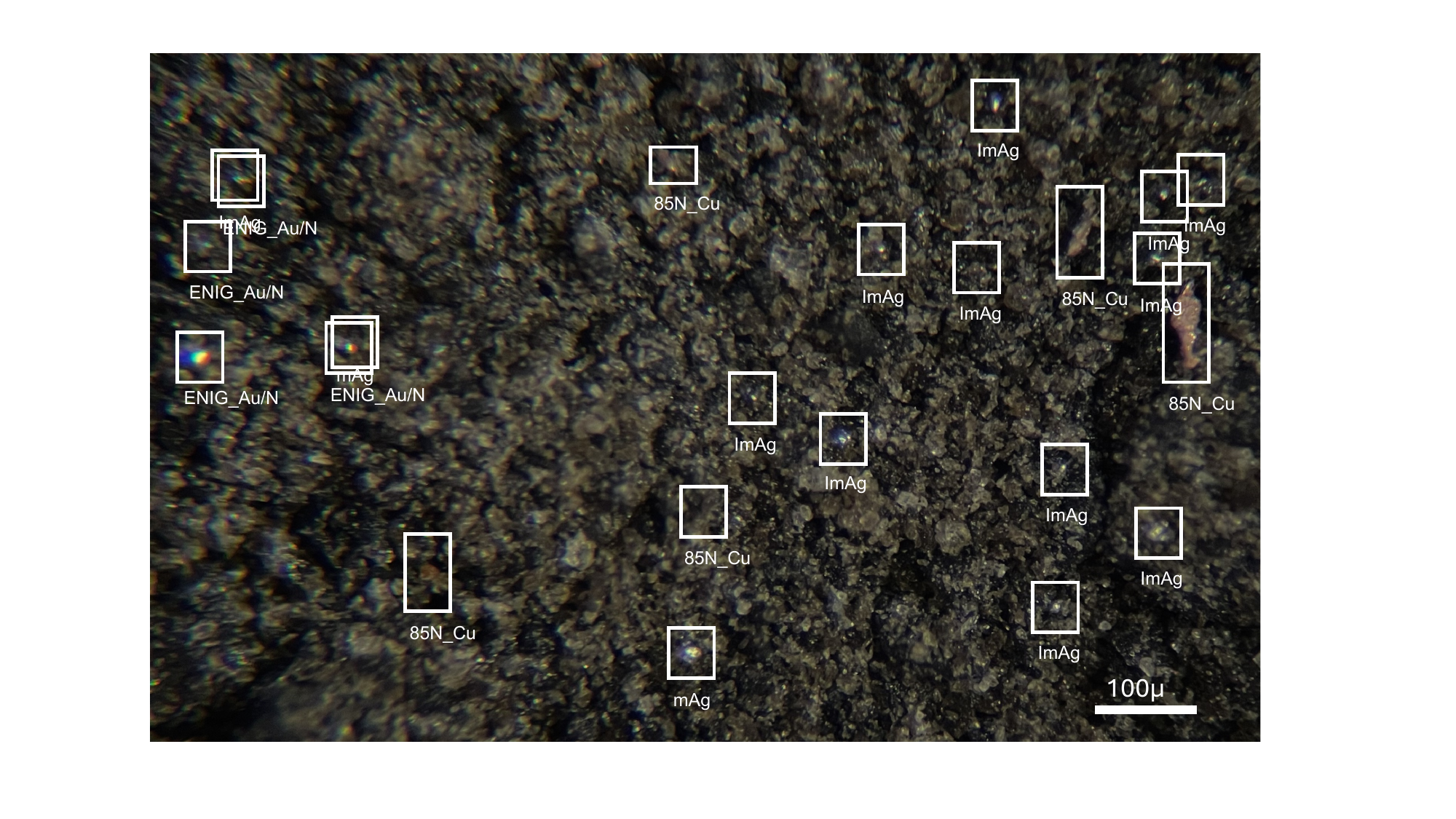}
  \caption{Illustrative application of the YOLO-ET machine-vision pipeline to a heterogeneous micron-scale particulate field. The image shows a prepared JSC-1 lunar regolith analogue sample containing mixed natural grains and engineered spacecraft-derived particulates. Bounding boxes indicate grain classes identified by the model, in this case, Cesium Astro's software-defined-radio (SDR) and Nightingale antenna products for low Earth orbit satellites, granulated and sieved to <80 microns, including Arlon 85N polyimide and copper layers, ENIG gold over electroless nickel plating, and Rogers 4305B with copper layers \changed{(CesiumAstro SDR-1000 and SAPA-1; \citealt{CesiumAstroSDR2104})}
    {(CesiumAstro SDR-1000 and SAPA-1)}. YOLO-ET was developed for extraterrestrial and analogue microparticle classification \citep{Pinault2024a,Pinault2024b} and has since been extended to multi-scale anomaly detection in orbital datasets via YOLO-ETA \citep{Pinault2026a}. Related applications to asteroid regolith and micrometeorite-feature correlations are in preparation \added{based on asteroid Ryugu analyses} \deleted{\citep{Pinault2026b}}\added{\citep{Genge2023Polar,Genge2024MetSoc,Genge2024NatAstron,Pinault2026b}}
. The present figure demonstrates the model’s capability to isolate candidate anomalous grains within a visually complex regolith matrix.}
\label{fig:yolo_demo}
\end{figure}

To highlight its implementation in practice, Figure~\ref{fig:yolo_demo} shows an example application of the YOLO-ET pipeline to a heterogeneous micron-scale particulate field in a lunar regolith analogue. The workflow illustrated shows how automated object detection can isolate candidate anomalous grains within visually complex regolith matrices, thereby reducing the search space for subsequent high-fidelity laboratory analysis. In this framework, machine vision functions as a variance amplifier: it does not determine artificial origin, but identifies candidate grains whose statistical, morphological, or contextual properties warrant targeted laboratory scrutiny.

\subsection{Forensic analysis of candidate APs and BPs}
\label{subsec:forensicanalysisofcandidategrains}

Candidate APs, BPs, or their residues flagged by machine-vision tools will require laboratory characterisation to determine their nature. Several complementary techniques could support this process. Scanning electron microscopy (SEM) provides high-resolution imaging of surface morphology and microcrater rims \citep{Keller2011}. Energy-dispersive X-ray spectroscopy (EDS) yields elemental compositions and can highlight non-chondritic alloys or engineered materials. Secondary ion mass spectrometry (SIMS) enables isotopic measurements at sub-micron scales, revealing non-solar isotopic patterns that may indicate artificial origin \citep{Zinner2014}. Focused-ion-beam (FIB) tomography and nano-computed tomography (nano-CT) permit volumetric reconstruction of internal structure, revealing layered fabrication, patterned voids, or other architectures inconsistent with natural lunar grains \citep{Jaeger2021}.

\added{Given the extraordinary evidential standard required for any claim of extraterrestrial technology, interpretability is as important as classification accuracy. AI systems should therefore produce traceable evidential pathways identifying the observations responsible for a given prediction, together with calibrated confidence estimates and explicit comparison against competing natural and anthropogenic explanations. Automated methods should be regarded as scientific decision-support tools rather than autonomous discovery engines \citep{Bommasani2021}.}

These methods would provide the evidential framework required to evaluate technosignatures robustly. Even when the primary particle is largely destroyed, diagnostic residues may remain identifiable through anomalous elemental abundances, non-solar isotopic ratios, or internal microstructures revealed by SEM, SIMS, or FIB–tomographic analysis.

% NOTE(LJP) This 5.3 supercedes the one %below it.
\subsection{Agglutinates as archives of technomaterial}
\label{subsec:agglutinatesasarchivesoftechnomaterial}

A substantial fraction of lunar soil consists of agglutinates: glass-rich grains produced by repeated micrometeorite impacts. Their formation involves rapid melting and quenching, during which heterogeneous inclusions, including nanophase iron droplets and foreign clasts, become entrained within a glass matrix \deleted{\citep[Chapters~5 and~7]{Heiken1991}}\added{\citep[Chapters~5 and~7]{Heiken1991}; see also \citep{KellerMcKay1993,KellerMcKay1997,KellerClemett2001}}. As emphasised by Russell Kerschmann (pers.\ comm.\ 2025), this process provides a natural mechanism by which refractory exogenous materials -- including potential technogenic particles -- may be encapsulated and preserved within the lunar regolith.

Impact temperatures inferred from lunar flash observations, vaporisation models, and laboratory heating experiments span approximately $1500$ to more than $5000~\mathrm{K}$ \deleted{\citep{Burchell1999,Yano1994,Yano1997}}\added{\citep{Burchell1999,Pokorny2019,Szalay2020,Yano1994,Yano1997}}. While such temperatures destroy most silicates, certain engineered materials -- including tungsten alloys, advanced ceramics, and rhenium-bearing superalloys -- can withstand these extremes. If such technomaterial becomes entrained within an agglutinate melt, it may survive as an inclusion exhibiting recognisable crystalline or compositional signatures.

Agglutinates therefore represent both a destructive and a preservational pathway. Their formation can destroy fragile grains, yet preserve refractory inclusions with high fidelity over long timescales. Importantly, agglutinates are readily isolated by magnetic or density-based separation techniques, facilitating efficient screening. Once isolated, they may be examined using confocal microscopy, spectroscopy, or high-resolution imaging to identify anomalous inclusions indicative of artificial origin.

Agglutinates may also provide a natural amplification of spatial and compositional variance: rare or anomalous grains, once encapsulated, may become statistically overrepresented within agglutinates, improving detectability even when the overall technomaterial flux remains extremely low. This distinction between variance and bulk enrichment is developed further in the following section.

\subsection{Additional pathways and diagnostic environments}
\label{subsec:additionalpathwaysanddiagnosticenvironments}

Although grain-scale detection remains central to particulate SETA, the lunar surface is not expected to behave as a perfectly uniform archive at all spatial scales, which may offer further pathways to investigation. Even under spatially uniform delivery, stochastic arrival, regolith processing, and preservation effects will introduce local variance in the abundance, exposure state, or detectability of particulate material. These effects do not require any active concentration mechanism, nor do they alter the global flux constraints developed in \secref{subsec:fluxformalismandconstraintmodelling}; rather, they influence where technomaterial may be most readily detected or preserved.

Several well-established lunar processes can modulate the near-surface
distribution or exposure history of small particles. Electrostatic lofting may potentially redistribute fine grains near terminators, crater rims, and high-latitude terrain \citep{Grun2011}. Regolith migration and impact gardening can transport particles laterally and vertically over time, producing stratigraphic and spatial variability without changing the integrated surface fluence \citep{Costello2021}. Permanently shadowed regions at the lunar poles may further alter preservation conditions by suppressing thermal cycling and volatile loss, although their role in retaining refractory particulates remains an open question.

These sources of spatial variance are relevant primarily as \textit{diagnostic tools}. Orbital observations and surface imaging would not be expected to detect micron-scale technosignatures directly, but they could still prove useful by identifying regions whose regolith properties depart from local geological expectations, thereby informing the prioritisation of sampling locations. Orbital detection methods can thus complement laboratory analysis of returned or \textit{in situ} samples. And while the absence of %orbital anomalies 
anomalies detected from orbit does not weaken the particulate SETA framework, their presence would provide valuable guidance for targetted investigation.

Bulk-processing approaches could potentially provide an additional, laboratory-based pathway for interrogating large regolith volumes. Dissolution and leaching experiments on lunar soils demonstrate that glassy phases, nanophase iron, and high-surface-area grains can release metals and other constituents into solution under controlled conditions \citep{Kerschmann2021}. Applied judiciously, such assays may enable high-sensitivity screening for anomalous materials or engineered particulates without presupposing intact grain recovery. These techniques also have ancillary value for astrobiology and contamination studies, further broadening their utility.

\subsection{Advanced technomaterial analogues.}
\label{subsec:advancedtechnomaterialanalogues}

Beyond analogue studies tied to present-day materials, a future experimental programme should broaden into advanced engineered technomaterial analogues. Since our dynamical modelling suggests that a non-negligible fraction of interstellar grains may reach the Moon at sub–$5\,\mathrm{km\,s^{-1}}$ encounter velocities owing to solar radiation-pressure deceleration, intact or partially preserved technomaterial may be more common than earlier estimates assumed. This prospect argues for a second phase of laboratory work in which hypervelocity experiments are performed not only with simple engineered analogues \added{\citep{Hibbert2017}}, but also with more complex materials relevant to speculative technosignatures  --  including high–temperature ceramics, multilayer composites, crystalline doping lattices, inscribed-matter substrates \citep{RoseWright2004}, quantum–dot arrays, and programmable–matter prototypes. Such materials may also be deliberately embedded as inclusions within experimental agglutinates to test survivability and signature retention across a range of melting and quenching conditions. These data, in turn, will support the next generation of machine-learning models by providing training sets that encompass both natural inclusions and engineered structures. Collectively, these laboratory and modelling efforts define a pathway toward interpreting a wide spectrum of potential AP and BP residues, from mundane engineered fragments to highly structured information-bearing grains.

\subsection{Practical reference volumes for micron-scale searches}

\deleted{A reference sampling volume of approximately $1~\mathrm{m^{3}}$ of lunar regolith provides a tractable operational starting point for particulate technosignature searches. For context, the combined Apollo missions returned $\sim382$\,kg of lunar material, corresponding to roughly $0.25$--$0.3\,\mathrm{m^{3}}$ of loose regolith equivalent \citep{Heiken1991}. The cumulative mass returned to Earth from the Apollo, Luna, and Chang’e programmes to an order of magnitude approximates to the material contained within a cubic metre of regolith.}

\added{The Apollo missions collectively returned approximately \(382\,\mathrm{kg}\) of lunar material, corresponding to roughly one quarter to one third of a cubic metre of loose-regolith equivalent \citep{Heiken1991}. An initial dedicated particulate-technosignature search might therefore begin with a smaller volume, perhaps of order \(0.1\,\mathrm{m^3}\). We nevertheless adopt \(1\,\mathrm{m^3}\) as the reference scale throughout this paper because our objective is not to predict the capacity of the next individual mission, but to establish a simple and readily scalable benchmark against which future searches may be compared.}

\deleted{Future robotic and commercial lunar missions are expected to return comparable or larger quantities, and to begin emplacing \textit{in situ} capabilities for larger sample processing and analyses. What may become routine excavations associated with landing pads, construction activities, and resource utilisation may disturb substantially larger volumes. We estimate that a cubic metre represents a scale of material that could plausibly be extracted, curated, and systematically examined for APs, BPs, their residues and impact signatures during the initial scientific phases of renewed lunar surface activity, prior to any large-scale industrial operations.}

\added{The choice of one cubic metre should also be viewed in the context of the rapidly evolving lunar exploration architecture. Future heavy-cargo landers, sustained human surface operations, \textit{in situ} resource-utilisation activities, landing-pad preparation and excavation for construction are all expected to disturb regolith volumes substantially larger than those returned by previous sample-return missions. A cubic metre therefore represents a practical scientific reference scale for the emerging era of extended lunar surface operations, while the formalism developed below remains directly applicable to smaller or larger sampled volumes through the explicit dependence on sampled area, effective integration time and analytical completeness.}

We interpret this cubic metre reference volume as the physical reservoir of grains within which a search must ultimately be conducted. Because the null-detection limits derived in \secreftwo{sec:undirectedreleasesintotheinterstellarmediumconstraints}{sec:deliberatetargetingoftheinnersolarsystem} assume that no technogenic grains are present within this sampled volume, a complete search in principle requires examination of all grains and fine material contained within that reservoir down to the relevant micron and submicron scales.

At present such exhaustive examination would be extremely demanding.
Preparation of SEM samples from lunar soils typically proceeds through milligram-scale subsampling and manual handling steps that represent significant throughput bottlenecks. However, several technological developments suggest a pathway toward substantially higher analysis rates. Automated imaging pipelines, machine-vision triage systems such as YOLO-ET
\deleted{\citep{Pinault2024b}}\added{\citep{Pinault2023Hayabusa,Pinault2024b,Pinault2025}}, Fig. \ref{fig:yolo_demo}, and related anomaly-detection architectures
\changed{\citep{LesnikowskiAngerhausen2023}}
    {\citep{Lesnikowski2024}} demonstrate that grain-scale image analysis can be scaled to very large datasets. The rapid expansion of GPU-accelerated image analysis and distributed machine-learning pipelines further suggests that petascale microscopy datasets will become increasingly tractable within the operational timeframe of upcoming lunar missions. Comparable automated inspection systems already operate in semiconductor manufacturing and industrial quality control, where billions of microscopic features must be screened routinely.

Recent high-resolution SEM studies illustrate the scale of features that must be detected. Submicron impact craters and melt splashes on individual lunar grains have been documented in Chang’e-5 soils \citep{Gu2025}, with characteristic diameters of $\sim0.1$–$1\,\mu$m. Similar microtextures have been identified on grains returned from asteroid Ryugu, including indications of possible interplanetary or interstellar dust impact \citep{Matsumoto2024}. These observations confirm that diagnostically relevant structures can occur in and amongst the grains, at the grain-surface level, and within the fine interstitial fraction between grains.

Progress in automated imaging, high-throughput microscopy, and machine learning therefore suggests a credible pathway toward large-scale analysis of lunar regolith volumes. While the complete examination of a cubic metre of regolith remains beyond present-day laboratory practice, ongoing developments in automated microscopy, semiconductor-style inspection systems, CT scanning and machine-vision pipelines indicate that such searches may become feasible on the timescale of forthcoming lunar exploration programmes.

\subsection{A comprehensive detection framework}
\label{subsec:acomprehensivedetectionframework}

The detection of micron-scale technosignatures thus requires an integrated
methodological framework that combines statistical screening, physical
characterisation, and contextual interpretation. Rather than relying on strong concentration mechanisms, the approach developed here exploits variance across multiple observational domains: grain morphology, composition, microstructural context, and spatial patterning. In this sense, agglutinates and other regolith components function not as traps, but as variance-enhancing archives in which rare inclusions may become statistically visible even when the global flux of technomaterial is extremely low.

At the grain scale, machine-vision pipelines such as YOLO-ET enable rapid triage of large particle populations, flagging anomalous grains, microcraters, and melt features for detailed follow-up. Laboratory analyses using SEM, EDS, SIMS, nano-CT, and FIB tomography then provide compositional and structural discrimination between natural materials, anthropogenic contaminants, and candidate technosignatures. This division of labour between automated screening and high-fidelity analysis allows surveys of unprecedented scale while preserving rigorous evidential standards.

Ongoing work extends this approach to curated extraterrestrial datasets, including micron- and submicron-scale microcrater and impact-splatter imagery from Chang’e returned samples and asteroid Ryugu material \citep{Pinault2026b}, where the emphasis is on calibrated detection performance, cross-validation, and physical interpretability. These efforts are designed to establish robust grain-scale anomaly detection prior to any technosignature-specific application.

A key supporting requirement is the construction of a calibrated reference library linking projectile properties to impact outcomes. %To this end, we propose a coordinated programme of two-stage light–gas-gun experiments, 
To this end, laboratory hypervelocity impact experiments—such as two-stage light–gas-gun studies of micron-scale projectiles—provide a practical means of calibrating projectile properties against crater morphology and residue chemistry, e.g. by firing $10$--$100\,\mu\mathrm{m}$ projectiles of natural silicates, glasses, and engineered analogues (e.g.\ titanium, stainless steel, tungsten carbide) into polished targets at $3$--$7\,\mathrm{km\,s^{-1}}$ \citep{Pinault2024a}. These experiments could help quantify how density, melting point, and brittleness control crater morphology, spallation, melt textures, and residue chemistry, establishing the experimental baselines required to interpret candidate AP signatures. They also enable the controlled fabrication of analogue agglutinates containing refractory
inclusions, following the insight that certain engineered materials may survive encapsulation during micrometeorite melting (R.~Kerschmann, pers.\ comm., 2025).

%At larger spatial scales, orbital analysis serves primarily as a contextual and prioritisation tool. Particularly 
Should foreign inclusions in agglutinates prove viable detection targets, variations in surface soil maturity or other regolith properties may help identify sites %and where variations in microcrater density, albedo, surface maturity, or regolith properties may identify sites 
for targeted sampling, particularly in the instances described in \secref{sec:deliberatetargetingoftheinnersolarsystem}, while deliberately emplaced macroscopic artefacts—if present—would constitute an independent technosignature class. Unsupervised anomaly-detection techniques applied to orbital imagery have demonstrated the recovery of known landing sites without labelled training data \changed{\citep{LesnikowskiAngerhausen2023}}
    {\citep{Lesnikowski2024}}, and frameworks such as YOLO-ETA extend this approach to systematic detection of candidate surface anomalies \citep{Pinault2026a}.

\added{Particular caution is warranted regarding human technological materials. Any candidate must first be demonstrated to be inconsistent with known terrestrial manufacturing techniques, documented spacecraft materials and lunar mission hardware. Available mission records, contamination-knowledge archives and spacecraft-material inventories should therefore be treated as a reference set against which candidate particles are routinely compared \citep{DaviesWagner2013,DworkinContamination2018,McCubbinCuration2019}.}

\added{Future sample-return and surface missions would benefit from maintaining openly accessible materials inventories describing spacecraft alloys, composites, coatings, polymers, adhesives, lubricants and nanomaterials carried to the lunar surface. Such archives would substantially strengthen the exclusion of human provenance, and would remain valuable even where construction, mining or industrial inspection takes place in operational environments that cannot be maintained to sample-return clean-room standards.}

\added{Within this framework, the principal objective of AI is not simply particle recognition, but technological provenance analysis. Candidate particles are evaluated by combining multiple independent lines of evidence -- including morphology, composition, isotopic characteristics, crystallography, contextual geology and comparison with reference collections -- to estimate the probability of alternative origins. This evidential approach mirrors established practice in archaeology, planetary sample science and forensic investigation while providing a scalable methodology for analysing very large datasets \citep{LingamFrameworks2023,DworkinContamination2018,McCubbinCuration2019}.}

These elements define a hierarchical detection architecture: orbital analyses may help constrain areas of interest, while machine vision accelerates grain-scale triage, and laboratory forensics provide compositional and structural verification. The framework relies not on strong concentration mechanisms, but on identifying statistically anomalous material within a well-characterised natural background.

\section{Undirected releases into the interstellar medium: constraints}
\label{sec:undirectedreleasesintotheinterstellarmediumconstraints}

In \secref{sec:transportprocessesthroughtheinterstellarmedium} we outlined the physical processes governing the transport, survival, and velocity filtering of micron-scale grains in the interstellar medium and during entry into the Earth-Moon system. Here we use those results to frame a scenario for the production of undirected technogenic particles, and to assess the constraints implied by their absence in sampled lunar regolith.

Large engineered orbital systems -- such as Dyson swarms or related distributed infrastructures \citep{Dyson1960,Wright2020} -- would plausibly involve masses comparable to planetary bodies. Independent of intent, such systems could undergo fragmentation through impacts, dynamical instabilities, or decommissioning, generating collisional cascades that convert a fraction of their mass into micron-scale debris \citep{Lacki2025}. Stars of sufficient luminosity can expel a portion of this debris into the ISM, providing a natural source of APs. This class of technosignature is distinct from waste-heat or stellar-obscuration searches and is constrained by the time-integrated particulate record preserved in the lunar regolith.

The two fundamental variables in this scenario are the technomaterial mass released into the ISM per event and the frequency of such releases. \changed{We consider models in which each star is a potential source of megaswarm ejecta.}{We consider models in which a fraction \(\eta_{\mathrm{ej}}\) of the stellar population is eligible to contribute megaswarm ejecta of the relevant grain size.} Swarm-disintegration events are treated as instantaneous and occur as a Poisson process with mean rate \(\Gamma_S\) per \added{eligible} star. \added{Thus, \(\Gamma_S\) is measured in events star\(^{-1}\) time\(^{-1}\).} \changed{On average, the number of megaswarms destroyed per Sun-like star over its main-sequence lifetime is of order \(\sim\Gamma_S(10\,\Gyr)\), a number that may be far below unity if most stars never host one of these structures.}{Over a \(10\,\Gyr\) reference interval, the expected number of events per eligible star is \(\Gamma_S(10\,\Gyr)\), which may be far below unity if most eligible stars never host such a structure.} \changed{The amount of mass released per event is $M_S$.}{We assume that each swarm-destruction event releases a technomaterial mass \(M_S\).} The pair \((M_S,\Gamma_S)\) therefore defines the parameter space constrained by AP searches, waste-heat surveys and material-budget limits. \added{The product \(M_S\Gamma_S\) is the mean technomaterial production rate per eligible star and is the principal source parameter entering the lunar technograin flux.}

For the constraint modelling developed here, a regolith sample is represented as a column of area \(A_s\) extending to depth \(z_s\). The sampled area sets the intercepted surface flux, while the depth enters through the effective sampling time \(t_s(z_s)\) determined by regolith mixing. If \(N_{\mathrm{obs}}\) technogenic grains are observed within this volume, the corresponding flux constraint follows from the expected number of recognised particles.

\subsection{Flux formalism and constraint modelling}
\label{subsec:fluxformalismandconstraintmodelling}

Because the lunar regolith integrates particulate influx over gigayear timescales, even a null detection within volumes as small as a well-characterised \(\sim1\,\mathrm{m^3}\) permits upper bounds to be placed on the cumulative technomaterial output of technologically capable civilisations. As discussed in \secref{sec:detectionstrategies}, a cubic metre also represents a realistic medium-term reference volume not only for lunar sampling operations, but for the development and deployment of computer-vision and machine-learning techniques optimised for screening large numbers of grains and impact features. Constraints derived at this scale therefore provide a baseline that can be systematically strengthened as both larger regolith volumes become accessible and analytical capabilities expand.

We adopt a simple flux model that incorporates (1) the fraction of stars luminous enough to eject micron-scale grains into the ISM, (2) the expected mass distribution of fragments produced by collisional cascades in large-scale swarms, (3) the slowing of such grains to the velocity of the local ISM before heliospheric entry, (4) heliospheric and lunar modulation of the incoming flux, and (5) the effective sampling time associated with regolith mixing. For our baseline constraints, we use relatively conservative assumptions for these factors, before considering how loosening these assumptions can improve the limits.

\changed{The expected number of detectable particles in the regolith volume \(A_s z_s\) is}{The expected number of recognised grains is the received surface flux multiplied by the sampled area, the effective sampling time and the aggregate completeness factor:}
\begin{equation}
\label{eqn:expectedundirectedcount}
\Mean{N_{\mathrm{obs}}}
=
\eta_{\mathrm{comp}}\,\FluxRec\,A_s\,t_s(z_s),
\end{equation}
where \(t_s(z_s)\) is the effective sampling time down to depth \(z_s\). \added{This is the expectation value of the Poisson count used below; it separates physical delivery, \(\FluxRec\), from the geometry and completeness of the search.}

The completeness factor \(\eta_{\mathrm{comp}}\) is the fraction of grains or grain relics \changed{in the regolith that are actually identified as technosignatures. We take it as $1$, but a small completeness can be compensated by scaling the survey area up.}{that both retain a recognisable technogenic signature and are subsequently identified by the analytical workflow. It therefore incorporates impact survival, preservation, distinguishability and instrumental or classification efficiency, as decomposed conceptually in Equation~\ref{eqn:completenessdecomposition}. We use \(\eta_{\mathrm{comp}}=1\) to present an optimistic reference limit and retain the factor explicitly so that constraints can be rescaled as material-specific survival and detection efficiencies become available. Values below unity weaken the constraint in inverse proportion; equivalent sensitivity may be recovered by increasing the effective surveyed area \(\eta_{\mathrm{comp}}A_s\).}

\changed{The}{As described in \secref{subsec:velocityfilteringandslowarrivaldynamics}, the} flux of recognisable grains received per unit lunar surface area is
\begin{equation}
\label{eqn:receivedundirectedflux}
\FluxRec
=
\frac{1}{4}\,
\eta_{\mathrm{mod}}\,
\shortchanged{\ninf}{\eta_m\frac{\rhoinf}{\mRef}}\,
\Mean{\vinf},
\end{equation}
where \(\shortchanged{\ninf}{\rhoinf}\) is the \changed{number}{mass} density of \added{the swarm-generated techno}grain population in the ISM\deleted{ of the relevant size}, \(\Mean{\vinf}\) is its mean approach speed at infinity, and \(\eta_{\mathrm{mod}}\) is the fraction of the interstellar reference flux admitted through the slow-arrival channel after heliospheric and lunar modulation. \added{The reference mass \(\mRef\) approximates the mass of a typical grain arriving through this channel, while \(\eta_m\) is the dimensionless fraction of \(\rhoinf\) contained per unit logarithmic mass interval around \(\mRef\). Thus, \(\eta_m\rhoinf/\mRef\) is an estimate of the number
density of grains relevant to the fiducial search. The factor \(1/4\) is geometric: an isotropic population intersects a spherical collector through its projected area \(\pi R^2\), and division by the total surface area \(4\pi R^2\) gives the mean flux per unit surface area.}
For the baseline constraint we retain only grains arriving through this slow-arrival channel and adopt \(\eta_{\mathrm{mod}}\Mean{v_\infty}=0.01\,\kms\), the conservative value selected in \secref{subsec:velocityfilteringandslowarrivaldynamics}.

\changed{In turn,}{Only a fraction \(\eta_{\mathrm{ej}}\) of the total stellar population is eligible to contribute grains of the size admitted by the slow-arrival model. We therefore take \(n_\star\) to be the total local stellar number density and \(\eta_{\mathrm{ej}}n_\star\) to be the local number density of eligible stars. Their volumetric technomaterial injection rate is \(\eta_{\mathrm{ej}}n_\star M_S\Gamma_S\). If the grains remain in the ISM for a mean residence time \(\tau_{\mathrm{ISM}}\), the steady-state} \changed{number}{mass} density of swarm-generated technoparticles is
\begin{equation}
\label{eqn:swarmtechnograinmassdensity}
\shortchanged{\ninf}{\rhoinf}
=
\eta_{\mathrm{ej}}\,n_\star\,
\shortchanged{\eta_m\frac{M_S\Gamma_S\tau_{\mathrm{ISM}}}{m_G},}{M_S\Gamma_S\tau_{\mathrm{ISM}}}.
\end{equation}
\deleted{where $n_\star$ is the local stellar density, $\tau_{\mathrm{ISM}}$ is the mean residence time of grains in the ISM (assumed here to be 100\,Myr), and $m_G$ is the mass of a single grain.}
\added{This expression assumes a statistical steady state in which production and destruction have operated for longer than the adopted residence time; we use \(\tau_{\mathrm{ISM}}=100\,\Myr\) in the fiducial calculation.}

We approximate the reference grain as a sphere of radius \(r_G\) and density \(\rho_G\), so that
\begin{equation}
\label{eqn:referencegrainmass}
\added{\mRef
=
\frac{4\pi}{3}\,\rho_G r_G^3.}
\end{equation}
We use \(r_G=0.3\,\um\) and \(\rho_G=3\,\gram\,\cm^{-3}\), for which \(\mRef\simeq3.4\times10^{-16}\,\kg\) and \(\beta=0.7\). This value of \(\beta\) allows grains entering the heliosphere at \(v_\infty=20\,\kms\) to occupy the slow-arrival channel and reach the Moon at comparatively low encounter speeds.

The debris created in a collisional cascade has a broad mass distribution. For rapid collisions, \(\mathrm{d}N/\mathrm{d}m_G\propto m_G^{-2}\) over several decades in mass, so approximately equal mass lies within each logarithmic mass interval \citep{Rossi1994}. We \changed{define the factor \(\eta_m\) as the fraction of mass contained over a range \(\Delta\ln m_G=1\)}{define \(\eta_m\) as the fraction of the total technograin mass density contained per unit logarithmic mass interval} around \(\mRef\):
\begin{equation}
    \label{eqn:massfractionlogbin}
    \added{
    \eta_m
    \equiv
    \left.
    \frac{1}{\rhoinf}
    \frac{\mathrm{d}\rhoinf}{\mathrm{d}\ln m_G}
    \right|_{m_G=\mRef}
    \approx
    \frac{1}
    {3\ln(r_{G,\mathrm{max}}/r_{G,\mathrm{min}})}.
    }
    \end{equation}

\deleted{and is then given by $3 \ln(r_{G,\mathrm{max}}/r_{G,\mathrm{min}})$.}
Here \(r_{G,\mathrm{max}}\) is the largest grain radius ejected into the ISM by radiation pressure, while \(r_{G,\mathrm{min}}\) may approach molecular scales. An illustrative interval from \(0.1\,\mathrm{nm}\) to \(1\,\um\) gives \(\eta_m\simeq0.036\). To allow conservatively for vaporised material and deviations from the ideal cascade, we adopt \(\eta_m=0.02\).

Most stars are substantially fainter than the Sun and cannot expel grains of the relevant size by radiation pressure; only much smaller grains, if any, are blown out. Such grains would have \(\beta\gg1\) on entering the Solar System and could occupy the slow-arrival channel only at unusually large \(v_\infty\). They are also more strongly excluded from the heliosphere by electromagnetic forces. We therefore restrict the baseline source population to stars brighter than the Sun that can eject grains with \(\beta\sim0.5\)--1. As an additional conservative assumption, we include only dwarfs that remain on the main sequence for at least \(4.5\,\Gyr\), reducing the adopted mass range to \(1\)--\(1.25\,\Msun\) \added{\citep[see also][]{Zuckerman2022}}. We take the resulting eligible fraction to be \(\eta_{\mathrm{ej}}=0.01\); with the adopted total local stellar density, this gives \(\eta_{\mathrm{ej}}n_\star\simeq0.001\,\pc^{-3}\) \citep{Chabrier2003,Bovy2017}.

The number of grains observed in the sample, \(N_{\mathrm{obs}}\), is Poisson distributed\changed{, with a null detection implying an upper limit \(\Mean{N_{\mathrm{obs}}}\leq\bar N\). Combining the above equations, we find:}{. For zero observed grains, \(P(0\mid\lambda)=e^{-\lambda}\); the conventional 95\% upper limit is \(\bar N=-\ln(0.05)=2.996\simeq3\). A null detection therefore implies \(\Mean{N_{\mathrm{obs}}}\leq\bar N\). Substituting Equations~\ref{eqn:receivedundirectedflux}--\ref{eqn:massfractionlogbin}, using the spherical reference mass in Equation~\ref{eqn:referencegrainmass}, and solving for the mean technomaterial production rate per eligible star gives}
\begin{equation}
    \label{eqn:constraintundirectedvariables}
    M_S\Gamma_S
    \shortchanged{<}{\lesssim}
    \frac{
    16\pi\bar N\rho_G r_G^3
    }{
    3\eta_{\mathrm{ej}}n_\star\eta_m\tau_{\mathrm{ISM}}
    \eta_{\mathrm{mod}}\Mean{v_\infty}
    \eta_{\mathrm{comp}}A_s t_s
    }.
    \end{equation}
\added{The cubic dependence on \(r_G\) reflects the declining number of particles available for a fixed released mass as the mass of each spherical grain increases.}

\begin{figure}[h!]
    \centering
    \includegraphics[width=\columnwidth]{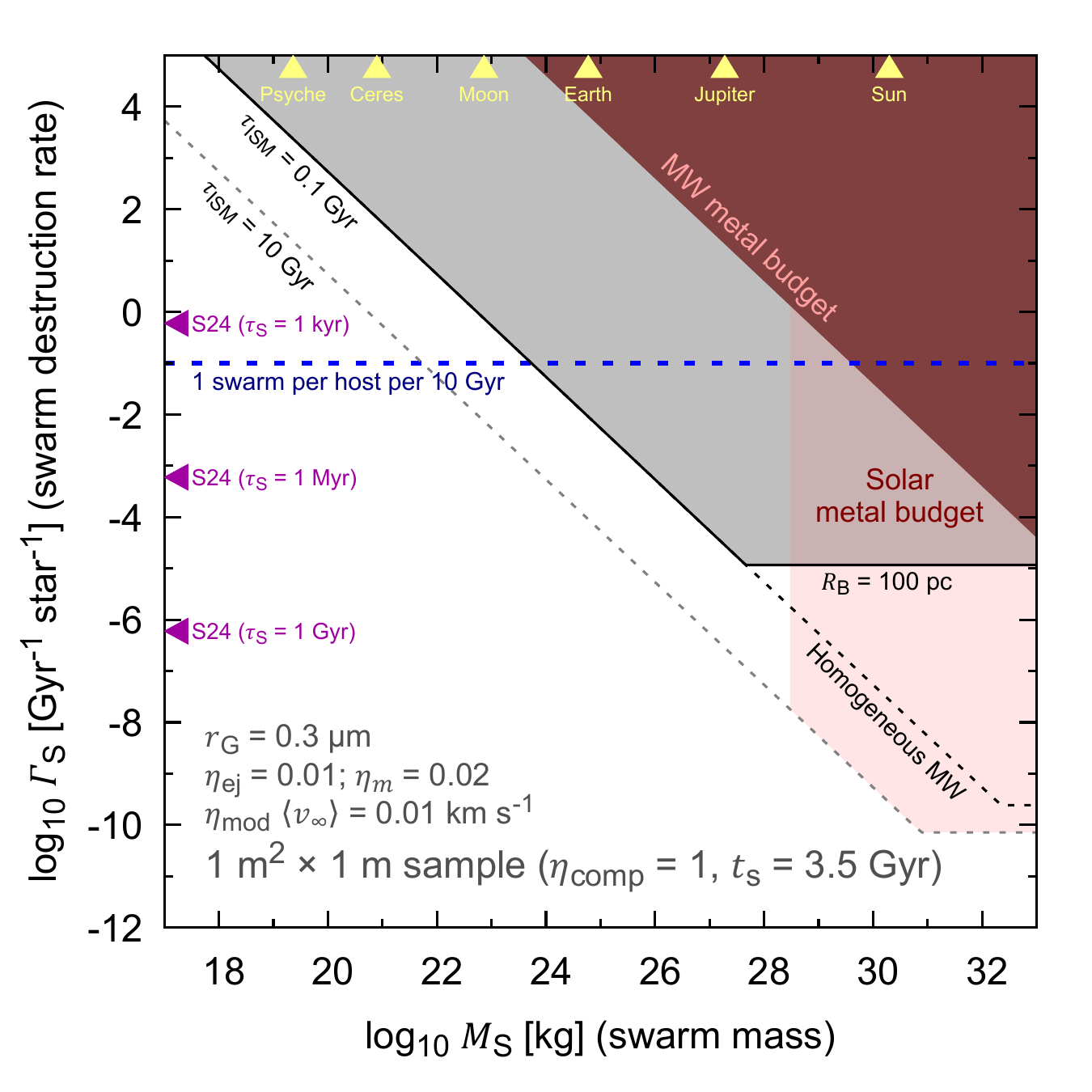}
    \caption{Constraints on the \((M_S,\Gamma_S)\) megaswarm parameter space derived from a null detection of micron-scale technomaterial in a cubic metre of lunar regolith, for the conservative case of undirected, collisionally generated debris from large engineered swarms subsequently ejected into the ISM by radiation pressure. The grains are assumed to arrive through the slow-arrival channel using the fiducial parameters stated in the text. The grey shaded region bounded by the solid black line indicates the parameter combinations excluded when grains remain confined to bubbles of radius \(R_B=100\,\pc\). The dashed boundaries show the stronger constraints obtained if grains are homogeneously mixed throughout the Milky Way. The crimson and pink regions indicate parameter combinations that exceed the total Galactic and Solar metal budgets, respectively. Purple markers on the left axis show the Hephaistos null limits for partial Dyson spheres from \citet{Suazo2024}, under different assumed structure lifetimes \(\tau_S\). The blue dashed line denotes one swarm-destruction event per host star per \(10\,\Gyr\). These constraints apply only to the undirected megaswarm-debris scenario; directed or engineered delivery pathways are treated separately in \secref{sec:deliberatetargetingoftheinnersolarsystem}. \added{For comparison, sustaining the present human Solar-escape rate \(\sim10^2\,\kg\,\yr^{-1}\) for one billion years would produce only \(\sim10^{-14}\,M_\oplus\).}}
    \label{fig:LimitsMegaswarms}
\end{figure}

We adopt \(\bar N=3\) for the 95\% Poisson upper limit from a null detection, finding
\begin{equation}
\label{eqn:constraintundirectednumerical}
M_S\Gamma_S
\shortchanged{<5.4}{\lesssim5}\times10^{23}\,\kg\,(10\,\Gyr)^{-1}.
\end{equation}
\begingroup
\raggedright
for the fiducial values
\(\rho_G=3\,\gram\,\cm^{-3}\),
\(r_G=0.3\,\um\),
\(\eta_{\mathrm{ej}}n_\star=0.001\,\pc^{-3}\),
\(\eta_m=0.02\),
\(\tau_{\mathrm{ISM}}=100\,\Myr\),
\(\eta_{\mathrm{mod}}\Mean{v_\infty}=0.01\,\kms\),
\(\eta_{\mathrm{comp}}=1\),
\(A_s=1\,\meter^2\), and
\(t_s=3.5\,\Gyr\).
\par
\endgroup
\added{The numerical limit is obtained by evaluating the analytic expression at these fiducial grain, stellar-population, transport, modulation and sampling parameters; it is not an independent empirical fit. Over the \(10\,\Gyr\) normalisation interval, the bound corresponds to a cumulative released mass of approximately \(5\times10^{23}\,\kg\), or \(\sim0.1\,M_\oplus\), per eligible star.} The resulting excluded region in the \((M_S,\Gamma_S)\) parameter space for undirected megaswarm debris is shown in Figure~\ref{fig:LimitsMegaswarms}.
\footnote{The appearance of a \(10\,\Gyr\) timescale in the numerical normalisation serves only as a convenient reference interval for expressing cumulative mass processing and does not imply that individual stars persist, or remain active, for that duration. The constraint therefore applies to the time-integrated particulate output of a stellar population and is independent of the detailed main-sequence lifetimes of individual stars.}

\added{The scale of this bound is illustrated by comparison with present human technology. The constrained quantity is the cumulative mass of long-lived particulate material ultimately escaping into interstellar space, for which no direct terrestrial analogue presently exists. Five spacecraft are on Solar-escape trajectories -- Pioneer~10 and 11, Voyager~1 and 2, and New Horizons -- with a combined mass of order a few thousand kilograms \citep{Siddiqi2018,NASAVoyagerFAQ2026}. Several associated spent kick stages and despin masses are also expected to remain on Solar-escape trajectories. Including this additional hardware changes the inventory by only a factor of order 2 and does not alter the estimate below. Averaged over several decades, the effective human Solar-escape rate remains of order}
\[
\added{\dot M_{\mathrm{human}}\sim10^2\,\kg\,\yr^{-1},}
\]
\added{while most human technological material remains gravitationally bound to the Earth or the Solar System.}

\added{Maintaining this rate for \(1\,\Gyr\) would yield approximately}
\[
\added{M_{\mathrm{human}}\approx10^{11}\,\kg
\approx1.7\times10^{-14}\,M_\oplus,}
\]
\added{roughly thirteen orders of magnitude below the fiducial constraint. At the same rate, accumulating \(0.1\,M_\oplus\) would require approximately \(6\times10^{21}\,\yr\).}

\added{An alternative comparison may be made using industrial throughput. Distributing \(0.1\,M_\oplus\) over \(10\,\Gyr\) corresponds to an average mass-processing rate of approximately}
\[
\added{6\times10^{10}\,\mathrm{tonnes}\,\yr^{-1}.}
\]
\added{If each heavy-lift launch delivered \(100\,\mathrm{tonnes}\), sustaining this throughput would require approximately \(6\times10^8\) launches per year, or about twenty launches every second for ten billion years. Even a sustained programme of one thousand such launches annually for ten thousand years would process only approximately \(2\times10^{-13}\,M_\oplus\).}

\added{These comparisons show that the present constraints become significant only for technological systems operating many orders of magnitude beyond present-day human activity.}

Although these values are large in isolation, they constrain the cumulative mass processed or mobilised by technological civilisations over Galactic history, for which direct empirical bounds are otherwise sparse. Future lunar mining or construction operations that screen regolith during processing could strengthen the limits substantially. An effective surveyed area \(\eta_{\mathrm{comp}}A_s=0.1\,\km^2\) would give
\[
M_S\Gamma_S\lesssim5\times10^{18}\,\kg\,(10\,\Gyr)^{-1},
\]
sufficient, under the same assumptions, to exclude the dispersal of a mass comparable to asteroid (16) Psyche once per Sun-like star over \(10\,\Gyr\). The strongest formal limit would result from complete examination of the lunar surface with \(\eta_{\mathrm{comp}}=1\):
\[
M_S\Gamma_S\lesssim1\times10^{10}\,\kg\,(10\,\Gyr)^{-1}.
\]

The global \(M_S\Gamma_S\) bound is supplemented by a lower event-rate cutoff set by propagation statistics. If grains remain confined to discrete bubbles, a lunar null result can constrain only populations for which the expected number of intercepted regions is at least the adopted Poisson threshold. Using Equation~\ref{eqn:NBubblesEncountered}, we therefore require \(\Mean{N_{B,\odot}}\geq\bar N\). For \(R_B=100\,\pc\), this corresponds to \(\Gamma_S\sim10^{-5}\,\Gyr^{-1}\) per eligible star, or roughly one event per \(10^4\) eligible stars over \(10\,\Gyr\). This floor is independent of the regolith area examined because the entire Moon is exposed during each bubble encounter; increasing the sampled area only causes the main \(M_S\Gamma_S\) boundary to meet the floor at a lower event mass.

A lower cutoff applies if the grains are homogeneously distributed throughout the Milky Way. \changed{Then, we only need to be confident that there has been at least one swarm destruction during the entire history probed by the sample ($\tau_{\mathrm{ISM}} + t_s$):}{Regardless of the mass released per event, the Moon cannot constrain a population so rare that no eligible source event is likely to have contributed during the accessible time interval. Grains in the sampled regolith were either emitted during the effective sampling time or were already present in the ISM, for no longer than \(\tau_{\mathrm{ISM}}\), when that interval began. Requiring the expected number of source events to exceed \(\bar N\) gives}
\begin{equation}
\label{eqn:undirectedeventratefloor}
\Gamma_S
\geq
\frac{\bar N}
{\eta_{\mathrm{ej}}N_{\star,\mathrm{MW}}
(\tau_{\mathrm{ISM}}+t_s)}.
\end{equation}
\added{This is the minimum event rate accessible to a null search, regardless of the mass released per event or the amount of regolith examined.} With \changed{around}{approximately} \(\eta_{\mathrm{ej}}N_{\star,\mathrm{MW}}=3\times10^9\) eligible stars in the Galaxy, the floor is \(\Gamma_S\gtrsim\shortchanged{2.8}{3}\times10^{-9}\,(10\,\Gyr)^{-1}\) per eligible star.

The constraints derived here apply to stars capable of ejecting micron-scale grains into interstellar space, which primarily requires sufficient stellar luminosity rather than any particular evolutionary stage. Apart from the Poisson event-rate model, no assumption is made about the detailed timing of individual ejection episodes; the constraint applies to the time-integrated particulate output of the eligible stellar population.

\subsection{Alternate models}
\label{subsec:alternateundirectedmodels}

The limits derived here apply specifically to undirected, collisionally generated technomaterial produced by the disruption of large engineered swarms and subsequently ejected into the ISM by radiation pressure. Other undirected release scenarios -- including smart dust engineered to operate within the ISM, unguided self-replicating probes, or technomaterial produced in distant Oort-cloud or asteroid-mining environments \added{\citep{ForganElvis2011}} -- need not be restricted to the \(\eta_{\mathrm{ej}}\) fraction of stars capable of radiative blowout. Removing the fiducial \(\eta_{\mathrm{ej}}=0.01\) penalty would strengthen the mass-production constraint by two orders of magnitude.

The host stars of megaswarms also need not be the systems in which the responsible civilisations evolved. If already spacefaring populations settle stars with main-sequence lifetimes shorter than the adopted \(4.5\,\Gyr\) evolutionary interval, the eligible stellar fraction may increase. Even within the radiative-blowout scenario, broadening the host population in this way could raise \(\eta_{\mathrm{ej}}\) by a factor of \(\sim2\)--3.

Intentionally dispersed or actively directed particulate probes (\secreftwo{subsec:weaklyintentionalparticulatesystemsandsmartdust}{subsec:deliberatelydispersedbracewellparticles}) need not follow the isotropic, time-averaged flux assumed here. The null constraints in this section should therefore be interpreted as conservative limits on undirected technogenic debris; engineered delivery may produce substantially higher lunar surface densities and is treated separately.

These distinctions underscore that the limits derived here reflect a specific, explicitly stated set of transport and survival assumptions, developed in \secref{sec:transportprocessesthroughtheinterstellarmedium} and \hyperref[sec:appendix:derivationsoffluxoflowspeedgrains]{Appendix~B}. They provide a conservative baseline for undirected technogenic particle production rather than a universal limit on all particulate technosignatures.

\subsection{Comparison with other limits on megaswarms}
\label{subsec:comparisonwithotherlimitsonmegaswarms}

Large-scale orbital infrastructures such as Dyson swarms are already subject to significant constraints from mid- and far-infrared searches for waste heat \citep{Dyson1960,Carrigan2009,Wright2014,Suazo2024,Huang2026}. Such ``Dysonian SETI'' approaches probe the reprocessing of stellar luminosity into thermal emission, thereby constraining the presence of intact, radiatively efficient megastructures. In contrast, the technograin limits derived here constrain the cumulative solid-mass processing and fragmentation history of such systems, independent of their instantaneous radiative output. They therefore provide a complementary—and in some regimes orthogonal—constraint on megaswarm-producing civilisations.

Detectable waste heat requires an intact swarm that absorbs and reprocesses a substantial fraction of the host star’s luminosity, but aside from weak limits imposed by radiation pressure, the total solid mass of such a structure cannot be inferred directly from its thermal emission. A canonical Dyson swarm completely enclosing a Solar-type star has a mass of order one Jupiter mass \citep{Dyson1960}, for which even a cubic metre of regolith would yield comparatively strong technograin constraints. At the opposite extreme, the minimum mass is set by the requirement $\beta < 1$; for the Sun this implies a surface density $\gtrsim 0.8\,\gram\,\meter^{-2}$, corresponding to a swarm mass of approximately \(2\times10^{20}\,\kg\) at \(1\,\mathrm{AU}\), for which a cubic metre of regolith has negligible sensitivity. Swarms confined to smaller radii would further reduce the required mass \citep{Wright2023}.

The strongest current waste-heat limits on megaswarms come from the Hephaistos project. \citet{Suazo2024} (hereinafter S24) examined infrared photometry for approximately five million stars and identified seven candidate systems.\footnote{All seven candidates orbit red dwarfs, whose luminosities are too low for radiation pressure to eject grains in the slow-arrival size range considered in our baseline model.} At least one candidate is now known to be a false positive caused by a background galaxy \citep{Ren2025}. If all seven candidates are treated as false positives, the 95\% Poisson upper limit on the instantaneous fraction of stars hosting detectable megaswarms is \(f_S=6.7\times10^{-7}\).

A direct comparison with the technograin limits requires \changed{a conversion between the fraction of stars currently hosting a megaswarm and the fraction that have ever hosted one in the past several Gyr. Given a megaswarm lifetime $\tau_S$,}{relating \(f_S\) to the swarm-destruction rate. For a statistically stationary population in which a detectable swarm remains in that state for a characteristic lifetime \(\tau_S\), the low-occupancy relation is \(f_S=\Gamma_S\tau_S\):}
\begin{equation}
\label{eqn:hephaistosrateconversion}
\ifreview
\begin{aligned}
\Gamma_S
&=
\frac{f_S}{\tau_S}
\\[-1pt]
&\deleted{=0.0006}
\\[-2pt]
&\added{\simeq 7\times10^{-4}\,\Gyr^{-1}}
\\[-1pt]
&\quad\times
\left(
\frac{f_S}{6.7\times10^{-7}}
\right)
\left(
\frac{\tau_S}{1\,\Myr}
\right)^{-1}.
\end{aligned}
\else
\begin{aligned}
\Gamma_S
&=
\frac{f_S}{\tau_S}
\simeq
7\times10^{-4}\,\Gyr^{-1}
\\
&\quad\times
\left(
\frac{f_S}{6.7\times10^{-7}}
\right)
\left(
\frac{\tau_S}{1\,\Myr}
\right)^{-1}.
\end{aligned}
\fi
\end{equation}
\deleted{for a population in equilibrium,}
\added{This conversion assumes statistical stationarity and negligible overlap between multiple detectable episodes around the same star.} For comparison with the eligible-star rate used above, we additionally assume that \(f_S\) is representative across stellar types. If the structures are stable for much of a host star's lifetime, the S24 limits are substantially stronger than the technograin limit from a single cubic metre of regolith. Their lifetime could, however, be millennia or shorter \citep[cf.][]{Lacki2025}; in that case, the Hephaistos observations remain consistent with nearly every star hosting a megaswarm at some point, whereas a cubic-metre regolith search could exclude that possibility for planetary-mass swarms. Representative S24 limits for several \(\tau_S\) values are shown in Figure~\ref{fig:LimitsMegaswarms}.

The availability of heavy elements imposes the remaining resource limits in Figure~\ref{fig:LimitsMegaswarms}. The total metal inventory of a stellar system is much smaller than the mass of its host star. In the Solar System, the Solar metal mass fraction is \(\sim0.014\) \citep{Asplund2021}, corresponding to approximately \(2.8\times10^{28}\,\kg\) of heavy elements. We use this as a deliberately generous guide to the material potentially available for megaswarm construction in one system. Still larger inventories would require importing material from other stellar systems.

A second ceiling follows from the total metal budget of the Milky Way. Over a production interval \(t_{\mathrm{MW}}\), the cumulative ejected technomaterial mass is \(M_S\Gamma_S t_{\mathrm{MW}}\eta_{\mathrm{ej}}N_{\star,\mathrm{MW}}\), where \(N_{\star,\mathrm{MW}}\) is the Galactic stellar population. The total stellar heavy-element inventory is \(Z_\star M_{\star,\mathrm{MW}}=Z_\star\Mean{m_\star}N_{\star,\mathrm{MW}}\). In the absence of extensive recycling, \changed{the metal budget implies}{an absolute upper bound follows by requiring the cumulative ejected technomaterial mass not to exceed the entire Galactic heavy-element inventory, even though only the fraction \(\eta_{\mathrm{ej}}\) of stars contributes events:}
\begin{multline}
\label{eqn:galacticmetalbudget}
M_S\Gamma_S
\lesssim
\frac{Z_\star\Mean{m_\star}}
{t_{\mathrm{MW}}\eta_{\mathrm{ej}}}
=
4\times10^{29}\,\kg\,(10\,\Gyr)^{-1}
\left(\frac{Z_\star}{0.01}\right)
\\
\times
\left(\frac{\Mean{m_\star}}{0.2\,\Msun}\right)
\left(\frac{t_{\mathrm{MW}}}{10\,\Gyr}\right)^{-1}
\left(\frac{\eta_{\mathrm{ej}}}{0.01}\right)^{-1}.
\end{multline}
Here \(\Mean{m_\star}\) is the mean stellar mass. \added{This is a deliberately permissive resource ceiling: it allows the full Galactic heavy-element inventory to be converted into technomaterial and dispersed into the ISM without losses or competing uses.}

\section{Deliberate targeting of the inner Solar System: constraints}
\label{sec:deliberatetargetingoftheinnersolarsystem}

The foregoing analyses address undirected, Galaxy-wide dispersal of technogenic particles through natural interstellar transport processes. A distinct class of scenarios arises if a civilisation deliberately delivers particulate artefacts into the inner Solar System. Such delivery may take several forms: release into heliocentric orbits within the inner Solar System, direct low-velocity insertion onto the lunar surface, or targeted arrival from nearby stellar systems using engineered trajectories. In contrast to the undirected case, the arrival statistics of directed delivery depend primarily on the geometry and intent of the delivery strategy rather than on ISM transport, heliospheric filtering or population-level occurrence rates. The parallel question in this regime is therefore: what mass must be released for at least one grain to be detectable in a sampled volume of lunar regolith?

The directed-delivery regime defines its own two-parameter space, analogous in structure to the \((M_S,\Gamma_S)\) space used for undirected megaswarm debris in \secref{subsec:fluxformalismandconstraintmodelling}. We model visits that release particulate artefacts as a Poisson process with mean rate \(\Gamma_V\). During each visit, a mass \(M_V\) is released into the local environment, of which some fraction is ultimately deposited on the Moon. The number of particles recovered from a regolith sample therefore constrains the product \(M_V\Gamma_V\), subject to delivery geometry, physical deposition and analytical completeness. These engineered-delivery scenarios are orthogonal to the undirected constraints considered above and illustrate how deliberate targeting can produce detectable BP concentrations with material budgets many orders of magnitude below those associated with megastructure-scale debris.

We distinguish two limiting transport regimes. In the first, release occurs on the Moon or in low lunar orbit. As a quiescent, geologically stable body close to the life-bearing Earth but without a likely biosphere of its own to disturb, the Moon may offer an attractive target for BP emplacement. Direct surface delivery can make the physical deposition fraction high, although it may confine the particles to relatively small patches. In the second regime, release occurs elsewhere in the Solar System. Only a small fraction of the particles then intersects the Moon, and only a fraction of the impacting material may be retained in the regolith.

Let \(M_{\mathrm{dep}}\) denote the mass physically deposited and retained in the lunar regolith per visitation after transport and impact-retention losses. \changed{It is related to \(M_V\) as}{To avoid double-counting material survival and analytical recognition, we define \(\eta_{\mathrm{dep}}\) only as the physical deposition efficiency. The recognisable-survival, recognition and detection probabilities remain contained in the separate completeness factor \(\eta_{\mathrm{comp}}\) introduced in \secref{subsec:hypervelocityimpactandpartialsurvivability}. Thus}
\begin{equation}
\label{eqn:directeddepositionefficiency}
M_{\mathrm{dep}}=\eta_{\mathrm{dep}}M_V.
\end{equation}
This definition allows the following constraints to be expressed either in terms of deposited mass or, after division by \(\eta_{\mathrm{dep}}\), the original released mass.

Additional symbols specific to directed-delivery scenarios are defined in \hyperref[sec:appendix:notationandsymboldefinitions]{Appendix~C}.

\subsection{Constraint modelling: globally distributed BPs}
\label{subsec:constraintmodellinggloballydistributedbps}

\changed{Let \(\Mean{N_{\mathrm{obs}}}\) be the expected number of grains in a randomly sampled location on the Moon's surface.}{For globally uniform deposition, a sample of area \(A_s\) intercepts the fraction \(A_s/A_{\mathrm{Moon}}\) of the particles deposited in each event. Multiplying the deposited particle number \(M_{\mathrm{dep}}/m_G\) by the visitation rate \(\Gamma_V\), the effective sampling time \(t_s\) and the analytical completeness \(\eta_{\mathrm{comp}}\) gives}
\begin{equation}
\label{eqn:directedglobalexpectedcount}
\begin{aligned}
\Mean{N_{\mathrm{obs}}}
&=
\eta_{\mathrm{comp}}
\frac{M_{\mathrm{dep}}}{m_G}
\frac{A_s}{A_{\mathrm{Moon}}}
t_s\Gamma_V
\\
&=
\frac{3\eta_{\mathrm{comp}}\eta_{\mathrm{dep}}
M_V\Gamma_V A_s t_s}
{4\pi\rho_G r_G^3 A_{\mathrm{Moon}}},
\end{aligned}
\end{equation}
where the second form follows from \(M_{\mathrm{dep}}=\eta_{\mathrm{dep}}M_V\) and the spherical-grain mass \(m_G=(4/3)\pi\rho_G r_G^3\). The Moon's surface area is \(A_{\mathrm{Moon}}\simeq3.8\times10^7\,\km^2\).

\changed{When each visit results in particles being laid down homogeneously across the entire Moon's surface, the observed number of grains is Poisson distributed.}{Strictly, if the number of visits is Poisson distributed and the number of particles contributed by each visit is also Poisson distributed, the total count is a compound Poisson variate of the Neyman Type~A form \citep{Neyman1939}. The simple Poisson approximation with mean \(\Mean{N_{\mathrm{obs}}}\) is accurate in the diffuse-arrival limit relevant to the sloping product constraint, where the expected number of recognised particles contributed to the sample by any one visit is much less than unity. Each visit then acts approximately as a Bernoulli trial. For rare, high-yield visits the null constraint instead saturates at the separate event-rate floor discussed below.} Applying the adopted null-result condition \(\Mean{N_{\mathrm{obs}}}\leq\bar N\) in the diffuse-arrival regime gives
\begin{equation}
\label{eqn:Constraint_Directed_Global_Variables}
M_V\Gamma_V
\lesssim
\frac{4\pi\bar N\rho_G r_G^3 A_{\mathrm{Moon}}}
{3\eta_{\mathrm{comp}}\eta_{\mathrm{dep}}A_s t_s}.
\end{equation}
The factor \(4\pi/3\) enters through the mass of a spherical grain, and the \(r_G^3\) scaling reflects the declining number of particles available for a fixed released mass as grain size increases.

\begin{figure}[h!]
    \centering
    \includegraphics[width=\columnwidth]{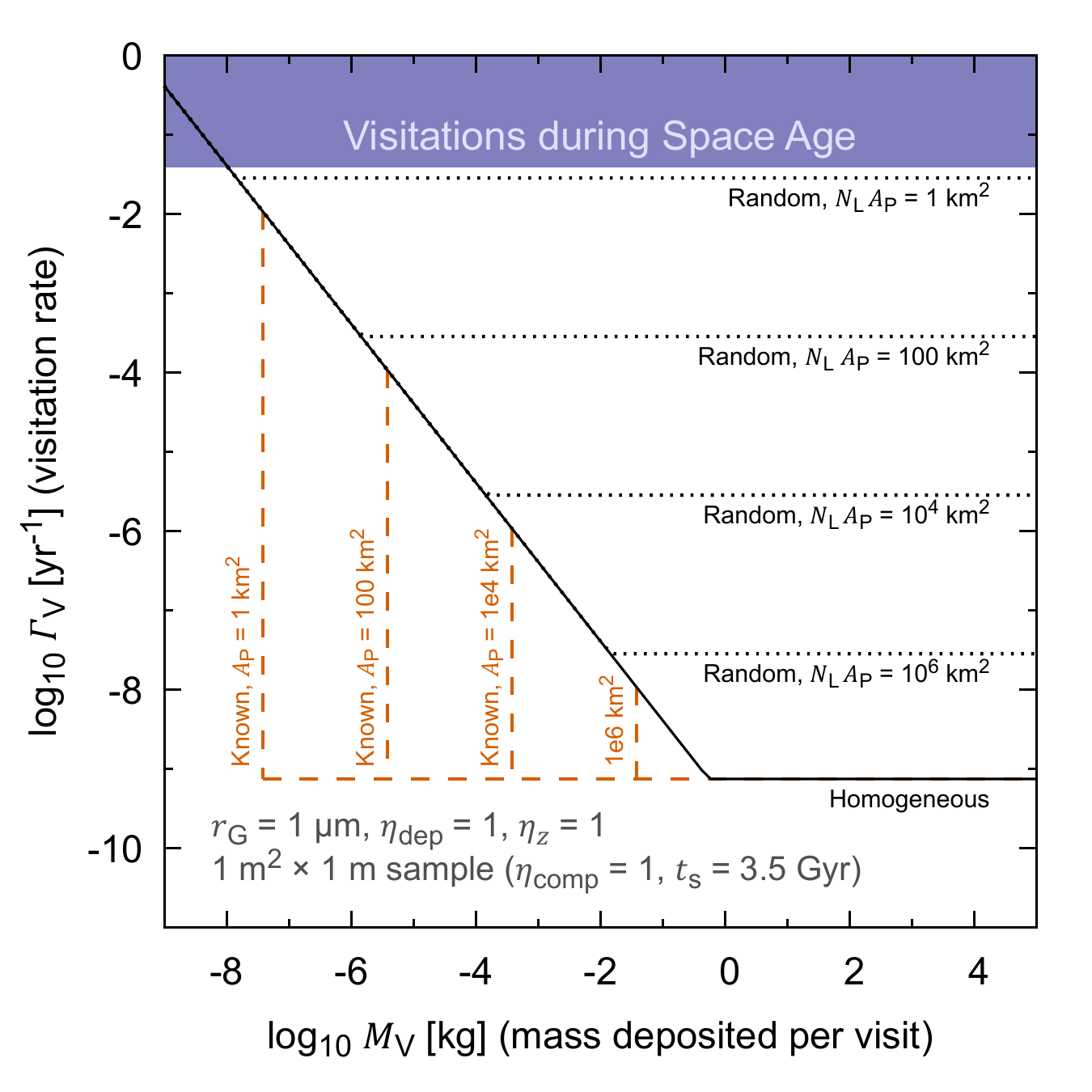}
    \caption{Constraints on intentionally dispersed Bracewell Particles (BPs) for micron-sized grains (\(r_G=1\,\um\)), a \(1\,\meter^2\) search area and an effective sampling time \(t_s=3.5\,\Gyr\). The axes show visitation rate \(\Gamma_V\) versus mass released per visitation \(M_V\). The plotted case adopts \(\eta_{\mathrm{dep}}=1\), so released and deposited masses coincide. Different assumptions about BP clustering and sampling produce different constraint curves; parameter combinations above and to the right of each curve are excluded for that model. The solid black curve corresponds to uniform deposition over the entire lunar surface. The family of black dotted curves represents patch-seeding scenarios in which deposition is confined to discrete surface areas and the search samples \(N_L\) independent random lunar locations. In the sparse-patch regime, the probability that one random location intersects deposited material is approximately \(A_{\mathrm{dep}}/A_{\mathrm{Moon}}\), so confinement to smaller total areas weakens blind-search limits by the reciprocal area factor. The orange dashed curves instead assume that a patch has been independently identified and that the regolith sample is taken from within it. These conditional known-patch limits benefit from the enhanced local grain density but should not be interpreted as the sensitivity of a randomly located sample. The shaded blue region shows visitation rates that would imply at least one event during the modern Space Age, here taken as 1957--2035. At values of \(\Gamma_VM_V\) far beyond the plotted range, technogenic material would dominate the lunar soil composition.}
    \label{fig:ABP_GrainLimits}
\end{figure}

For BPs laid gently across the Moon's surface, with \(\eta_{\mathrm{dep}}=1\), a null result in a completely examined \(1\,\meter^2\)-by-\(1\,\meter\) column (\(\eta_{\mathrm{comp}}A_s=1\,\meter^2\), \(t_s=3.5\,\Gyr\)) implies
\begin{equation}
\label{eqn:Constraint_Directed_Global_Numerical}
M_V\Gamma_V
\changed{\leq0.41}{\lesssim0.4}
\,\kg\,\Gyr^{-1}\,
\eta_{\mathrm{dep}}^{-1}
\left(\frac{r_G}{1\,\um}\right)^3.
\end{equation}
This is the ``homogeneous'' line in Figure~\ref{fig:ABP_GrainLimits}. The coefficient is the reader-facing rounded value of the same unrounded calculation used to plot the constraint.

\changed{The ultimate floor to this limit is set by the requirement that there probably has been at least one visit to the Moon over the integration time.}{A separate floor follows from the possibility that no visitation episode occurred during the full integration interval. Applying the same Poisson upper limit to the event count gives \(\Gamma_Vt_{\mathrm{int}}\leq\bar N\), or}
\begin{equation}
\label{eqn:Constraint_Directed_Global_Floor}
\Gamma_V
\leq
\frac{\bar N}{t_{\mathrm{int}}}
=
0.75\,\Gyr^{-1}
\left(\frac{t_{\mathrm{int}}}{4\,\Gyr}\right)^{-1},
\end{equation}
for the adopted 95\% Poisson upper limit \(\bar N=3\). The coefficient \(0.75\) is retained because it is the exact consequence of the adopted \(3/4\,\Gyr\) convention, rather than an uncertain model output. This limit is independent of particle mass and sampling area because it concerns the occurrence of visits rather than the number of particles recovered from each one.

A much weaker consistency ceiling follows from the requirement that relic BPs constitute no more than a fraction \(\zeta_{\mathrm{BP}}\) of the regolith mass. The mass of a globally mixed layer of depth \(\Lambda_{\mathrm{reg}}\) is \(\rho_{\mathrm{reg}}A_{\mathrm{Moon}}\Lambda_{\mathrm{reg}}\). Dividing the allowed BP mass by the accumulation interval \(t_{\mathrm{reg}}\) and correcting for the physical deposition fraction gives
\begin{equation}
\begin{aligned}
M_V\Gamma_V
&\lesssim
\frac{\zeta_{\mathrm{BP}}\rho_{\mathrm{reg}}
A_{\mathrm{Moon}}\Lambda_{\mathrm{reg}}}
{\eta_{\mathrm{dep}}t_{\mathrm{reg}}}
\\
&\lesssim
\changed{1.4}{1}\times10^{16}\,
\kg\,\Gyr^{-1}\,
\zeta_{\mathrm{BP}}\eta_{\mathrm{dep}}^{-1},
\end{aligned}
\end{equation}
where \(\rho_{\mathrm{reg}}\sim1.5\times10^3\,\kg\,\meter^{-3}\), \(\Lambda_{\mathrm{reg}}\sim1\,\meter\) and \(t_{\mathrm{reg}}\sim4\,\Gyr\). This ceiling uses no grain-count information and serves only to exclude the physically trivial regime in which the lunar regolith would itself be dominated by technomaterial.

If the Moon has been monitored sufficiently well that an arriving ETI would have been recognised during a specified interval \(t_{\mathrm{obs}}\), the absence of such an observation implies \(\Gamma_V\leq\bar N/t_{\mathrm{obs}}\), independently of grain survival. The corresponding comparison for the approximate duration of the modern Space Age is shown in Figure~\ref{fig:ABP_GrainLimits}.

\subsection{Targeted patches versus global seeding}
\label{subsec:targetedpatchesversusglobalseeding}

\deleted{A civilisation that instead brings capabilities to the Solar System to seed only a limited region of the surface can reduce the required mass proportionally -- if we know where to look. Locally, the density of grains is higher, but we are more likely to miss the grains altogether. The confinement of particles to these patches is akin to the confinement of grains in the ISM to bubbles in the undirected case, except that we have some choice of where to sample.}

\added{A civilisation that seeds only a limited region of the lunar surface can reduce the required deposited mass in direct proportion to the seeded area. This material advantage is accompanied by an equal and opposite search penalty. If the total area containing deposited technomaterial is \(A_{\mathrm{dep}}\ll A_{\mathrm{Moon}}\), a single location selected without prior information intersects that material with probability}
\begin{equation}
\added{p_{\mathrm{hit}}
\simeq
\frac{A_{\mathrm{dep}}}{A_{\mathrm{Moon}}}.}
\end{equation}
\added{Relative to global seeding, the probability of discovery is therefore reduced by the factor \(A_{\mathrm{Moon}}/A_{\mathrm{dep}}\). Equivalently, a blind search requires of order \(A_{\mathrm{Moon}}/A_{\mathrm{dep}}\) independent sampling locations to obtain an order-unity probability of intersecting the deposited region. The low material requirements derived for a known patch are consequently conditional on independent information that localises the deposition; they do not apply to a single randomly selected sample. The confinement of particles to surface patches is analogous to the confinement of undirected grains within ISM bubbles, except that lunar reconnaissance may in favourable cases provide some basis for choosing where to sample.}

\deleted{In this local delivery scenario, each visit deposits grains in a single patch with an area \(A_P\). The mean number of patches overlapping any given point on the Moon's surface is closely related to the covering factor, and is}
\added{In this local-delivery scenario, each visitation deposits grains within a patch of area \(A_P\). Over the interval \(t_{\mathrm{int}}\), the expected number of deposition events is \(\Gamma_Vt_{\mathrm{int}}\). If patch locations are statistically independent, the mean number of patches covering any specified point on the Moon is}
\begin{equation}
\label{eqn:patchcoveringfactor}
C_P
=
\Gamma_Vt_{\mathrm{int}}
\frac{A_P}{A_{\mathrm{Moon}}}.
\end{equation}
\added{For sparse, weakly overlapping patches, the total deposited area is approximately}
\begin{equation}
\added{A_{\mathrm{dep}}
\simeq
\Gamma_Vt_{\mathrm{int}}A_P,}
\end{equation}
\added{so that \(C_P\simeq A_{\mathrm{dep}}/A_{\mathrm{Moon}}\). More generally, statistically independent patch placement gives the covered surface fraction}
\begin{equation}
\label{eqn:patchcoveredfraction}
\added{f_{\mathrm{cov}}
=
1-\exp(-C_P),}
\end{equation}
\added{which approaches \(C_P\) in the sparse-patch limit and tends to unity once many patches overlap.}

\deleted{When \(C_P\gg1\), the population of grains in the regolith at any location is sourced from many visitations. While a smaller \(A_P\) means that fewer patches contribute to that population, the number of particles from each patch increases proportionately, so that the expected density of grains in a sample is constant. Even if we sample from a location that is known to be a patch from a specific visitation, the background of particles from geological history outnumbers the particles from that particular event. Thus the constraints from equations \ref{eqn:Constraint_Directed_Global_Variables}--\ref{eqn:Constraint_Directed_Global_Numerical} still apply. Of course, we could still target a specific patch for other reasons, perhaps to recover possibly still operating BPs from a very recent visitation.}

\added{When \(C_P\gg1\), almost every location has been covered by multiple deposition events because \(f_{\mathrm{cov}}\rightarrow1\). A smaller \(A_P\) reduces the number of patches covering a particular location but increases the particle surface density within each patch by the same proportion, leaving the long-term expected grain density in a random sample unchanged. The homogeneous constraints from Equations~\ref{eqn:Constraint_Directed_Global_Variables}--\ref{eqn:Constraint_Directed_Global_Numerical} therefore continue to apply. Targeting an individually recognised recent patch may nevertheless remain advantageous for recovering unusually concentrated, less degraded or potentially still functional material.}

\deleted{When \(C_P\lesssim1\), however, the constraints that we can set depend strongly on our search strategy. Any examination of a few random locations on the Moon's surface will probably just turn up empty spaces between all the patches and thus \(N_{\mathrm{obs}}=0\) as a matter of course even if \(M_V\) is very big. This is analogous to the undirected case when the Moon misses every grain-laden bubble (\(\Mean{N_{B,\odot}}\ll1\)). A major difference is that we have only one trajectory for the Moon through the ISM over its location in the undirected case. A sample from a location distant enough from one patch to be outside of its footprint serves as an independent trial, however, increasing the mean number of patches we have samples from proportionately. Let \(N_L\) be the number of independent random locations represented in our regolith sample. Then the floor on our constraints is given by \(C_PN_L\leq\bar N\), or}

\added{When \(C_P\lesssim1\), the interpretation depends strongly on the search strategy. A random location lies within deposited material with probability}
\begin{equation}
\added{p_1
=
f_{\mathrm{cov}}
=
1-\exp(-C_P)
\simeq C_P}
\end{equation}
\added{in the sparse-patch limit. The probability that at least one of \(N_L\) independent locations intersects a patch is}
\begin{equation}
\label{eqn:atleastonepatchintersection}
\added{p_{\geq1}
=
1-(1-f_{\mathrm{cov}})^{N_L}
=
1-\exp(-N_LC_P).}
\end{equation}
\added{Equivalently, the probability that all \(N_L\) locations miss deposited material is \(\exp(-N_LC_P)\). When \(N_LC_P\ll1\), the intersection probability reduces to \(p_{\geq1}\simeq N_LC_P\). A null result from only a few randomly selected locations is therefore expected whenever the deposited area occupies a small fraction of the lunar surface, even if the particle density and mass within each patch are large.}

\added{This is analogous to the undirected case in which the Moon misses every grain-rich ISM bubble, \(\Mean{N_{B,\odot}}\ll1\), but lunar sampling permits multiple spatially independent trials. Applying the same Poisson upper limit \(\bar N\) to the expected number \(N_LC_P\) of patch intersections gives}
\begin{equation}
\label{eqn:randompatchratefloor}
\Gamma_V
\leq
\frac{\bar N A_{\mathrm{Moon}}}
{N_LA_Pt_{\mathrm{int}}}.
\end{equation}
\added{This expression makes the area penalty explicit. At fixed visitation rate, reducing the patch area by the factor \(A_{\mathrm{Moon}}/A_P\) reduces the probability that any one blind sample intersects that visitation by the same factor. Increasing the mass deposited inside the patch raises the local grain density but does not improve the probability of locating the patch.}

\changed{Far more optimistic is the case in which we know (or suspect) where a patch is located.}{A different, conditional limit applies when independent evidence identifies or strongly localises a deposited patch before the regolith is examined. In that case the areal interception penalty has already been overcome, and the relevant question is the number of particles expected within a sample taken from inside the patch.} For a sparse patch population, the expected count from the identified visitation is
\begin{equation}
\label{eqn:knownpatchexpectedcount}
\Mean{N_{\mathrm{obs}}}
=
\changed{\eta_{\mathrm{comp}}}{\eta_{\mathrm{comp}}\eta_{\mathrm{dep}}}
\frac{M_V}{m_G}
\frac{A_s}{A_P}
\eta_z,
\end{equation}
where \(\eta_z\) is the fraction of deposited particles lying within the sampled depth after regolith mixing. Unlike the blind-search calculation, this conditional equation does not include the probability that the sample lies inside the patch; that geometric penalty has already been overcome by prior localisation. Substituting the spherical-grain mass and applying the Poisson null limit gives
\begin{equation}
\label{eqn:knownpatchmassconstraint}
M_V
\lesssim
\frac{4\pi\bar N\rho_G r_G^3 A_P}
{3\eta_{\mathrm{comp}}\eta_{\mathrm{dep}}\eta_zA_s}.
\end{equation}
The required mass scales linearly with patch area because distributing the same particle surface density over a larger region requires proportionally more material, and it scales as \(r_G^3\) because the mass per grain does likewise.

For a completely examined \(1\,\meter^2\)-by-\(1\,\meter\) column with \(\eta_{\mathrm{comp}}=\eta_z=1\),
\begin{equation}
M_V
\changed{\leq3.8}{\lesssim4}\,\mg\,
\eta_{\mathrm{dep}}^{-1}
\left(\frac{r_G}{1\,\um}\right)^3
\left(\frac{A_P}{100\,\km^2}\right).
\end{equation}
A lower aggregate completeness would weaken this limit in inverse proportion. For example, \(\eta_{\mathrm{comp}}=0.1\) would raise the required initial budget to order tens of milligrams even before any additional physical deposition loss.

This low mass should not be interpreted as the requirement for discovery through blind lunar sampling. For \(A_P=100\,\km^2\), a single patch occupies only
\begin{equation}
\frac{A_P}{A_{\mathrm{Moon}}}
\simeq
2.6\times10^{-6}
\end{equation}
of the lunar surface. Absent prior localisation, approximately
\[
\frac{A_{\mathrm{Moon}}}{A_P}
\simeq
3.8\times10^5
\]
independent sampling locations would be required for an order-unity probability of intersecting that one patch. The milligram-scale constraint is powerful precisely because it is conditional on sampling the correct region.

The sensitivity to visitation rate when sampling a suspected patch depends on how its location was identified. If visible impact or emplacement evidence persists for only a few million years, orbital reconnaissance would not constrain visitation rates much below order \(1\,\Myr^{-1}\). At the other extreme, if every visitation targets one particular lunar location, the absence of any patch there is governed by the global event-rate floor in Equation~\ref{eqn:Constraint_Directed_Global_Floor}. The lunar south pole, with its known water-ice resources, might be one such ``Schelling point'' at which extraterrestrial and human salience converge even without communication.

\deleted{These numbers can be compared with the macroscopic inscribed-matter scenario of \citep{RoseWright2004}, who argue that a single \(\sim100\,\gram\) information-bearing probe constitutes an energetically efficient interstellar communication channel.}

\added{These estimates may be viewed as a microscopic extension of the inscribed-matter paradigm developed by \citet{RoseWright2004} and quantified further by \citet{Hippke2018}. Rose and Wright argued that a single information-bearing probe of order \(100\,\gram\) could constitute an energetically efficient interstellar communication channel. Taking \(\Gamma_V\simeq1/t_s\) for one effective global seeding event during the sampled interval, Equation~\ref{eqn:Constraint_Directed_Global_Numerical} corresponds at unit completeness to approximately \(1.4\,\kg\,\eta_{\mathrm{dep}}^{-1}\) of released material; physical deposition efficiencies of \(0.01\)--\(0.1\) raise this to order \(10\)--\(100\,\kg\). By contrast, a known \(100\,\km^2\) patch requires only a few milligrams before the same efficiency correction. Both regimes are modest compared with large-scale industrial or megastructure mass budgets. The principal trade-off is therefore delivery geometry and recoverability: global seeding maximises the probability that an arbitrary future sample encounters a BP, while patch seeding requires far less material but depends on identifying and sampling the correct location.}

\subsection{Remote spraying of the lunar surface}
\label{subsec:globalseedingofthelunarsurface}

The Moon may also be seeded with technograins following a visit elsewhere in the Solar System that releases particles into interplanetary space. A collisional cascade that destroys a visitor-built swarm in heliocentric or geocentric orbit is one possibility; asteroid mining that releases recognisable particulate by-products is another \citep{ForganElvis2011}.

\changed{For material released elsewhere in the Solar System, successful deposition requires two sequential outcomes: a particle must first intersect the Moon, with probability \(\eta_{\mathrm{hit}}\), and must then survive the impact in recognisable form, with probability \(\eta_{\mathrm{imp}}\).}{For material released remotely, physical deposition requires two sequential outcomes: a particle must first intersect the Moon, with probability \(\eta_{\mathrm{hit}}(R)\), and a fraction of the impacting mass must then be retained in the lunar regolith, \(\eta_{\mathrm{ret}}\). We therefore write}
\begin{equation}
\label{eqn:remotedepositionefficiency}
\eta_{\mathrm{dep}}(R)
=
\eta_{\mathrm{hit}}(R)
\changed{\eta_{\mathrm{imp}}}{\eta_{\mathrm{ret}}}.
\end{equation}
\added{This factorisation separates orbital-delivery losses from impact retention. Loss of recognisable technological signature and failure of the analytical workflow are applied separately through \(\eta_{\mathrm{comp}}\), preventing the same survival loss from being counted twice.}

We define \(\eta_{\mathrm{hit}}(R)\) as the fraction of particles released at heliocentric distance \(R\) that ultimately intersects the lunar surface. The representative values used here assume isotropic release into heliocentric orbits and incorporate geometric interception, gravitational focussing and solar-radiation effects at the order-of-magnitude level rather than through a detailed trajectory integration. A full trajectory-dependent calculation is deferred to future work.

Under the illustrative interception efficiencies adopted here, release distances corresponding to the orbits of Mercury, Venus and the Earth-Moon system require initial masses of order \(10^9\)--\(10^{10}\,\kg\). Although substantial, these quantities remain far below the material budgets associated with megastructures or planetary-scale industry. A civilisation capable of sustained space operations could, in principle, deploy such a signature deliberately.

\section{Positioning APs within technosignature frameworks}
\label{sec:positioningapswithintechnosignatureframeworks}

Having established the distinct constraint regimes associated with undirected
and directed particulate delivery, it is useful to situate APs within broader
technosignature frameworks \added{\citep{Tarter2007,Wright2021}}. Different technosignature approaches probe
different fundamental dimensions of technological activity: the Kardashev scale
classifies civilisations by their characteristic energy throughput
\citep{Kardashev1964}, while Zubrin’s framework emphasises their spatial reach
and operational domain \citep{Zubrin2000}. By contrast, the particulate
technosignatures considered here constrain a civilisation’s \emph{material}
footprint—the cumulative mass of engineered material processed, mobilised, or
released into circumstellar or interstellar space over gigayear timescales.
This axis of technological expression is largely orthogonal to energy usage or
spatial presence and provides a complementary means of assessing both extant
and long-extinct civilisations.

\added{The present framework should be viewed as extending the existing technosignature landscape rather than replacing established approaches. Contemporary technosignature research increasingly encompasses a diverse portfolio of complementary observational strategies, each probing different technological manifestations, temporal regimes and astrophysical environments \citep{LingamLoeb2021,WrightSETI2026}. The particulate approach developed here is intended to complement these methods by providing direct observational access to the deep archaeological record of technological activity.}

\added{The principal astrophysical implication of these limits is therefore not that present-day technological civilisations should already be detectable through particulate debris. Rather, they constrain cumulative material processing on scales associated with planetary-scale industry, extensive astroengineering or very long-lived technological activity. Present human technological throughput remains many orders of magnitude below these levels, illustrating both the conservatism of the assumptions and the scale of the technological systems to which this search is most sensitive.}

\begin{figure*}[t!]
\centering
\includegraphics[width=\textwidth]{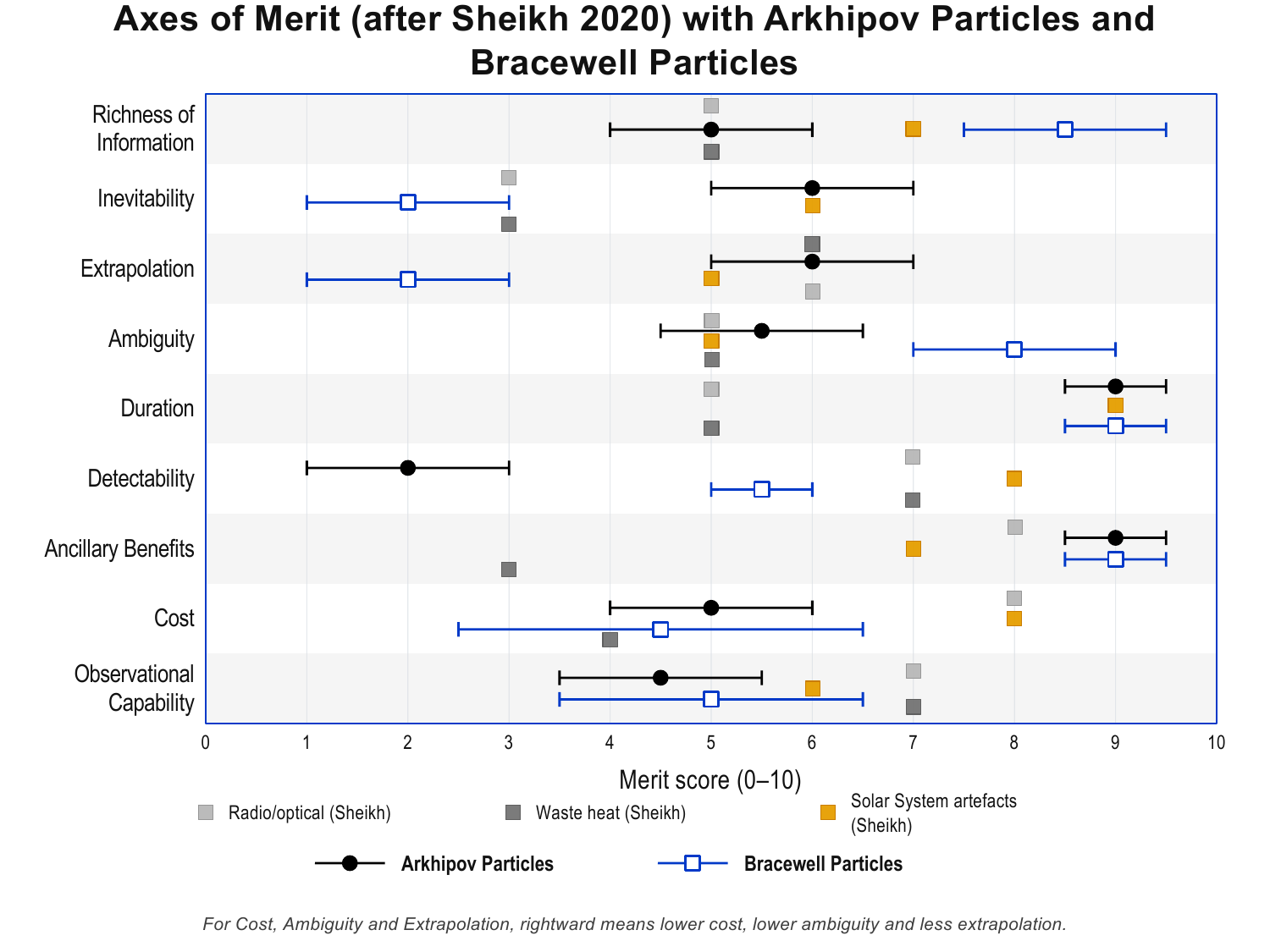}
\caption{%
\deleted{Axes-of-merit comparison of technosignature classes using the nine-dimensional framework of Sheikh (2020). Green curves reproduce Sheikh's assessments for radio SETI, purple for optical SETI, grey for infrared Dyson-sphere searches, and red for Solar System artefacts. Gold and navy curves show the qualitative placements of Arkhipov Particles (APs) and Bracewell Particles (BPs) introduced in this work: the AP trace reflects the undirected, technodust scenario shaped by Galactic transport and heliospheric filtering (\secref{subsec:fluxformalismandconstraintmodelling}), whereas the BP trace corresponds to directed or engineered delivery pathways with fewer astrophysical constraints (\secref{sec:deliberatetargetingoftheinnersolarsystem}). These curves are illustrative rather than prescriptive, indicating mode-dependent merit trends rather than absolute scores, and highlight how particulate technosignatures occupy a distinct region of Sheikh's technosignature landscape.}%
\added{Qualitative placement of Arkhipov Particles and Bracewell Particles within Sheikh's Nine Axes of Technosignature Merit. The indicated reference profiles approximate qualitative appraisals presented by \citet{SheikhAxes2020} for radio/optical communication, waste heat and Solar System artefacts, mapped here onto a \(0\)--\(10\) scale in which higher values denote greater merit. The AP and BP markers show corresponding fiducial assessments, while the horizontal bars indicate plausible ranges across alternative production, delivery, preservation, sampling and analytical scenarios; they are not statistical confidence intervals. The figure is intended to illustrate comparative strengths and trade-offs rather than to provide an objective ranking, and the AP and BP ranges are expected to evolve as transport modelling, laboratory calibration and observational experience improve.}%
}
\label{fig:axesmerit}
\end{figure*}

Sheikh’s ``Nine Axes of Technosignature Merit'' -- observational capability, cost, ancillary benefits, detectability, duration, ambiguity, extrapolation, inevitability and information -- provide a useful framework for situating APs and BPs within broader technosignature studies. In her analysis, Solar System artefacts already score highly on several axes relative to classical radio or waste-heat searches \citep{SheikhAxes2020}. \deleted{Building on this, we sketch in Fig.~\ref{fig:axesmerit} our qualitative placement of Arkhipov Particles (yellow markers) and Bracewell Particles (red markers) along the same axes.}
\added{Building on this, Figure~\ref{fig:axesmerit} reconstructs Sheikh's qualitative appraisals of radio/optical communication, waste-heat and Solar System artefact searches, and places Arkhipov Particles and Bracewell Particles within the same nine-axis framework. The light grey, dark grey and orange filled squares reproduce Sheikh reference profiles for Radio/optical, Waste heat and Solar System artefacts respectively, while APs and BPs are shown as filled black circles and open blue boxes indicating fiducial placements with horizontal ranges representing plausible variation across alternative scenarios. These ranges are heuristic and scenario dependent, not statistical confidence intervals, and no aggregate score is implied.}

\deleted{On observational capability and search cost, the limiting factor is not analytical instrumentation or computation but access to curated lunar samples or robust \textit{in situ} microscopy at scale. Where such access exists, the relevant measurements can be carried out with existing or near-term laboratory methods, augmented by machine-learning pipelines for high-throughput screening.
Both classes score strongly in detectability and duration: once deposited, solid artefacts can persist in the lunar regolith for gigayear timescales, and their signatures can be sought retrospectively.}
\added{On observational capability and cost, the principal limitations are access to representative lunar samples, high-throughput sample preparation and validated analytical pipelines. The underlying microscopy and materials-analysis instrumentation exists, but its application at the required scale remains immature. Duration is a strong attribute of both APs and BPs because solid residues may persist in the lunar regolith for gigayear timescales. Detectability is more scenario dependent: it is limited for diffuse AP populations, but may be higher for deliberately concentrated or information-bearing BPs.}

Interpreting these merit axes requires distinguishing the undirected and
directed cases described in \secreftwo{subsec:fluxformalismandconstraintmodelling}{sec:deliberatetargetingoftheinnersolarsystem}.
Undirected APs, arising as inert
technodust from megaswarm debris, necessarily inherit Galactic-transport and
heliospheric-filtering uncertainties, and their ratings on axes such as
observational capability, extrapolation and ambiguity reflect these
limitations. Directed particulate probes or BPs, by contrast, benefit from
engineered delivery pathways and potentially low-velocity emplacement, leading
to higher merit scores on detectability, duration and ambiguity, but lower
scores on extrapolation owing to the greater technological assumptions involved.
\deleted{In all cases the achievable observational capability is constrained primarily by
access to curated or \textit{in situ} lunar samples rather than by
computational or analytical resources, and the ratings shown in
Fig.~\ref{fig:axesmerit} should therefore be read as mode-dependent placements
within this broader merit landscape rather than as universal maxima.}
\added{The ranges shown in Figure~\ref{fig:axesmerit} reflect these mode-dependent uncertainties and should not be interpreted as universal maxima.}

\changed{Ambiguity remains a challenge, particularly for first detections in imagery, but isotopic analyses and detailed microstructural studies should allow many candidate APs or BP remnants to be distinguished from terrestrial contaminants or spacecraft hardware.}{Ambiguity remains a central challenge, particularly for first-stage detections in imagery. In favourable cases, convergent isotopic, compositional and microstructural evidence may distinguish candidates from natural materials, terrestrial contamination and spacecraft hardware, but no single property is expected to be decisive.}
\changed{On the extrapolation axis, APs score higher because we already produce microscopic debris that escapes the Solar System, whereas BPs represent a plausible, but still hypothetical, extrapolation of current smart-dust and autonomous-probe concepts.}{On extrapolation, APs require less departure from demonstrated technology than BPs because human space activity already generates microscopic and macroscopic technological debris, although sustained interstellar dispersal remains speculative. BPs represent a larger extrapolation from current smart-dust and autonomous-probe concepts.}
\changed{For inevitability, APs fare relatively well: any technological spacefaring civilisation is likely to generate microscopic ``trash'' as a by-product of its activity, even in the absence of deliberate signalling, whereas the deployment of BPs requires an additional level of intent.}{For inevitability, APs are rated more favourably than BPs because sustained spacefaring activity may generate microscopic debris even without deliberate signalling, whereas BP deployment requires an additional decision to manufacture and disperse such carriers.}
Finally, richness of information is moderate for APs -- microscopic debris may still encode advanced materials engineering -- but potentially very high for BPs, which in extreme cases could contain substantial embedded data or even more complex autonomous systems.
\changed{In this sense the AP and BP search space occupies a distinct, and in several dimensions attractive, region of Sheikh’s merit landscape.}{In this sense, APs and BPs occupy distinct and complementary regions of Sheikh’s merit landscape.}

\added{Several attributes distinguish particulate searches from traditional technosignature programmes. Because physical artefacts may survive for geological timescales, they provide access to technological activity that no longer exists and are not limited to civilisations transmitting during the present epoch. Conversely, the approach presently relies upon less mature transport models, limited laboratory calibration of candidate technomaterials and comparatively little observational experience. Radio and optical SETI benefit from more than six decades of theoretical development, instrument optimisation and survey practice. Accordingly, although Solar System exo-archaeology offers access to a complementary region of technosignature parameter space, some aspects of its practical performance remain more uncertain.}

\added{The comparative framework presented here should therefore be viewed as a snapshot of the present state of the field rather than a permanent evaluation. As laboratory studies, lunar sample analyses, transport modelling and analytical techniques mature, several of the criteria considered in Figure~\ref{fig:axesmerit} may change substantially. In this sense, the figure represents a research roadmap as much as a comparison of existing capabilities.}

\section{Conclusion and methodological outlook}
\label{sec:conclusionandmethodologicaloutlook}

We have examined the plausibility, transport, survivability, and
\changed{means of potential detection}{potential detection} of micron-scale
extraterrestrial artefacts, described here as Arkhipov Particles (APs) and
Bracewell Particles (BPs). By treating the Moon as a long-duration collector
of exogenous material, we have outlined an empirically grounded approach to
possible technosignature discovery based on planetary sampling, materials
analysis, and astrophysical transport modelling. The analysis presented here is prospective: it neither claims that artificial particles of non-human origin exist on the Moon nor reports evidence of their existence. Its contribution is a framework for testing this possibility and for interpreting future detections or non-detections under explicitly stated assumptions.

Unlike electromagnetic SETI\added{ \citep{CocconiMorrison1959,Drake1961}},
which \changed{probes}{primarily probes} contemporaneous transmitters, this
particulate approach integrates technological activity over gigayear timescales
and is therefore sensitive to technomaterial released by civilisations that may
no longer exist. \changed{A particle-focussed SETA}{Particulate SETA} can thus
function as a form of exo-archaeology: a search for durable material traces of
technological activity preserved within planetary archives.

Our analysis \changed{shows}{indicates} that artificial
\changed{sub-micron to micron-scale}{grains in the submicron-to-micron size
range} can survive transport through the interstellar medium over hundreds of
Myr to Gyr, become dynamically coupled to ISM flows, and traverse
sub-kiloparsec to kiloparsec scales.
\deleted{A narrow but physically well-defined slow-arrival channel allows a
fraction of such grains to reach the lunar surface at relative velocities
compatible with partial survival or diagnostic residue formation.}
\added{Solar radiation pressure also defines a narrow slow-arrival channel in
which lower encounter speeds make recognisable preservation more favourable.
This channel is not a universal, material-independent survival boundary:
intact particles, fragments and diagnostic residues will survive with
efficiencies that depend on composition, impact geometry, target conditions
and subsequent regolith processing.}

\changed{A verified detection of even a single AP}{A verified detection
establishing an extraterrestrial technological origin for even a single AP or
BP} would indicate the past existence of at least one other technological
civilisation in the history of the Galaxy. Such a conclusion would not depend
on whether the release was intentional or incidental, nor on the specific
scenario modelled here: it would simply but profoundly imply that
\changed{large-scale astroengineering or sustained technological mass
processing has occurred}{technological activity beyond Earth has produced
durable engineered material capable of entering a planetary archive} at some
time in Galactic history.

Quantitatively, the absence of detectable artificial grains within a
well-characterised \(\sim1\,\mathrm{m^3}\) regolith volume places upper bounds
on the time-integrated cumulative undirected technomaterial output of
large-scale spacefaring civilisations: as shown in
Figure~\ref{fig:LimitsMegaswarms}, a null detection excludes scenarios in which
\added{eligible }Solar-type stars typically disperse more than
\changed{\ensuremath{\sim0.09\,M_{\oplus}}}
{\ensuremath{\sim0.1\,M_{\oplus}}} of long-lived artificial particulate debris
over Galactic history. These bounds constrain the combination of release mass
and event rate per eligible star under explicitly stated transport assumptions.
\added{The one-cubic-metre volume is a scalable scientific benchmark rather
than a threshold mission capability: initial searches may analyse volumes of
order \(0.1\,\mathrm{m^3}\), with progressively stronger constraints
accumulating as independent sites and larger regolith volumes become
accessible.}

Directed delivery defines a second, formally distinct constraint regime.
Rather than constraining the cumulative particulate output of an eligible
stellar population, null results on the Moon can also be interpreted as bounds
on the product of visitation rate and released mass per event,
\(M_V\Gamma_V\), for civilisations that deliberately place particulate
artefacts within the inner Solar System. In this case, detectability depends
primarily on delivery geometry: globally distributed BPs maximise the chance
that an arbitrary regolith sample intercepts a relic grain, whereas confinement
to small patches can either weaken or strengthen the resulting limits depending
on whether sampling locations are chosen at random or guided by independent
evidence of where deposition occurred. \added{Confinement raises the local
grain density but imposes a reciprocal area penalty on blind sampling, so the
milligram-scale known-patch limits are conditional on independent localisation
of the deposited region.} Remote \changed{spraying}{release} from elsewhere in
the inner Solar System introduces a further attenuation through the deposition
efficiency \(\eta_{\mathrm{dep}}\), so that large released masses may still
yield only modest physically deposited inventories on the Moon.

The methodology developed here is experimentally tractable.
\changed{Machine-vision pipelines such as YOLO-ET provide a scalable triage
layer for grain-scale analysis, enabling the rapid identification of
statistically unusual particles within large regolith datasets prior to
detailed compositional and structural characterisation. SEM, EDS, SIMS,
nano-CT, and FIB tomography could provide forensic discrimination between
natural and engineered materials. Hypervelocity impact experiments using
engineered analogues are essential for calibrating survivability and residue
morphology. Improved modelling of ISM transport, heliospheric filtering,
regolith mixing, and directed-delivery geometries will refine both detections
and null constraints.}{Machine-vision pipelines can provide high-throughput
triage of regolith grains and impact features, while SEM, EDS, SIMS, nano-CT
and FIB tomography provide the compositional and structural evidence needed to
distinguish natural, anthropogenic and potentially engineered materials.}

\added{A priority experimental programme should combine controlled
hypervelocity impacts of representative advanced technomaterials and plausible
biological or molecular information-storage media into well-characterised
lunar-regolith simulants with openly available reference libraries of natural
lunar materials, spacecraft contaminants and experimentally generated residues
\citep{DworkinContamination2018,McCubbinCuration2019}. Measurements should
quantify intact and fragment survival, changes in composition and
microstructure, and false-negative regimes in which engineering signatures
become unrecognisable, thereby providing empirical estimates of
\(\eta_{\mathrm{surv}}\), \(\eta_{\mathrm{rec}}\) and
\(\eta_{\mathrm{det}}\). These data would support high-throughput microscopy
and multimodal AI as transparent decision-support tools, validated against
held-out materials, replicated measurements and independent laboratories
\citep{Bommasani2021}.}

The inference space is notably asymmetric. A
\changed{detection would immediately establish the existence of large-scale
technological activity beyond Earth}{verified detection would establish
extraterrestrial technological activity}, independent of detailed model
assumptions. A null result does not imply that such activity has never
occurred, but it does progressively restrict the allowed parameter space of
undirected particulate-release scenarios as sampled regolith volumes and
analytical sensitivity increase, while also constraining classes of deliberate
delivery scenarios under explicitly stated emplacement assumptions.

Within the broader landscape of technosignature searches, particulate SETA
probes a dimension largely orthogonal to energy-based (Kardashev-type) or
communication-based searches: the cumulative material footprint of
technological civilisations. Because the lunar regolith integrates over
billions of years and multiple Galactic environments, it provides a uniquely
stable archive in which such traces may persist.

The Moon already represents a scientifically compelling target for planetary
science and astronomy \citep{Crawford2012}. Its surface preserves an
unparalleled record of early Solar System evolution and aspects of the Galactic
environment of the Sun, including variations in cosmic-ray flux and episodic
events such as nearby supernovae or dense-cloud encounters
\citep{Crawford2021}. The additional possibility that it may also preserve
material evidence of past extraterrestrial technology strengthens the case for
systematic, well-characterised regolith sampling. Whether through positive
detection or increasingly stringent null constraints, the search for Arkhipov
Particles and Bracewell Particles offers a physically grounded and
experimentally accessible route to addressing one of astrobiology's central
questions: whether technological life has ever arisen elsewhere in our Galaxy.

\ack[Acknowledgements] {\added{We wish to thank the anonymous Reviewer for the many lucid and relevant contributions to the evolution of our thinking in the context of Solar System technosignatures.} LJP thanks the CEO and Chairman of the SETI Institute, Bill Diamond and Dan Lankford, together with the Director of the Carl Sagan Center there, Nathalie Cabrol, and their many  colleagues, for their deep engagement and encouragement in exploring this new and highly complementary modality for SETI. Drs. David Deamer and Russell Kerschmann were also both instrumental in challenging and developing ideas about how dense information and even self-replicating capabilities can be contained at the micron scale, and where and how they might best be detected.  BCL and APVS thank the Breakthrough Listen program for their support. Funding for Breakthrough Listen research is sponsored by the Breakthrough Prize Foundation (https://breakthroughprize.org/).
During preparation and revision of this manuscript, LJP used ChatGPT (OpenAI), accessed through a ChatGPT Pro subscription, from approximately January 2025 to October 2026 for editorial assistance. The models and modes used included o1 pro mode, o3-pro, GPT-5 Pro through GPT-5.6 Sol Pro and GPT-6 Astra Pro. The exact model version was not recorded for every interaction. This assistance was limited to simple language editing, basic manuscript organisation, identification of redundant passages, and LaTeX formatting. The tool also assisted with checking bibliographic details and matching arXiv preprints to corresponding peer-reviewed publications. The authors reviewed and revised the AI-assisted material, cross-verified all sources, and take full responsibility for the manuscript’s content, interpretations and references.}

\appendix
\section*{Appendix A. Derivations of the number of ``bubbles'' encountered}
\addcontentsline{toc}{section}{Appendix A. Derivations of the number of ``bubbles'' encountered}
\label{sec:appendix:derivationsofthenumberofbubblesencountered}

\changed{In this appendix, we present a derivation and more complete expressions for how many release events have contributed to the Moon's AP population in the bubble model. For simplicity, each bubble is assumed to be static and spherical over the duration of the encounter.}{In this appendix, we derive the mean number of grain-rich regions that intercept the Solar System during the interval sampled by the lunar regolith. For simplicity, each bubble is assumed to remain spherical with a fixed radius during an encounter, while its centre moves at constant relative speed.}

Working in the Solar System's reference frame, consider a bubble whose centre follows a straight trajectory with closest-approach distance, or impact parameter, \(b\). Suppose that the bubble forms at time \(\vartheta\), when its centre is at longitudinal coordinate \(X\), and moves with speed \(v>0\). Its centre is then at
\begin{equation}
x(t)=X+v(t-\vartheta).
\label{eqn:bubblecentretrajectory}
\end{equation}
\changed{its leading edge is at $x(t) + \sqrt{R^2 - b^2}$ and its trailing edge is at $x(t) - \sqrt{R^2 - b^2}$.}{For a bubble of radius \(R\), the half-chord intercepted by a trajectory of impact parameter \(b\) is}
\begin{equation}
h(R,b)=\sqrt{R^2-b^2},
\qquad 0\leq b\leq R.
\label{eqn:bubblehalfchord}
\end{equation}
\changed{Detectable grains come from bubbles that have contained the Moon in the past integration time $-t_{\mathrm{int}} \le t \le 0$. For a bubble with lifetime $\tau$, this implies that it must have formed between $-(t_{\mathrm{int}} + \tau)$ and $0$.}{The bubble contains the origin whenever \(\lvert x(t)\rvert\leq h\). Additionally, detectable grains can come only from bubbles whose lifetimes overlap the integration interval \(-t_{\mathrm{int}}\leq t\leq0\). For a fixed bubble lifetime \(\tau_B\), the formation time must lie in}
\[
-(t_{\mathrm{int}}+\tau_B)\leq\vartheta\leq0.
\]

\added{For a given \(\vartheta\), the overlap interval begins at \(t_-=\max(\vartheta,-t_{\mathrm{int}})\) and ends at \(t_+=\min(\vartheta+\tau_B,0)\). Because \(x(t)\) increases monotonically for \(v>0\), interception requires \(x(t_-)\leq h\) and \(x(t_+)\geq-h\). These conditions are equivalent to the longitudinal bounds}
\begin{equation}
\begin{aligned}
X_-(R,b,v,\vartheta)
&=
-h(R,b)-v\min(\tau_B,-\vartheta),
\\
X_+(R,b,v,\vartheta)
&=
h(R,b)-v\max(0,-t_{\mathrm{int}}-\vartheta).
\end{aligned}
\label{eqn:bubblelongitudinallimits}
\end{equation}
\added{These limits select precisely the bubble trajectories whose finite spatial and temporal extents overlap the Solar System's worldline during \(t_{\mathrm{int}}\).}

The bubbles form at a volumetric rate \(n_{\star,B}\Gamma_B\). \changed{They have a velocity distribution $f_v(v)$ and the radius at interception has a distribution $f_R(R)$.}{Let \(f_v(v)\) and \(f_R(R)\) be normalised distributions of relative speed and bubble radius, respectively, and assume that \(R\) and \(v\) are statistically independent.} \added{The annular area element for trajectories with impact parameters between \(b\) and \(b+\mathrm{d}b\) is \(2\pi b\,\mathrm{d}b\).} The exact mean number of intercepted bubbles is then
\begin{multline}
\Mean{N_{B,\odot}} = n_{\star,B}\Gamma_B \int_0^\infty  
\int_0^\infty \int_0^R \int_{-(t_{\mathrm{int}}+\tau_B)}^0 \int_{X_-}^{X_+}  2\pi b \\
\times f_R(R) \times f_v(v)\,\mathrm{d}X\,\mathrm{d}\vartheta\,\mathrm{d}b\,\mathrm{d}v\,\mathrm{d}R.
\label{eqn:exactbubblecountintegral}
\end{multline}

%\added{Integrating first over the permitted longitudinal offsets and formation times gives}
%\begin{equation}
%\int_{-(t_{\mathrm{int}}+\tau_B)}^0
%\left(X_+-X_-\right)\,\mathrm{d}\vartheta
%=
%2h(R,b)(t_{\mathrm{int}}+\tau_B)
%+
%v\,t_{\mathrm{int}}\tau_B.
%\label{eqn:bubblelongitudinalintegral}
%\end{equation}
%\added{The subsequent \(b\)-integration converts the first term into the spherical volume \(\tfrac{4\pi}{3}R^3\), while the second gives the projected cross-section \(\pi R^2\). Thus}
The solution to this integral is:
\begin{equation}
\Mean{N_{B,\odot}}
=
n_{\star,B}\Gamma_B
\left[
\frac{4\pi}{3}\Mean{R^3}(\tint+\tau_B)
+
\pi\Mean{R^2}\Mean{v}\,\tint\tau_B
\right].
\label{eqn:exactbubblecount}
\end{equation}
This result is valid regardless of whether \(t_{\mathrm{int}}\) or \(\tau_B\) is larger. 
\changed{The first term dominates when bubbles are large and slow, popping in around the Solar System.}{The first term is an ``end-cap'' contribution, dominating at slow velocities: it counts the bubbles that form with the Solar System inside, essentially popping in place around it.}\changed{The latter term dominates when the bubbles are small and fast, so that the Solar System easily crosses them in their lifespans.}{The latter term is the swept-path contribution used in the main text, and dominates when the relative velocities are high and the Solar System easily crosses them over their lifespans. It is the geometric ``optical depth'', the product of number density, cross section, and path length.}

%\changed{The first term dominates when bubbles are large and slow, popping in around the Solar System. The latter term dominates when the bubbles are small and fast, so that the Solar System easily crosses them in their lifespans.}{The first term is a finite-volume, or ``end-cap'', contribution: it counts bubbles that form or remain extant while the Solar System already lies within their volume. The second is the swept-path contribution used in the main text: the instantaneous number density of extant bubbles, \(n_{\star,B}\Gamma_B\tau_B\), is multiplied by their mean projected cross-section, \(\pi\Mean{R^2}\), and by the path length \(\Mean{v}t_{\mathrm{int}}\).}

\added{For a radius distribution, define the effective radius \(R_{\mathrm{eff}}\equiv\Mean{R^3}/\Mean{R^2}\). Equating the two terms in Equation~\ref{eqn:exactbubblecount} gives the transition scale}
\begin{equation}
\begin{aligned}
R_{\mathrm{tr}} 
&= \frac{3}{4}\Mean{v} \frac{t_{\mathrm{int}}\tau_B}{t_{\mathrm{int}}+\tau_B}
\\
& \approx 1.5\,\kpc \left(\frac{\Mean{v}}{20\,\kms}\right)  \times
\left[ \frac{t_{\mathrm{int}}\tau_B/(t_{\mathrm{int}}+\tau_B)}{100\,\Myr} \right].
\end{aligned}
\label{eqn:bubbletransitionradius}
\end{equation}
\added{The swept-path term dominates when \(R_{\mathrm{eff}}\ll R_{\mathrm{tr}}\). For the fiducial values, the transition radius is approximately \(1.5\,\kpc\), so the swept-path contribution exceeds the finite-volume term by a factor of order \(15\). This quantitatively supports the high-relative-velocity approximation used in the main text.}

\changed{Note that for the Maxwellian distribution,}{When a velocity average is required, we adopt either a monovelocity distribution,}
\begin{equation}
\added{f_v(v) = \delta(v - v_0)}
\end{equation}
\added{or an isotropic three-dimensional Maxwellian distribution,}
\begin{equation}
f_v(v)
=
\sqrt{\frac{2}{\pi}}
\frac{v^2}{\sigma_v^3}
\exp\left(-\frac{v^2}{2\sigma_v^2}\right),
\qquad v\geq0,
\label{eqn:maxwellianspeeddistribution}
\end{equation}
with mean speed
\begin{equation}
\Mean{v}
=
2\sqrt{\frac{2}{\pi}}\,\sigma_v.
\label{eqn:maxwellianmeanspeed}
\end{equation}
\added{The factor \(\sqrt{2/\pi}\) normalises the distribution over non-negative speeds, while the factor \(v^2\) is the spherical velocity-space volume element.}

\section*{Appendix B. Derivations of the flux of low-speed grains}
\addcontentsline{toc}{section}{Appendix B. Derivations of the flux of low-speed grains}
\label{sec:appendix:derivationsoffluxoflowspeedgrains}
\label{sec:AppendixSlowgrains}

%\begin{notefrombrian}
%I've only started working on the corrections for Appendix B.
%\end{notefrombrian}

%In this Appendix, we consider grain trajectories that enter the Earth-Moon system at slow relative velocities. Solar radiation pressure enables this possibility, as it modifies the Solar gravitational potential. In these calculations, we assume the Earth-Moon system has a circular orbit around the Sun with speed $\vEarth = \sqrt{G \Msun / \rEarth}$, where $\rEarth$ is 1~AU. We also ignore Lorentz forces on the grain, which dominate for the smallest grains.

In this appendix, we \changed{consider grain trajectories}{derive the density and lunar surface flux of grains} that enter the Earth-Moon system with small relative velocities. Solar radiation pressure 
\changed{enables this possibility, as it modifies the Solar gravitational potential.}{allows for this possibility by replacing the Solar gravitational parameter with the effective value \(G\Msun(1-\beta)\).} \changed{In these calculations, we assume}{We approximate} \changed{has a circular orbit around the Sun with speed $\vEarth = \sqrt{G \Msun / \rEarth}$, where $\rEarth$ is 1~AU.}{the Earth-Moon system as moving on a circular heliocentric orbit of radius \(\rEarth=1\,\mathrm{AU}\) and speed}
\[
\vEarth=\sqrt{G\Msun/\rEarth}.
\]
\changed{We also ignore Lorentz forces on the grain, which dominate for the smallest grains.}{Lorentz forces are neglected; our calculations thus are unreliable for the smallest charged grains.}

%A grain enters the Solar System with speed $\vinf$ and at ecliptic latitude $\iinf$\footnote{Note that $\iinf$ is not the same as the orbital inclination of the grain.}. The grain's trajectory is as described in \citet{Sterken2012}, entering the Solar System in the $+x$ direction. \added{It encounters the Earth-Moon system when its distance from the Sun is 1 AU, with a heliocentric velocity of $\vEncVec$.} The Earth's position is

A grain enters the Solar System with \added{asymptotic} speed \(\vinf\) and \added{from} ecliptic latitude \(\iinf\). \changed{Note that $\iinfOLD$}{The latitude \(\iinf\) describes the orientation of the incoming direction and} is not the same as orbital inclination. The grain's trajectory is as described in \citet{Sterken2012}, entering the Solar System in the $+x$ direction\added{, using the same inflow-oriented coordinate system}. \changed{It encounters the Earth-Moon system when its distance from the Sun is 1 AU, with a heliocentric velocity of $\vEncVec$.}{At heliocentric distance \(\rEarth\), its velocity is \(\vEncVec\), and its speed relative to the Earth-Moon system is}
%is not generally the osculating orbital inclination of the grain. Following the coordinate convention of \citet{Sterken2012}, the grain enters in the \(+x\) direction. At heliocentric distance \(\rEarth\), its velocity is \(\vEncVec\), and its speed relative to the Earth-Moon system is
\[
\added{\Delta\vEnc=\left|\vEncVec-\vEarthVec\right|.}
\]
The Earth's position and velocity in this frame are
\begin{align}
\rEarthVec
&=
\rEarth\bigl[
\cos\thetaEarth\cos\iinf\,\xNorm
+
\sin\thetaEarth\,\yNorm
+
\cos\thetaEarth\sin\iinf\,\zNorm
\bigr],
\label{eqn:earthpositionincomingframe}
\\
\vEarthVec
&=
\vEarth\bigl[
-\sin\thetaEarth\cos\iinf\,\xNorm
+
\cos\thetaEarth\,\yNorm
-
\sin\thetaEarth\sin\iinf\,\zNorm
\bigr].
\label{eqn:earthvelocityincomingframe}
\end{align}

\subsection*{B.1 The optimal trajectory}
\label{subsec:appendix:theoptimaltrajectory}

%The optimal trajectory given these assumptions is one where
%(1) the grain reaches perihelion at Earth's orbit,
%(2) the grain's \added{encounter} speed $\vEnc$ is equal to $\vEarth$,
%(3) the encounter is in-plane ($\iinf = 0$) and prograde, and
%(4) the Earth-Moon happens to be at the right place in its orbit.
%On this trajectory, the grain's trajectory just grazes the Earth's orbit at perihelion (Figure~\ref{fig:optimalsolartrajectory}).

\begin{figure}[t]
    \centering
    \includegraphics[width=0.5\textwidth]{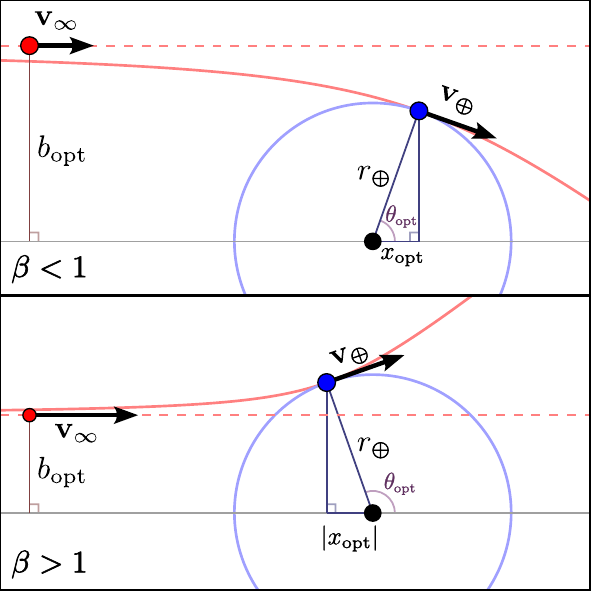}
    \caption{Geometry of the optimal slow-arrival trajectory (light red), for which the grain reaches perihelion at the circular Earth-Moon orbit (light blue) with heliocentric speed \(\vEarth\). Top: \(\vinf<\vEarth\), so \(\betaOpt<1\) and the net Solar force is attractive. Bottom: \(\vinf>\vEarth\), so \(\betaOpt>1\) and radiation pressure provides the net deceleration required to match the orbital speed.}
    \label{fig:optimalsolartrajectory}
\end{figure}

\added{Under these assumptions}, the optimal \added{slow-arrival} trajectory \deleted{given these assumptions} satisfies four conditions: (1) the grain reaches perihelion at \added{the} Earth's \changed{orbit}{orbital radius}; (2) \changed{the grain's speed $\vEnc$ is equal to $\vEarth$}{its heliocentric speed there equals \(\vEarth\)}; (3) the encounter is \changed{in-plane}{coplanar} ($\iinf = 0$) and prograde; \added{and} (4) the Earth-Moon system \changed{happens to be at the right place in its orbit}{is at the perihelion longitude when the grain arrives, resulting in a collision}. \added{On this trajectory,} the grain \deleted{just} grazes the Earth's orbit at perihelion\added{, as illustrated in} Figure~\ref{fig:optimalsolartrajectory}.

%Under these assumptions, the optimal slow-arrival trajectory satisfies four conditions: the grain reaches perihelion at the Earth's orbital radius; its heliocentric speed there equals \(\vEarth\); the encounter is coplanar and prograde; and the Earth-Moon system is at the perihelion longitude when the grain arrives. The grain then grazes the Earth's orbit at perihelion, as illustrated in Figure~\ref{fig:optimalsolartrajectory}.

%\changed{From energy conservation, the speed of the grain at $\rEarth$ is}{Conservation of specific orbital energy between infinity and \(r_{\!\oplus}\), with Solar gravity reduced by radiation pressure,gives}

\changed{From energy conservation,}{Conservation of effective specific orbital energy between infinity and \(\rEarth\) gives} the speed of the grain at $\rEarth$ \changed{is}{as}
\begin{equation}
\vEnc^2
=
\vinf^2
+
\added{2} \dfrac{G\Msun}{\rEarth}(1-\beta).
\label{eqn:encounterspeedenergy}
\end{equation}
with $\beta$ being the ratio of radiation and gravitational forces on the grain. \added{For \(\beta>1\), the effective force is repulsive and this term is negative.}
\changed{In order for}{On the optimal encounter trajectory, with} $\vEnc = \vEarth$, this ratio \changed{should take its optimal value}{must be}
%\added{The first term is the kinetic energy retained from infinity, while the second is twice the change in the effective Solar potential. For \(\beta>1\), the effective force is repulsive and this term is negative.} Setting \(\vEnc=\vEarth\) and using \(\vEarth^2=G\Msun/\rEarth\) gives
\begin{equation}
\betaOpt
=
\frac{1}{2}
\left[
1+
\left(\frac{\vinf}{\vEarth}\right)^2
\right].
\label{eqn:betaoptimal}
\end{equation}

The impact parameter $b$ measures how far off-axis the grain is as it first approaches the Sun. \added{Each value of $b$ selects an axisymmetric collection of trajectories, forming a ring at each value of $x$ attained (see appendix~B.4), although for now we are only interested in those lying on the ecliptic plane.} \added{The optimal impact parameter is determined by} the tangential (non-radial) speed \added{$\vEncTan$} of the grain at \changed{Earth's orbit}{$\rEarth$}, \added{which} is \deleted{additionally} constrained by angular momentum conservation. \changed{When the grain's perihelion is at Earth's orbit, its velocity there is}{At perihelion, the velocity must be} entirely tangential, just like an object in circular orbit. \added{Therefore,}
%to be $b \vinf/\rEarth$. The impact parameter $b$ measures how far off-axis the grain is as it first approaches the Sun. When the grain's perihelion is at Earth's orbit, its velocity there is entirely tangential, just like an object in circular orbit. This establishes an optimal impact parameter:
%At perihelion, the velocity is tangential. Equality of specific angular momentum between infinity and the Earth-orbit crossing therefore requires
\begin{equation}
\bOpt\vinf
=
\rEarth\vEarth,
\qquad
\bOpt
=
\frac{\rEarth\vEarth}{\vinf}.
\label{eqn:boptimal}
\end{equation}

%\changed{The angle between the grain's perihelion, the Sun and the downstream \((+x)\) direction is found from \citet{Sterken2012}.}{The angle between the downstream \((+x)\) direction and the optimal perihelion follows from the conic geometry \citet{Sterken2012}:}
The \added{optimal perihelion longitude $\thetaOpt$ is the} angle between the grain's perihelion, the Sun, and the downstream ($+x$) direction for the optimal trajectory\added{.} \changed{is found from}{It is calculated using the hyperbolic trajectory in} \citet{Sterken2012}, equations 5--11:
\begin{equation}
\sin\thetaOpt = \frac{2\vinf\vEarth}{\vEarth^2+\vinf^2},
\label{eqn:thetaoptimal}
\end{equation}
Thus \(0\leq\thetaOpt<\pi/2\) for \(\betaOpt<1\), while \(\pi/2<\thetaOpt\leq\pi\) for \(\betaOpt>1\).

\changed{The Sun's influence concentrates grains as they pass through
perihelion by gravitational-radiative focussing.}
{As a result of the effective Solar potential, grain trajectories are
concentrated and grain densities are higher in the encounter region than in
the ISM.}
\added{For grains entering at exactly $\vinf$, this compression factor is}
%The Sun's influence concentrates grains as they pass through perihelion by gravitational-\changed{radiatiative}{radiative} focussing. \changed{Given a fixed entry speed of $\vinf$, this compression factor,}{For grains entering at exactly $\vinf$, this compression factor is}
%\added{The Solar potential concentrates trajectories near the optimal encounter. For grains with fixed \(\vinf\), the corresponding density Jacobian is \citet{Sterken2012}
\begin{equation}
\frac{\mathrm{d}\rhoOpt}{\mathrm{d}\rhoinf}
=
\frac{1}{4}\left(\frac{\vEarth}{\vinf} + \frac{\vinf}{\vEarth}\right)^2,
\label{eqn:CompressionFactor}
\end{equation}
\changed{can be calculated to be}{from} \citet{Sterken2012} equation 15 using $\added{r = }\rEarth$ and $\added{\theta =} \thetaOpt$, and only including grains passing through perihelion \added{\citep[see also Appendix B of][]{Landgraf1999}}.

In actual encounters, these conditions do not hold exactly and so the relative speed $\Delta \vEnc$ is non-zero. \deleted{The tolerance in these parameters sets the fraction of grains that leave detectable signatures.} We now consider \changed{each of the contributing factors in turn.}{how much of the incoming population lies sufficiently close to this optimal trajectory.}

\subsection*{B.2 Tolerance in \(\beta\)}
\label{subsec:appendix:toleranceinbeta}

When a grain has $\beta \ne \betaOpt$, it reaches the Earth-Moon system with a nonzero relative velocity. The maximum \added{allowable} relative \added{encounter} speed\added{, $\dvEncMax$,} defines a window in $\beta$, and thus, grain size. Consider a grain with $\beta = \betaOpt + \Delta\beta$. Its speed at 1~AU is, by energy conservation \added{(equation~\ref{eqn:encounterspeedenergy})},
\begin{equation}
\vEnc = \vEarth \sqrt{1 - 2 \Delta\beta} \changed{\approx}{\simeq} \vEarth (1 - \Delta\beta),
\label{eqn:betaperturbedspeed}
\end{equation}
with the approximation holding if $|\Delta\beta| \ll 1$.

\added{For any $\beta$ value,} the \changed{best case is still when the grain}{trajectory with the smallest relative speed at the Earth-Moon system} reaches perihelion at \changed{the Earth-Moon's orbit}{$\rEarth$}, \changed{in the prograde direction with}{while retaining the coplanar prograde geometry} ($\changed{i_\infty}{\iinf} = 0$). \deleted{,, for which} The relative speed is $\sim |\Delta \beta| \vEarth$. \added{Imposing} an upper limit on this relative speed, $\dvEncMax$, \changed{imposes a limit on}{in turn limits} how much $\beta$ can diverge from $\betaOpt$:
\begin{equation}
|\Delta\beta| \added{~\le \Delta \beta_{\mathrm{max}}} \changed{\lesssim \frac{2 (\dvEncMax/\vEarth)}{1 + (\vinf/\vEarth)^2}.}{\simeq \frac{\dvEncMax}{\vEarth}} .
\label{eqn:betatolerance}
\end{equation}

%Let \(\beta=\betaOpt+\Delta\beta\), while retaining the optimal impact parameter and coplanar prograde geometry. Substitution into Equation~\ref{eqn:encounterspeedenergy} gives
%\begin{equation}
%\vEnc
%=
%\vEarth\sqrt{1-2\Delta\beta}
%\simeq
%\vEarth(1-\Delta\beta),
%\label{eqn:betaperturbedspeed}
%\end{equation}
%for \(|\Delta\beta|\ll1\). The resulting speed mismatch is therefore \(\Delta\vEnc\simeq\vEarth|\Delta\beta|\). Requiring it to remain below \(\dvEncMax\) gives
%\deleted{The earlier tolerance expression retained an unnecessary incoming-speed factor after the energy relation had been linearised about \(\betaOpt\).}
%\added{Requiring the linearised speed mismatch to remain below \(\dvEncMax\) gives}
%\begin{equation}
%|\Delta\beta|
%\leq
%\Delta\beta_{\max}
%\simeq
%\frac{\dvEncMax}{\vEarth}.
%\label{eqn:betatolerance}
%\end{equation}
%\added{This is a local linearisation about the optimum; it is adequate for the order-of-magnitude phase-space estimate used here.}

If there is a technograin population in the ISM, it plausibly has a wide size distribution. The debris from collisional cascades has a power-law distribution in mass. Additionally, because stars have different luminosities and masses, grains dispersed via radiation pressure should have a wide range in mass. \changed{When the density of grains is already normalized by $\eta_m$, the fraction of grains within the $\beta$ window is}{For each $\vinf$, the $\beta$ criterion only selects a narrow ``window'' centred on $\betaOpt$ out of the wide distribution in mass.}

%\added{A spherical grain's $\beta$ ratio is related to its mass. The radiative force on the grain is, from the momentum flux of sunlight, $L_{\odot} / (4\pi d^2 c) \times A_{\rm eff}$, where $d$ is the grain's distance from the sun and $c$ is the speed of light. We parameterise the effective optical cross section of the grain as $\eta_{\mathrm{opt}} \pi r_G^2$, with $\eta_{\mathrm{opt}}$ depending on its optical properties. For the grains we are considering, $\eta_{\mathrm{opt}} \sim 1$, although very small grains have $\eta_{\mathrm{opt}} \ll 1$ because they couple poorly to optical light of much longer wavelengths. The gravitational force on the grain is $G M_{\odot} m_G/d^2$. Their ratio is}

\added{For a spherical grain, the ratio of radiation pressure to gravity is}
\begin{equation}
\beta
=
\frac{3\Qrad L_\odot}
{16\pi G\Msun c\rho_G r_G},
\label{eqn:betagrainsize}
\end{equation}
\added{where \(\Qrad\) parameterises the effective radiation-pressure cross-section as $\Qrad \pi r_G^2$ (cf. \citealt{Draine2011}, chapter 22). If \(\Qrad\) varies slowly across the accepted interval, then \(\beta\propto r_G^{-1}\propto m_G^{-1/3}\):}
\begin{equation}
%\added{\mathrm{d}\ln m_G = -3\,\mathrm{d}\ln\beta.}
\added{|\mathrm{d}\ln m_G| = 3\,\mathrm{d}\ln\beta.}
\label{eqn:betamassdifferential}
\end{equation}

\added{As defined in Equation~\ref{eqn:massfractionlogbin}, \(\eta_m\) is the local mass fraction per unit logarithmic mass interval around the representative grain mass. The actual mass fraction falling within the accepted \(\beta\) window is therefore}
\begin{equation}
\begin{aligned}
\added{\eta_{\beta}(\vinf)} & \added{\equiv \frac{1}{\rhoinf} \int_{\beta_o(\vinf) - \Delta \beta_{\mathrm{max}}(\vinf)}^{\beta_o(\vinf) + \Delta \beta_{\mathrm{max}}(\vinf)} \frac{\mathrm{d}\rhoinf}{\mathrm{d}\beta} \mathrm{d}\beta}\\
 & = \added{3 \eta_m \ln \left(\frac{1 + \Delta \beta_{\mathrm{max}} / \betaOpt}{1 - \Delta \beta_{\mathrm{max}} / \betaOpt}\right) },
\end{aligned}
\end{equation}
\added{where the $\eta_m$ factor, defined in section~\ref{subsec:fluxformalismandconstraintmodelling} equation~\ref{eqn:massfractionlogbin}, supplies the normalization for the distribution of $\beta$. When \(\Delta\beta_{\max}\ll\betaOpt\),}
\begin{equation}
\label{eqn:etaBeta}
\eta_{\beta} \added{\simeq 6 \eta_m} \left(\frac{\Delta \beta_{\mathrm{max}}}{\betaOpt}\right)  = \frac{\changed{4}{12 \eta_m} (\dvEncMax/\vEarth)}{1 + (\vinf/\vEarth)^2} \shortchanged{,}{.}
\end{equation}
The additional factor of two \changed{arising}{arises in the last representation} because $\beta$ may be either smaller or larger than $\betaOpt$. \added{The representative mass satisfying \(\beta(m_G)\simeq\betaOpt(\vinf)\) varies with incoming speed, and the calculation approximates the mass spectrum and optical efficiency as locally smooth over that variation.}

%\added{The permitted interval narrows as $\vinf$ grows, because faster grains require more precise radiative cancellation.}

%\added{As defined in Equation~\ref{eqn:massfractionlogbin}, \(\eta_m\) is the local mass fraction per unit logarithmic mass interval around the representative grain mass. The actual mass fraction falling within the accepted \(\beta\) window is therefore}
%\begin{equation}
%\begin{aligned}
%\eta_\beta(\vinf)
%&=
%3\eta_m
%\ln\!\left(
%\frac{\betaOpt+\Delta\beta_{\max}}
%{\betaOpt-\Delta\beta_{\max}}
%\right)
%\\
%&\simeq
%6\eta_m
%\frac{\Delta\beta_{\max}}{\betaOpt}
%\\
%&=
%\frac{\changed{4}{12\eta_m}(\dvEncMax/\vEarth)}
%{1+(\vinf/\vEarth)^2}.
%\end{aligned}
%\label{eqn:etaBeta}
%\end{equation}
%\added{The last two lines assume \(\Delta\beta_{\max}\ll\betaOpt\). This convention includes \(\eta_m\) in \(\eta_\beta\); it cancels when the final modulation is normalised to the same local logarithmic mass interval. The representative mass satisfying \(\beta(m_G)\simeq\betaOpt(\vinf)\) varies with incoming speed, and the calculation approximates the mass spectrum and optical efficiency as locally smooth over that variation.}

\subsection*{B.3 Tolerance in impact parameter and orbital phase}
\label{subsec:appendix:toleranceinbandthetaearth}

The optimal trajectory requires the Moon to be located exactly at the grain's
perihelion ($\thetaEarth = \thetaOpt$), a condition that is
unlikely to happen in practice. \changed{If $b < \bOpt$ and $\iinf = 0$, the grain trajectory has a}{Consider a coplanar ($\iinf = 0$) trajectory with \(\beta=\betaOpt\) and
\[
b=\bOpt-\Delta b~\text{with}~\Delta b>0.
\]} 
\added{The trajectory reaches} perihelion closer to the Sun and crosses the Earth-Moon's orbit twice. \added{Each of these crossings is an opportunity for these suboptimal trajectories to impact the Moon, but $\Delta \vEnc$ rises with $\Delta b$.}

%For a fixed $\beta$ \added{value of $\betaOpt$}, 
\changed{For a fixed $\beta$}{With $\beta$ fixed to $\betaOpt$}, the grain speed \added{when it crosses Earth orbit} is still \deleted{given by} $\vEnc \added{= \vEarth}$, but now there is a radial component to the grain's velocity during the encounter. This results in a relative encounter speed
\begin{equation}
\Delta \vEnc
= \sqrt{2\vEarth^2 - 2 \vEncTan \vEarth} \added{= \sqrt{2 \vinf \vEarth \frac{\Delta b}{\rEarth}}}.
\end{equation}
\added{where $\vEncTan = \vinf b / \rEarth = \vEarth - \vinf \Delta b/\rEarth$ by angular momentum conservation.}

The difference in \changed{angle}{longitude} between perihelion and these orbit crossings, $\Delta\theta$, is found by using equations 5--6 in \citet{Sterken2012} \added{for radiation pressure-modified Keplerian orbits} to find the angle \changed{of the orbit crossing}{where the grain trajectory crosses $r = \rEarth$}:\footnote{The perihelion angle also changes as $b$ decreases, \added{which could expand the total longitude range where a low velocity encounter can happen,} but it is only a second-order effect.}
\begin{equation}
\rEarth
=
\frac{b^2 \vinf^2}{G \Msun (1 - \betaOpt)}
\frac{1}{1 + \sgn(1-e) \cdot e \cos(\Delta\theta)}
\end{equation}
\added{where the eccentricity is found using equation 7 of \citet{Sterken2012},}
\begin{equation}
e = \sqrt{1 + \left[\frac{\vinf^2 b}{G \Msun (1 - \beta)}\right]^2}.
\end{equation}

%The optimal trajectory also requires the Earth-Moon system to be at the grain's perihelion longitude. Consider a coplanar trajectory with \(\beta=\betaOpt\) and
%\[
%b=\bOpt-\Delta b,
%\qquad
%\Delta b>0.
%\]
%Its speed at \(\rEarth\) remains \(\vEarth\), but its tangential component is
%\begin{equation}
%\vEncTan
%=
%\frac{\vinf b}{\rEarth}
%=
%\vEarth
%-
%\frac{\vinf\Delta b}{\rEarth}.
%\label{eqn:tangentialspeedimpactparameter}
%\end{equation}
%\added{The general relative-speed expression is}
%\begin{equation}
%(\Delta\vEnc)^2
%=
%\changed{\vEarth^2}{\vEnc^2+\vEarth^2}
%-2\vEncTan\vEarth.
%\label{eqn:generalrelativespeed}
%\end{equation}
%\added{Using \(\vEnc=\vEarth\) and Equation~\ref{eqn:tangentialspeedimpactparameter} gives}
%\begin{equation}
%\Delta\vEnc
%=
%\left(
%2\vinf\vEarth
%\frac{\Delta b}{\rEarth}
%\right)^{1/2}.
%\label{eqn:impactparametervelocitymismatch}
%\end{equation}

\deleted{By applying conservation of angular momentum, it can be shown that:}
\[
\deleted{\cos^2 \Delta\theta =
\frac{(\vEncTan^2 - \vEarth^2 (1 - \beta))^2}{\vEarth^4 (1 - \beta)^2 + \vinf^2 \vEncTan^2}.}
\]
\deleted{One can then solve for $\vEncTan$ in terms of $\Delta \theta$, assuming the optimal $\beta$. Generally, for a small $\Delta \vEnc$, the allowed angle is small enough to allow for Taylor-series approximation in $\Delta\theta$, which gives}
\[
\deleted{v_{\mathrm{enc,tan}} \approx \vEarth \left[1 - \frac{\Delta\theta^2}{8} \left(1 + \frac{\vinf^2}{\vEarth^2}\right)^2 \right].}
\]
%\added{The orbit then crosses \(r=\rEarth\) on either side of perihelion. Expanding the radiation-pressure-modified conic equations about \(b=\bOpt\) and the optimal perihelion longitude \citet{Sterken2012} gives, to leading non-zero order,}
\added{Solving for the angular displacement from perihelion, and keeping only terms of quadratic order, we eventually find}
\begin{equation}
\Delta\theta
\simeq
\pm
\frac{2\Delta\vEnc\vEarth}
{\vEarth^2+\vinf^2}.
\label{eqn:angulardeparture}
\end{equation}
%\added{There is no term linear in the impact-parameter displacement because the optimal encounter is an extremum of the tangential-speed match.} 
%\added{Requiring the tangential-speed mismatch to remain below the allowed encounter velocity thus defines a maximum angular deviation.} 
\changed{At any time during the year, only grains from a fraction $\eta_{\theta}$ of ecliptic longitudes have the required slow relative speed at any point in the year. For grains arriving in-plane with optimal $\beta$,}{Requiring \(|\Delta\vEnc|\leq\dvEncMax\) and dividing the resulting accepted longitude interval \(2\Delta\theta_{\max}\) by the full range \(2\pi\) gives}
\begin{equation}
\eta_\theta
\equiv
\frac{2\Delta\theta_{\max}}{2\pi}
\simeq
\frac{2}{\pi}
\frac{\dvEncMax}{\vEarth}
\frac{1}{1+(\vinf/\vEarth)^2}.
\label{eqn:etaTheta}
\end{equation}

\subsection*{B.4 Tolerance in incoming latitude}
\label{subsec:appendix:toleranceiniinfinity}

%\added{For non-zero \(\iinf\), trajectories with \(\beta=\betaOpt\) and \(b=\bOpt\) reach perihelion on a ring of radius \(\rEarth\) around the incoming \(x\)-axis.} 

The geometry is much more complicated when grains arrive off the ecliptic plane. The formerly optimal trajectories no longer even intersect with the Earth-Moon's orbit (unless $\beta = 1$), but there can exist trajectories with $\beta = \betaOpt$ and $b = \bOpt$ that cross the orbit, which is tilted from the perspective of the incoming grains. The trajectories of all grains with impact parameter $\bOpt$ lie along a cylinder that flares out or pinches inward near the Sun. They all reach perihelion at the same $x$ coordinate, \added{$x_{\mathrm{opt}} = \rEarth \cos \thetaOpt$,} with $\theta_{\mathrm{enc}} = \thetaOpt$. Thus the perihelia of these grains form a ring, each point of which is by construction one AU from the Sun\changed{and having $x_{\mathrm{opt}} = \rEarth \cos \thetaOpt$, lying along}{. Parameterising this ring by azimuth \(\phi\), the perihelion position and velocity are}
\begin{align}
\mathbf{r}
&=
\rEarth\bigl[
\cos\thetaOpt\,\xNorm
+
\sin\thetaOpt\cos\phi\,\yNorm
+
\sin\thetaOpt\sin\phi\,\zNorm
\bigr],
\label{eqn:optimalringposition}
\\
\vEncVec
&=
\vEarth\bigl[
-\sin\thetaOpt\,\xNorm
+
\cos\thetaOpt\cos\phi\,\yNorm
+
\cos\thetaOpt\sin\phi\,\zNorm
\bigr].
\label{eqn:optimalringvelocity}
\end{align}
\deleted{with $\phi$ defining an azimuth rolling around the $x$ axis. Because the trajectories still reach perihelion , a grain passing through this locus has no radial velocity, and its velocity vector is \(\vEncVec  = \vEarth \Bigl[ (-\sin \thetaOpt)\,\xNorm + (\cos \thetaOpt \cos \phi)\,\yNorm + (\cos \thetaOpt \sin \phi)\,\zNorm \Bigr]\) .}{The velocity vector is tangential and orthogonal to the radius vector by construction.}
As long as the Earth-Moon's orbit reaches that $x$ coordinate, it must
intersect with this ring; there exists some $\phi$ such that a grain with $\beta = \betaOpt$ and $b = \bOpt$ reaches perihelion at Earth's orbit. \added{However, $\thetaEarth$ is no longer equal to $\thetaOpt$ when $\iinf \ne 0$.} By comparing the coordinates of the Earth-Moon and the grain, the conditions for an intersection are

\changed{ By comparing the coordinates of the Earth-Moon and the grain, the conditions for an intersection are}{Equating the grain and Earth positions gives}
\begin{equation}
\cos\thetaEarth
=
\frac{\cos\thetaOpt}{\cos\iinf},
\qquad
\sin\phi
=
\cot\thetaOpt\tan\iinf.
\label{eqn:ringintersectionconditions}
\end{equation}
\changed{Note that this implies a maximum ecliptic latitude for the incoming grains, given by $|\cos \iinfmax| = |\cos \thetaOpt|$.}{The ring can intersect the Earth orbit only while \(|\cos\iinf|\geq|\cos\thetaOpt|\).}

%\deleted{The earlier form omitted the outer square root required to convert the squared velocity-vector separation into a speed.}
%\added{Subtracting the two equal-magnitude velocity vectors at an intersection yields}
\changed{Using}{Subtracting} the velocity vectors \added{$\vEncVec$ and $\vEarthVec$}, the relative speed is shown to be
\begin{equation}
\Delta\vEnc
=
\vEarth
\left\{
2\left[
1-
\cos\iinf
\sqrt{
1-
\cot^2\thetaOpt\tan^2\iinf
}
\right]
\right\}^{1/2}.
\label{eqn:inclinationvelocitymismatch}
\end{equation}
\added{For \(|\iinf|\ll1\),}
\begin{equation}
\Delta\vEnc
\simeq
\vEarth|\iinf|\csc\thetaOpt.
\label{eqn:inclinationvelocitysmall}
\end{equation}
When $\iinf$ is small, $\Delta \vEnc \approx \vEarth \iinf \csc \thetaOpt$. \changed{The fraction of these grains with a small enough $\iinf$ is then}{Grains in the allowed inclination range, with $|\iinf|$ less than $\added{\iinfmax \simeq} (\dvEncMax/\vEarth) \sin \thetaOpt$, come in near the ecliptic, within an equatorial belt on the celestial sphere. For an isotropic incident distribution, the fraction of the sky in this belt is \(\sin \iinfmax \simeq \iinfmax\):}
\begin{equation}
\label{eqn:etaIncl}
\etaIncl \shortchanged{\approx}{\simeq} 2 \frac{\dvEncMax}{\vEarth} \frac{1}{(\vEarth/\vinf) + (\vinf/\vEarth)} .
\end{equation}

%\added{The accepted incident directions form an equatorial belt of half-width}
%\[
%i_{\infty,\max}
%\simeq
%\frac{\dvEncMax}{\vEarth}\sin\thetaOpt.
%\]
%\added{For an isotropic incident distribution, the fraction of the sky in this belt is \(\sin i_{\infty,\max}\simeq i_{\infty,\max}\). Using Equation~\ref{eqn:thetaoptimal},}
%\begin{equation}
%\eta_i
%\simeq
%2
%\frac{\dvEncMax}{\vEarth}
%\frac{1}
%{(\vEarth/\vinf)+(\vinf/\vEarth)}.
%\label{eqn:etaIncl}
%\end{equation}

\subsection*{B.5 The \changed{flux}{density} of slow-arrival grains in the Earth-Moon system}
\label{subsec:appendix:thefluxofslowmovinggrainsenteringtheearthmoonsystem}
The grains that \changed{fulfill all the criteria}{falling in the slow-channel window of phase space} are drawn from a velocity distribution $f_{\vinf}$. At each incoming speed $\vinf$, \changed{only a small range of $\beta$, $b$, and $\iinf$ have the right kinematics to fall in this window.}{the factors \(\etaBeta\), \(\etaTheta\) and \(\etaIncl\) select the accepted regions in grain mass, orbital phase and incoming direction, while equation~\ref{eqn:CompressionFactor} accounts for the density enhancement near perihelion.} The sampled fraction changes with $\vinf$, so we integrate over the velocity to find the distribution:
\begin{equation}
\nEncSlow = \int_0^{\infty} \xi \etaBeta \etaTheta \etaIncl  \frac{\rhoinf}{\mRef} \frac{\mathrm{d}\shortchanged{\nOpt}{\rhoOpt}}{\mathrm{d}\shortchanged{\ninf}{\rhoinf}} f_{\vinf}(v) \mathrm{d}v,
\label{eqn:nslowgeneral}
\end{equation}
\added{where $\mRef$ is a reference mass typical of grains arriving by the slow channel.\footnote{\added{A more precise calculation would take into the fact that the grain mass with the optimal $\beta$ changes with $v$.}} We assume the three tolerances define an ellipsoidal region in parameter space, and that a deviation in one from the optimal case reduces the allowed tolerances in the other. The $\xi$ factor accounts for this, and we assume it to be $\sim 0.5$, the fraction of volume an ellipsoid fills in its bounding prism.}

%\added{At each incoming speed \(v\), the available technograin mass density is \(\rhoinf f_{\vinf}(v)\,\mathrm{d}v\).} \added{The factors \(\etaBeta\), \(\etaTheta\) and \(\etaIncl\) select the accepted regions in grain mass, orbital phase and incoming direction, while Equation~\ref{eqn:CompressionFactor} supplies the Solar focusing Jacobian.} \added{Because excursions in these coordinates are correlated, we multiply their rectangular product by an order-unity ellipsoidal correction \(\xi\simeq0.5\). The density of slow-arrival grains near the Earth-Moon system is then}
%\begin{equation}
%\nEncSlow
%=
%\int_0^\infty
%\xi\,
%\eta_\beta\eta_\theta\eta_i
%\frac{\rhoinf}{\mRef}
%\frac{\mathrm{d}\rhoOpt}{\mathrm{d}\rhoinf}
% f_{\vinf}(v)\,\mathrm{d}v.
%\label{eqn:nslowgeneral}
%\end{equation}
%\added{Here \(\mRef\) is a representative mass near the optimum \(\beta\) window. This approximation neglects the weak variation of the representative mass across the incoming speed distribution.}

Multiplying Equations~\ref{eqn:etaBeta}, \ref{eqn:etaTheta} and \ref{eqn:etaIncl} gives
%\deleted{The earlier tolerance product used the coefficient \(16/\pi\) and treated the mass-bin normalisation separately.}
%\added{With \(\eta_m\) included in \(\eta_\beta\), the product is}
\begin{equation}
\etaBeta \etaTheta \etaIncl
\simeq
\frac{48\eta_m}{\pi}
\left(\frac{\dvEncMax}{\vEarth}\right)^3
\frac{(\vinf/\vEarth)}
{[1+(\vinf/\vEarth)^2]^3}.
\label{eqn:toleranceproduct}
\end{equation}
\added{The cubic dependence on \(\dvEncMax\) comes from the three independent tolerance directions. Combining this product with Equation~\ref{eqn:CompressionFactor} gives}
%\deleted{The earlier expression was written as a flux with coefficient \(4/\pi\) and number density \(n_\infty\).}
\changed{From the previous derivations}{After applying the grain density enhancement factor, the slow-channel number density is}
\begin{multline}
\shortchanged{\FluxEnc}{\nEncSlow}
\simeq
\frac{\shortchanged{4}{12}}{\pi}\,
\xi\,
\shortchanged{\dvEncMax \ninf}{\eta_m \frac{\rhoinf}{\mRef}}
\left(\frac{\dvEncMax}{\vEarth}\right)^3
\\
\times
\int_0^\infty
\frac{\vEarth/v}{1+(v/\vEarth)^2}
 f_{\vinf}(v)\,\mathrm{d}v.
\label{eqn:nslowintegral}
\end{multline}

\changed{For a Maxwellian distribution of incoming $\vinf$,}{For a Maxwell speed distribution with one-dimensional dispersion \(\sigma_v\), the remaining integral can be evaluated analytically.} Defining
\[
w
=
\frac{\vEarth^2}{2\sigma_v^2},
\]
one obtains
\begin{equation}
\nEncSlow
\simeq
\sqrt{\frac{72}{\pi^3}}\,
\xi\,
\eta_m\frac{\rhoinf}{\mRef} \frac{\dvEncMax^3}{\sigma_v^3}
\exp(w)E_1(w).
\label{eqn:nslowmaxwell}
\end{equation}
where
\changed{$E_1(x)=\int_{-\infty}^{x} e^{x}/x\,\mathrm{d}x$ is the exponential integral.}{the exponential integral is defined by \(E_1(x)=\int_x^\infty \frac{e^{-t}}{t}\,\mathrm{d}t\), for \(x>0\).}
%\deleted{The submitted definition of \(E_1\) had incorrect limits and integrand signs.}
%\added{We use the standard definition}
%\begin{equation}
%E_1(x)
%=
%\int_x^\infty
%\frac{e^{-t}}{t}\,\mathrm{d}t,
%\qquad x>0.
%\label{eqn:exponentialintegraldefinition}
%\end{equation}
\added{The \(\dvEncMax^3\) scaling reflects the three-dimensional volume of the accepted local velocity space.}

\subsection*{B.6 Lunar focussing and the surface flux}
\label{subsec:appendix:lunarfocussingandtheheliosphericmodulationfactor}

\added{Near the Earth-Moon system, the slow-channel grains appear as a roughly isotropic density field with low relative velocity. Within the slow-arrival population, the cumulative local velocity-space volume below relative speed \(v\) scales as \(v^3\).\footnote{\added{The relative velocity $\Delta \vEncVec$ depends on how close to the optimal trajectory each grain comes, not the infalling velocity; grains of all $\iinf$ are mixed into all parts of the $\Delta \vEncVec$ distribution.}} Normalising at \(v=\dvEncMax\) gives}
\begin{equation}
\frac{\mathrm{d}\nEncSlow}{\mathrm{d}v}
=
\frac{3\nEncSlow v^2}{\dvEncMax^3},
\qquad
0\leq v\leq\dvEncMax.
\label{eqn:slowspeeddistribution}
\end{equation}

\deleted{The derivation in Subsection~A.5 yields the flux of slow-moving grains entering the Earth-Moon system.} Once inside \changed{this}{the Earth-Moon} system, additional gravitational focussing by the Moon \changed{modifies both the velocity distribution and the effective cross section for impact.}{enhances the flux by a factor $(1+v_{\mathrm{esc}}^2/v^2)$, where $v_{\mathrm{esc}}=2.38$~km\,s$^{-1}$ is the lunar escape velocity.}

\deleted{Additionally, not all of the grains have a speed of $\dvEncMax$, as assumed for $\FluxEnc$ in equation~22. Accounting for these factors introduces an additional local modulation factor $\eta_{\mathrm{lm}}$:}

\added{The flux incident on the Moon's surface is}
\begin{equation}
    \begin{aligned}
    \FluxRec
    &=
    \frac{1}{4}
    \int_0^{\dvEncMax}
    v
    \left(
    1+\frac{v_{\mathrm{esc}}^2}{v^2}
    \right)
    \frac{\mathrm{d}\nEncSlow}{\mathrm{d}v}
    \,\mathrm{d}v
    \\
    &=
    \frac{3}{16}
    \nEncSlow\dvEncMax
    \left[
    1+
    2\left(
    \frac{v_{\mathrm{esc}}}{\dvEncMax}
    \right)^2
    \right].
    \end{aligned}
    \label{eqn:lunarreceivedflux}
    \end{equation}
\added{The factor \(1/4\) is geometric: for an isotropic population,
    the projected collecting area is one quarter of the total lunar surface
    area. The explicit factor \(v\) converts the differential number density
    into an incident particle flux.}
%\added{The received flux on the Moon is therefore}
%\begin{equation}
%\added{
%\begin{aligned}
%\FluxRec & = \frac{3}{16} \nEncSlow \dvEncMax \left(1 + \frac{2 v_{\mathrm{esc}}^2}{\dvEncMax^2}\right) \\
%         & \approx \frac{9}{4\pi} \xi \frac{\rhoinf}{\bar{m}_G} \frac{\dvEncMax^4}{\vEarth^3} \left(1 + 2 \frac{v_{\mathrm{esc}}^2}{\dvEncMax^2}\right) \int_0^{\infty} \frac{(\vEarth/v)}{1 + (v/\vEarth)^2} f_{\vinf}(v)dv
%\end{aligned}
%}
%\end{equation}

%\added{For an isotropic population with speed \(w\), the mean flux per unit lunar surface area is \(w/4\) times the number density.} \added{Lunar gravity enhances the collision cross-section by \(1+v_{\mathrm{esc}}^2/w^2\), where \(v_{\mathrm{esc}}=2.38\,\kms\). The received flux is therefore}
%\begin{equation}
%\FluxRec
%=
%\frac{1}{4}
%\int_0^{\dvEncMax}
%\added{w}
%\left(
%1+
%\frac{v_{\mathrm{esc}}^2}{w^2}
%\right)
%\frac{\mathrm{d}\nEncSlow}{\mathrm{d}w}
%\,\mathrm{d}w.
%\label{eqn:lunarreceivedfluxintegral}
%\end{equation}
%\added{Substitution of Equation~\ref{eqn:slowspeeddistribution} gives}
%\begin{equation}
%\FluxRec
%=
%\frac{3}{16}
%\nEncSlow\dvEncMax
%\left[
%1+
%2\left(
%\frac{v_{\mathrm{esc}}}{\dvEncMax}
%\right)^2
%\right].
%\label{eqn:lunarreceivedflux}
%\end{equation}

\added{It is useful to define the focusing enhancement relative to the corresponding no-focusing flux as}
\begin{equation}
\eta_{\mathrm{lm}}
\equiv
1+
2\left(
\frac{v_{\mathrm{esc}}}{\dvEncMax}
\right)^2.
\label{eqn:etalunar}
\end{equation}
\added{For \(\dvEncMax=5\,\kms\), \(\eta_{\mathrm{lm}}\simeq1.45\). The separate mean-speed average contributes the factor \(3/4\), so the combined local correction relative to \(\nEncSlow\dvEncMax/4\) is \((3/4)\eta_{\mathrm{lm}}\simeq1.1\).}

\deleted{Combining the slow-arrival fraction derived in Subsection~A.5, the lunar focussing factor $\eta_{\mathrm{lm}}$, and the geometric correction
$\xi\approx0.5$, we define} The heliospheric modulation factor
$\eta_{\mathrm{mod}}$ used throughout the main text \added{is the ratio of the received flux with the expected net flux in the ISM of grains in one logarithmic mass interval:}
\begin{equation}
\added{\eta_{\mathrm{mod}} \equiv \frac{\FluxRec}{(1/4) \eta_m (\rhoinf/\mRef) \Mean{\vinf}}.}
\end{equation}
This quantity encapsulates the probability that an interstellar grain of appropriate size both reaches the inner Solar System and arrives at the lunar surface with a survivable impact speed. \added{It is equal to}
%\added{The reference isotropic flux in interstellar space is}
%\begin{equation}
%\Fluxinf
%=
%\frac{1}{4}\,
%\eta_m\frac{\rhoinf}{\mRef}
%\Mean{\vinf}.
%\label{eqn:interstellarreferencefluxappendix}
%\end{equation}
%\added{Defining \(\eta_{\mathrm{mod}}\equiv\FluxRec/\Fluxinf\) and substituting Equation~\ref{eqn:nslowintegral} yields}
\begin{equation}
\begin{aligned}
\eta_{\mathrm{mod}}
&=
\frac{9}{\pi}\,
\xi\,\eta_{\mathrm{lm}}
\frac{\dvEncMax^4}
{\Mean{\vinf}\vEarth^3} 
\int_0^\infty
\frac{\vEarth/v}{1+(v/\vEarth)^2}
 f_{\vinf}(v)\,\mathrm{d}v.
\end{aligned}
\label{eqn:etamodgeneral}
\end{equation}
\added{The factor \(\eta_m\) cancels because the numerator and denominator refer to the same local logarithmic mass interval. The fourth power of \(\dvEncMax\) comprises the three-dimensional tolerance volume and the additional speed factor required to convert density into flux.}

%\added{For a mono-kinetic incoming population with speed \(v_0\), Equation~\ref{eqn:etamodgeneral} becomes}
%\begin{equation}
%\eta_{\mathrm{mod}}
%=
%\frac{9}{\pi}\,
%\xi\,\eta_{\mathrm{lm}}
%\frac{\dvEncMax^4}{v_0\vEarth^3}
%\frac{\vEarth/v_0}
%{1+(v_0/\vEarth)^2}.
%\label{eqn:etamodmono}
%\end{equation}
%\added{With \(v_0=20\,\kms\), \(\dvEncMax=5\,\kms\), \(\xi=0.5\) and \(v_{\mathrm{esc}}=2.38\,\kms\), this gives \(\eta_{\mathrm{mod}}\simeq2.5\times10^{-3}\), quoted as \(0.003\) in the main text.}

\added{For a Maxwell speed distribution,}
\begin{equation}
\eta_{\mathrm{mod}}
=
\frac{9}{4\pi}\,
\xi\,\eta_{\mathrm{lm}}
\left(
\frac{\dvEncMax}{\sigma_v}
\right)^4
\exp(x_\oplus)E_1(x_\oplus).
\label{eqn:etamodmaxwell}
\end{equation}
%\added{For \(\sigma_v=40\,\kms\), the result is \(\eta_{\mathrm{mod}}\simeq1.6\times10^{-4}\), quoted as \(2\times10^{-4}\). Multiplication by the corresponding mean incoming speeds gives the effective inflow range}
%\[
%\eta_{\mathrm{mod}}\Mean{\vinf}
%\simeq
%0.01\text{--}0.05\,\kms,
%\]
%\added{from which the main text adopts \(0.01\,\kms\) as a conservative value.}

\subsection*{B.7 Additional complications}
\label{subsec:appendix:additionalcomplications}

\changed{This model neglects several factors.}{The derivation is intentionally idealised.} \changed{First, smaller grains are also affected by
the magnetic field of the solar wind.}{It neglects Lorentz forces and time-dependent charging, which become increasingly important for small grains}.
\changed{Second, once grains arrive at the
Earth-Moon system, they will be accelerated and focused by its gravitational
potential, increasing both their velocity and density. The same is true when
they enter the Moon’s potential well.}{It also ignores the eccentricity of the Earth's orbit, the Moon's orbital velocity about the Earth, the evolution of the Moon's orbit over geological time, and terrestrial gravitational perturbations.} \deleted{Additionally, the Moon was closer to the Earth billions of years ago, which would increase the grains’ speed still further.} \changed{Third,}{Additionally} the kinematics change over the Solar System’s history, because the young Sun would have been closer to the local standard of rest, reducing the effective $\sigma_v$\changed{. Fourth,}{, and} the Sun’s luminosity was dimmer in the early Solar
System, so \changed{it probes slightly}{$\betaOpt$ samples} different parts of the grain-size distribution. \added{It also approximates the correlations among the three tolerances with a single factor \(\xi\), treats the local mass spectrum and optical efficiency as smooth while the optimal \(\beta\) varies with \(\vinf\), and assumes an isotropic local velocity distribution inside the accepted slow-arrival region. These effects may change the numerical coefficient and detailed speed distribution, but they do not alter the principal result that the accepted density scales as \(\dvEncMax^3\) and the received flux as \(\dvEncMax^4\). The adopted modulation should therefore be interpreted as an order-of-magnitude kinematic estimate rather than a material-specific impact-survival probability.}

\deleted{These effects define the limits of the idealised derivation above. While not removing these caveats, the subsections that follow clarify how
the resulting slow-arrival fraction should be interpreted under the velocity distributions and heliospheric conditions adopted in the main text.}

%\changed{This model neglects several factors.}{The derivation is intentionally idealised.} \added{It neglects Lorentz forces and time-dependent charging, which become increasingly important for small grains; the eccentricity of the Earth's orbit, the Moon's orbital velocity about the Earth and terrestrial gravitational perturbations; and changes in the Sun's luminosity, velocity relative to the local standard of rest and lunar distance over Solar System history. It also approximates the correlations among the three tolerances with a single factor \(\xi\), treats the local mass spectrum and optical efficiency as smooth while the optimal \(\beta\) varies with \(\vinf\), and assumes an isotropic local velocity distribution inside the accepted slow-arrival region. These effects may change the numerical coefficient and detailed speed distribution, but they do not alter the principal result that the accepted density scales as \(\dvEncMax^3\) and the received flux as \(\dvEncMax^4\). The adopted modulation should therefore be interpreted as an order-of-magnitude kinematic estimate rather than a material-specific impact-survival probability.}

\section*{Appendix C. Notation and symbol definitions}
\addcontentsline{toc}{section}{Appendix C. Notation and symbol definitions}
\label{sec:appendix:notationandsymboldefinitions}

This appendix summarises the principal symbols used throughout this work. Symbols are grouped by physical context for clarity. Unless otherwise stated, quantities are defined in the Galactic, heliocentric or lunar frame appropriate to the surrounding discussion. Local integration variables used only within a single derivation are not repeated here.

% A breakable description-list layout is used rather than tabular material so
% that long definitions can flow naturally across columns and pages.
\newenvironment{symboldefinitions}
{\begin{description}[
  style=multiline,
  leftmargin=0.29\linewidth,
  labelwidth=0.25\linewidth,
  labelsep=0.04\linewidth,
  itemsep=0.35em,
  parsep=0pt,
  topsep=0.35em,
  font=\normalfont
]}
{\end{description}}

\subsection*{C.1 Grain properties}

\begin{symboldefinitions}
\item[\(r_G\)]
Characteristic grain radius. Fiducial values are \(0.1\)--\(1\,\um\), with \(r_G\simeq0.3\,\um\) adopted for numerical estimates involving undirected APs.

\item[\(m_G\)]
Mass of an individual spherical grain of radius \(r_G\) and bulk density \(\rho_G\), \(m_G=(4\pi/3)\rho_G r_G^3\).

\item[\added{\(\mRef\)}]
\added{Reference spherical grain mass, used to convert the mass density in the selected logarithmic mass interval into a grain number density. In the slow-arrival calculation it approximates the mass near the locally optimal \(\beta\) window.}
%\added{Reference spherical grain mass, \(\mRef=(4\pi/3)\rho_G r_G^3\), used to convert the mass density in the selected logarithmic mass interval into a grain number density. In the slow-arrival calculation it approximates the mass near the locally optimal \(\beta\) window.}

\item[\(\rho_G\)]
Bulk density of a grain. The fiducial value is \(\rho_G\simeq3\,\gram\,\cm^{-3}\).

\item[\(\beta\)]
Ratio of Solar radiation pressure to Solar gravitational force acting on a grain, \(\beta=F_{\mathrm{rad}}/F_{\mathrm{grav}}\).

% This is the former \eta_opt, where "Q" is in line with Draine (2011), chapter 22. Additionally, I use "rad" instead of "opt" because "opt" might be confused with optimum.
\item[\added{\(\Qrad\)}]
\added{Dimensionless optical-efficiency factor used to parameterise the effective radiation-pressure cross-section, \(A_{\mathrm{eff}}= \Qrad \times \pi r_G^2\).}
\end{symboldefinitions}

\subsection*{C.2 Interstellar medium and grain transport}

\begin{symboldefinitions}
\item[\added{\(\rhoinf\)}]
\added{Steady-state mass density of the complete technogenic-grain population in the interstellar medium. The number density in the selected logarithmic mass interval is \(\eta_m\rhoinf/\mRef\).}

\item[\(\rho_{\mathrm{ISM}}\)]
Mass density of the ambient interstellar gas.

\item[\added{\(v_{\mathrm{rel}}\)}]
\added{Relative drift speed between a grain and the surrounding interstellar gas.}

\item[\added{\(c_{s,t}\)}]
\added{Isothermal sound speed used in the neutral-gas drag interpolation.}

\item[\(v_\infty\)]
Grain speed at infinity with respect to the Sun, prior to heliospheric interaction.

\item[\(f_{v_\infty}(v)\)]
Normalised distribution of incoming grain speeds at infinity.

\item[\added{\(\sigma_v\)}]
\added{One-dimensional velocity dispersion of an isotropic Maxwellian speed distribution.}

\item[\(\tau_{\mathrm{ISM}}\)]
Mean residence time of grains in the interstellar medium before destruction.

\item[\(t_{\mathrm{drag}}\)]
Gas-drag coupling timescale for grain--ISM interaction.

\item[\(s_{\mathrm{drag}}\)]
Characteristic drag-coupling length, \(s_{\mathrm{drag}}\sim v_{\mathrm{rel}}t_{\mathrm{drag}}\).

\item[\added{\(t_{\mathrm{sput}}\)}]
\added{Characteristic sputtering-erosion timescale of a grain in hot plasma.}
\end{symboldefinitions}

\subsection*{C.3 Galactic source population and release parameters}

\begin{symboldefinitions}
\item[\(n_\star\)]
\changed{Number density of stars in the Solar neighbourhood.}{Total local stellar number density before applying the eligibility factor \(\eta_{\mathrm{ej}}\). Thus \(\eta_{\mathrm{ej}}n_\star\) is the local number density of stars capable of contributing grains in the fiducial slow-arrival size range.}

\item[\(\Gamma_S\)]
\changed{Mean rate of technomaterial-releasing events per eligible star.}{Mean swarm-destruction, or equivalent particulate-release, event rate per eligible star; units of \(\mathrm{events}\,\mathrm{star}^{-1}\,\mathrm{Gyr}^{-1}\).}

\item[\(M_S\)]
Technomaterial mass released in one swarm-destruction or equivalent particulate-release event.

\item[\added{\(M_S\Gamma_S\)}]
\added{Mean technomaterial production rate per eligible star; units of \(\kg\,\mathrm{star}^{-1}\,\mathrm{Gyr}^{-1}\). This is the principal source quantity constrained by the undirected AP analysis.}

\item[\(\eta_{\mathrm{ej}}\)]
Fraction of the total stellar population eligible to eject grains in the size range admitted by the fiducial slow-arrival model.

\item[\(\eta_m\)]
\changed{Fraction of released mass contained in grains of size near \(r_G\).}{Dimensionless fraction of the total technograin mass density contained per unit logarithmic mass interval around \(\mRef\). The fiducial calculation adopts \(\eta_m=0.02\).}

\item[\added{\(N_{\star,\mathrm{MW}}\)}]
\added{Total number of stars in the Milky Way; \(\eta_{\mathrm{ej}}N_{\star,\mathrm{MW}}\) is the number of eligible Galactic source stars.}

\item[\added{\(f_S\)}]
\added{Instantaneous fraction of stars hosting a detectable megaswarm in the waste-heat comparison.}

\item[\added{\(\tau_S\)}]
\added{Characteristic duration for which a megaswarm remains in the detectable state used to define \(f_S\).}

\item[\added{\(t_{\mathrm{MW}}\)}]
\added{Reference interval over which megaswarm production is integrated in the Galactic resource-ceiling estimate.}

\item[\added{\(Z_\star\)}]
\added{Mean stellar heavy-element mass fraction used in the Galactic metal-budget ceiling.}

\item[\added{\(\Mean{m_\star}\)}]
\added{Mean stellar mass used to express the Galactic heavy-element inventory per star.}
\end{symboldefinitions}

\subsection*{C.4 Spatial clustering (``bubble'') parameters}

\begin{symboldefinitions}
\item[\added{\(n_{\star,B}\)}]
\added{Number density of stars belonging to the source population used in the localised bubble model.}

\item[\added{\(\Gamma_B\)}]
\added{Mean bubble-generating release-event rate per source star in the localised model.}

\item[\(R_B\)]
Characteristic radius of a grain-rich region associated with one release event.

\item[\(\tau_B\)]
Lifetime of a grain-rich region before dispersal or destruction.

\item[\(\Phi_B\)]
Mean instantaneous volume filling factor of grain-rich regions.

\item[\(\Mean{N_{B,\odot}}\)]
Expected number of grain-rich regions intercepted by the Solar System over the integration time \(t_{\mathrm{int}}\).

\item[\added{\(f_R(R)\)}]
\added{Normalised distribution of bubble radii in Appendix~A.}

\item[\added{\(f_v(v)\)}]
\added{Normalised distribution of Solar-System--bubble relative speeds in Appendix~A.}

\item[\added{\(R_{\mathrm{eff}}\)}]
\added{Effective radius \(R_{\mathrm{eff}}=\Mean{R^3}/\Mean{R^2}\) used to compare the finite-volume and swept-path terms in the exact bubble-interception expression.}

\item[\added{\(R_{\mathrm{tr}}\)}]
\added{Transition radius at which the finite-volume and swept-path contributions to the bubble-interception count are equal.}
\end{symboldefinitions}

\subsection*{C.5 Heliocentric and lunar encounter parameters}

\begin{symboldefinitions}
\item[\added{\(r_\oplus\)}]
\added{Heliocentric radius of the circular Earth-Moon orbit adopted in Appendix~B, \(r_\oplus=1\,\mathrm{AU}\).}

\item[\(v_\oplus\)]
Orbital speed of the Earth-Moon system around the Sun, \(v_\oplus\simeq30\,\kms\).

\item[\added{\(v_{\mathrm{enc}}\)}]
\added{Heliocentric grain speed when the trajectory reaches the Earth-Moon orbital radius.}

\item[\(\Delta v_{\mathrm{enc}}\)]
Relative speed between the grain and the Earth-Moon system at encounter.

\item[\(\dvEncMax\)]
\changed{Previously denoted \(\Delta v_{\mathrm{max}}\): the maximum encounter speed compatible with partial or intact survival, \(\simeq5\,\kms\).}{Maximum relative encounter speed admitted by the slow-arrival phase-space calculation. The fiducial value is \(5\,\kms\); it is a kinematic threshold rather than a universal material-independent survival boundary.}

\item[\added{\(\betaOpt\)}]
\added{Radiation-pressure ratio that makes the grain's heliocentric speed at \(1\,\mathrm{AU}\) equal to \(v_\oplus\) for a specified \(v_\infty\).}

\item[\added{\(\bOpt\)}]
\added{Impact parameter for the optimal slow-arrival trajectory, \(\bOpt=r_\oplus v_\oplus/v_\infty\).}

\item[\added{\(\thetaOpt\)}]
\added{Downstream angle of the optimal perihelion in the incoming-grain coordinate system.}

\item[\added{\(\iinf\)}]
\added{Ecliptic latitude of the incoming grain direction in Appendix~B.}

\item[\added{\(\etaBeta\)}]
\added{Fraction of the full technograin mass density admitted by the radiation-pressure tolerance. Under the adopted convention it includes the local logarithmic mass fraction \(\eta_m\).}

\item[\added{\(\etaTheta\)}]
\added{Fraction of orbital longitudes admitted by the orbital-phase tolerance.}

\item[\added{\(\etaIncl\)}]
\added{Fraction of incoming directions admitted by the inclination tolerance.}

\item[\added{\(\nEncSlow\)}]
\added{Number density of grains in the accepted slow-arrival region near the Earth-Moon system.}

\item[\(v_{\mathrm{esc}}\)]
Lunar escape speed, \(2.38\,\kms\).

\item[\(\eta_{\mathrm{lm}}\)]
\changed{Enhancement factor due to lunar gravitational focussing.}{Lunar gravitational-focusing enhancement averaged over the adopted slow-speed distribution, \(\eta_{\mathrm{lm}}=1+2(v_{\mathrm{esc}}/\dvEncMax)^2\). For \(\dvEncMax=5\,\kms\), \(\eta_{\mathrm{lm}}\simeq1.45\); the separate mean-speed factor \(3/4\) gives a combined local correction of \(\simeq1.1\).}

\item[\(\xi\)]
Order-unity geometric correction for the correlated, approximately ellipsoidal slow-arrival tolerance volume; the fiducial value is \(\xi\simeq0.5\).

\added\item[\added{\(\Fluxinf\)}]
\added{Reference isotropic interstellar flux in the selected logarithmic
grain-mass interval.}
\par\nobreak\smallskip
\noindent
\added{\(\displaystyle
\Fluxinf
=
\frac{\eta_m\rhoinf}{4\mRef}\,
\Mean{v_\infty}.
\)}

\item[\added{\(\FluxRec\)}]
\added{Physical flux of slow-arrival grains received per unit lunar surface area before the analytical completeness factor is applied.}

\item[\(\eta_{\mathrm{mod}}\)]
\changed{Heliocentric modulation factor incorporating radiation-pressure deceleration, velocity filtering and gravitational focussing.}{Ratio \(\FluxRec/\Fluxinf\) of the lunar received flux to the reference isotropic interstellar flux in the same logarithmic grain-mass interval. It incorporates Solar trajectory filtering, the three phase-space tolerances, the correction \(\xi\), and lunar gravitational focussing; \(\eta_m\) cancels from this ratio.}

\item[\(\eta_{\mathrm{mod}}\Mean{v_\infty}\)]
Effective inflow speed entering the lunar flux after slow-arrival phase-space weighting and lunar modulation. The fiducial constraint adopts \(0.01\,\kms\).

\item[\added{\(E_1(x)\)}]
\added{Exponential integral \(E_1(x)=\int_x^\infty e^{-t}t^{-1}\,\mathrm{d}t\), for \(x>0\).}
\end{symboldefinitions}

\subsection*{C.6 Regolith sampling and detection}

\begin{symboldefinitions}
\item[\(A_s\)]
Area of regolith sampled; the fiducial column has \(A_s=1\,\meter^2\).

\item[\(z_s\)]
Depth of the sampled regolith column.

\item[\(t_{\mathrm{int}}\)]
Physical integration interval over which grains may accumulate in the lunar archive.

\item[\(\Lambda(t)\)]
Characteristic gardening depth reached by material of age \(t\).

\item[\(q\)]
Power-law exponent in the statistical gardening relation \(\Lambda(t)\propto t^q\).

\item[\added{\(S(t)\)}]
\added{Recognisable post-depositional survival fraction for grains deposited a time \(t\) ago. The baseline mixing calculation adopts \(S(t)=1\) and treats unmodelled losses of recognisability through \(\eta_{\mathrm{comp}}\), avoiding double counting.}

\item[\added{\(f_z(z\mid t)\)}]
\added{Probability density for the present depth \(z\) of material deposited a time \(t\) ago.}

\item[\added{\(n_{\mathrm{reg}}(z)\)}]
\added{Present grain number density at regolith depth \(z\).}

\item[\added{\(\Sigma_{\mathrm{reg}}(z_s)\)}]
\added{Number of retained grains per unit sampled area integrated from the surface to depth \(z_s\).}

\item[\(t_s(z_s)\)]
Effective sampling time corresponding to regolith mixing down to depth \(z_s\).

\item[\(F_r\)]
\changed{Flux of recognisable grains onto the lunar surface.}{Generic constant incident flux used in the regolith-mixing derivation. When the model is applied to the slow-arrival population, \(F_r\) is identified with the physical received flux \(\FluxRec\).}

% BCL 2026-08-13: I moved these next two (\Fluxinf, \FluxRef) to C.5
%\item[\added{\(\Fluxinf\)}]
%\added{Reference isotropic interstellar flux in the selected logarithmic grain-mass interval, \(\Fluxinf=(1/4)\eta_m(\rhoinf/\mRef)\Mean{v_\infty}\).}

%\item[\added{\(\FluxRec\)}]
%\added{Physical flux of slow-arrival grains received per unit lunar surface area before the analytical completeness factor is applied.}

\item[\added{\(\eta_{\mathrm{comp}}\)}]
\added{Aggregate completeness factor: the fraction of physically delivered grains or relics that retains a recognisable technogenic signature and is successfully identified.}

\item[\added{\(\eta_{\mathrm{surv}}\)}]
\added{Fraction of delivered material retaining at least one diagnostic technogenic signature after impact and subsequent processing.}

\item[\added{\(\eta_{\mathrm{rec}}\)}]
\added{Probability that a surviving diagnostic signature remains distinguishable from natural and anthropogenic backgrounds.}

\item[\added{\(\eta_{\mathrm{det}}\)}]
\added{Probability that the adopted analytical workflow identifies a recognisable signature. Conceptually, \(\eta_{\mathrm{comp}}=\eta_{\mathrm{surv}}\eta_{\mathrm{rec}}\eta_{\mathrm{det}}\).}

\item[\(\Mean{N_{\mathrm{obs}}}\)]
Expected number of recognised technogenic grains in a sampled regolith volume.

\item[\(\bar N\)]
\changed{Maximum value of \(\Mean{N_{\mathrm{obs}}}\) consistent with a null detection; nominally taken to be \(3.0\).}{Poisson upper limit applied to a null result. At 95\% confidence, \(\bar N=-\ln(0.05)=2.996\simeq3\).}
\end{symboldefinitions}

\subsection*{C.7 Directed-delivery parameters}

\begin{symboldefinitions}
\item[\(M_V\)]
Mass of particulate technomaterial released during one visitation or delivery event.

\item[\(\Gamma_V\)]
Mean rate of visitation or delivery events that release particulate technomaterial.

\item[\(M_{\mathrm{dep}}\)]
\changed{Mass of technogenic material deposited in recognisable form per visitation.}{Mass from one event physically deposited and retained in the lunar regolith after transport and impact-retention losses, before recognisable-survival and analytical-completeness factors are applied.}

\item[\added{\(\eta_{\mathrm{dep}}\)}]
\added{Physical deposition efficiency, \(\eta_{\mathrm{dep}}=M_{\mathrm{dep}}/M_V\). Recognisable-signature survival, recognition and analytical detection are treated separately through \(\eta_{\mathrm{comp}}\).}

\item[\(A_{\mathrm{Moon}}\)]
Total surface area of the Moon.

\item[\(A_P\)]
Area of the lunar-surface patch seeded during one visitation.

\item[\added{\(A_{\mathrm{dep}}\)}]
\added{Total lunar surface area covered by one or more deposited patches during the integration interval. For sparse, weakly overlapping patches, \(A_{\mathrm{dep}}\simeq\Gamma_Vt_{\mathrm{int}}A_P\).}

\item[\(C_P\)]
\changed{Mean number of patches overlapping a single point on the Moon's surface.}{Mean number of deposited patches covering a specified point, \(C_P=\Gamma_Vt_{\mathrm{int}}A_P/A_{\mathrm{Moon}}\). For \(C_P\ll1\), it is also approximately the probability that a random location lies within deposited material.}

\item[\added{\(f_{\mathrm{cov}}\)}]
\added{Fraction of the lunar surface covered by at least one deposited patch; for statistically independent placement, \(f_{\mathrm{cov}}=1-\exp(-C_P)\).}

\item[\added{\(N_L\)}]
\added{Number of spatially independent lunar sampling locations in a blind patch-interception search.}

\item[\added{\(p_{\mathrm{hit}}\)}]
\added{Probability that one blind sampling location intersects deposited material; for sparse coverage, \(p_{\mathrm{hit}}\simeq A_{\mathrm{dep}}/A_{\mathrm{Moon}}\).}

\item[\added{\(p_{\geq1}\)}]
\added{Probability that at least one of \(N_L\) independent locations intersects a patch, \(p_{\geq1}=1-\exp(-N_LC_P)\).}

\item[\added{\(\eta_z\)}]
\added{Fraction of particles deposited in a known patch that lies within the sampled depth after regolith mixing.}

\item[\added{\(\eta_{\mathrm{hit}}(R)\)}]
\added{Fraction of particles released at heliocentric distance \(R\) that ultimately intersects the Moon.}

\item[\added{\(\eta_{\mathrm{ret}}\)}]
\added{Fraction of impacting mass physically retained in the lunar regolith. For remote release, \(\eta_{\mathrm{dep}}(R)=\eta_{\mathrm{hit}}(R)\eta_{\mathrm{ret}}\).}

\item[\added{\(\zeta_{\mathrm{BP}}\)}]
\added{Maximum allowed fraction of the globally mixed regolith mass composed of relic Bracewell Particles.}

\item[\added{\(\rho_{\mathrm{reg}}\)}]
\added{Bulk density of the lunar regolith used in the global mass-consistency ceiling.}

\item[\added{\(\Lambda_{\mathrm{reg}}\)}]
\added{Representative depth of the globally mixed regolith layer used in the mass-consistency ceiling.}

\item[\added{\(t_{\mathrm{reg}}\)}]
\added{Accumulation interval used in the global regolith-mass consistency ceiling.}

\item[\added{\(t_{\mathrm{obs}}\)}]
\added{Specified interval over which an arriving visitation would be expected to have been recognised independently of grain recovery.}
\end{symboldefinitions}

All numerical values adopted in the main text are intended as order-of-magnitude fiducial choices rather than precise measurements, except where a value is explicitly identified as a measured quantity, a published empirical fit or an exact geometrical or statistical consequence. The constraint framework is readily rescaled for alternative assumptions about grain size, material properties, source populations, ISM residence times, delivery efficiencies, analytical completeness or sampled regolith volume.

%%%%%%%%%%%% Biography text %%%%%%%%%%%%%%%%%%%%%%%%%%

% Biography text
%\lpbio

% Second author (no photo yet)

%\blbio

%\iacbio

%\apsbio

\end{document}